\documentclass{aa} 

\usepackage[varg]{txfonts}
\usepackage{natbib}
\usepackage{graphicx}
\usepackage[most]{tcolorbox}
\usepackage{float}
\usepackage{stfloats}
\usepackage{caption} % subfigure over more pages
\usepackage{subcaption}
\usepackage{mwe}
\usepackage{mathtools}
\usepackage{booktabs} %Tables
\usepackage{multicol} %Tables
\usepackage{multirow} %Tables
\usepackage{array} %Equations
\usepackage{mathabx} %Symbols
\usepackage{arydshln}
\usepackage{verbatim} 
\usepackage{lscape} % landscape page
\usepackage{pdflscape} % landscape page
\usepackage[export]{adjustbox} %Load graphicx
\usepackage{color}
\usepackage{epstopdf} %Include eps 
\usepackage{amsmath}
\newcommand{\RN}[1]{\textup{\uppercase\expandafter{\romannumeral#1}}}

\begin{document}

\title{Low levels of methanol deuteration in the high-mass star-forming region NGC 6334\RN{1}}

%\titlerunning{}

%\subtitle{} 

\author{Eva G. B\o gelund\inst{\ref{inst1}}
  \and Brett A. McGuire\inst{\ref{inst2}}
  \and Niels F. W. Ligterink\inst{\ref{inst1}, \ref{inst3}}
  \and Vianney Taquet\inst{\ref{inst4}}
  \and Crystal L. Brogan\inst{\ref{inst2}}
  \and  Todd R. Hunter\inst{\ref{inst2}}
  \and John C. Pearson\inst{\ref{inst5}}
  \and Michiel R. Hogerheijde\inst{\ref{inst1}, \ref{inst6}}
  \and Ewine F. van Dishoeck\inst{\ref{inst1}, \ref{inst7}}
    } 

%\offprints{}

\institute{Leiden Observatory, Leiden University, PO Box 9513, 2300
  RA Leiden, The Netherlands\label{inst1} \newline \email{bogelund@strw.leidenuniv.nl}
  \and National Radio Astronomy Observatory, 520 Edgemont Rd, Charlottesville, VA 22903, USA\label{inst2}
  \and Sackler Laboratory for Astrophysics, Leiden Observatory, Leiden University, PO Box 9513, 2300 RA Leiden, The Netherlands\label{inst3}
  \and INAF, Osservatorio Astrofisico di Arcetri, Largo E. Fermi 5, 50125 Firenze, Italy \label{inst4}
  \and Jet Propulsion Laboratory, 4800 Oak Grove Drive, Pasadena, CA 91109\label{inst5}, USA
  \and Anton Pannekoek Institute for Astronomy, University of Amsterdam, Science Park 904, 1098 XH Amsterdam, The Netherlands\label{inst6}
  \and Max-Planck Institut für Extraterrestrische Physik, Giessenbachstr. 1, 85748 Garching, Germany\label{inst7}
} 

\date{Submitted: 02/02/2018 / Accepted}

\abstract
{The abundance of deuterated molecules in a star-forming region is sensitive to the environment in which they are formed. Deuteration fractions, i.e, the ratio of a species containing D to its hydrogenated counterpart, therefore provide a powerful tool for studying the physical and chemical evolution of a star-forming system. While local low-mass star-forming regions show very high deuteration ratios, much lower fractions are observed towards Orion and the Galactic Centre. Astration of deuterium has been suggested as a possible cause for low deuteration in the Galactic Centre.}
{We derive methanol deuteration fractions at a number of locations towards the high-mass star-forming region NGC 6334\RN{1}, located at a mean distance of 1.3 kpc, and discuss how these can shed light on the conditions prevailing during its formation.} 
{We use high sensitivity, high spatial and spectral resolution observations obtained with the Atacama Large Millimeter/submillimeter Array to study transitions of the less abundant, optically thin, methanol-isotopologues: $^{13}$CH$_3$OH, CH$_3^{18}$OH, CH$_2$DOH and CH$_3$OD, detected towards NGC 6334\RN{1}. Assuming LTE and excitation temperatures of $\sim$120--330 K, we derive column densities for each of the species and use these to infer CH$_2$DOH/CH$_3$OH and CH$_3$OD/CH$_3$OH fractions.}
{We derive column densities in a range of (0.8--8.3)$\times$10$^{17}$ cm$^{-2}$ for $^{13}$CH$_3$OH, (0.13--3.4)$\times$10$^{17}$ cm$^{-2}$ for CH$_3^{18}$OH, (0.03--1.63)$\times$10$^{17}$ cm$^{-2}$ for CH$_2$DOH and (0.15--5.5)$\times$10$^{17}$ cm$^{-2}$ for CH$_3$OD in a $\sim$1$^{\second}$ beam. Interestingly, the column densities of CH$_3$OD are consistently higher than those of CH$_2$DOH throughout the region by factors of 2--15. We calculate the CH$_2$DOH/CH$_3$OH and CH$_3$OD/CH$_3$OH ratios for each of the sampled locations in NGC 6334\RN{1}. These values range from 0.03\% to 0.34\% for CH$_2$DOH and from 0.27\% to 1.07\% for CH$_3$OD if we use the $^{13}$C isotope of methanol as a standard; using the $^{18}$O-methanol as a standard, decreases the ratios by factors 2--3.}
{All regions studied in this work show CH$_2$DOH/CH$_3$OH as well as CH$_2$DOH/CH$_3$OD ratios that are considerably lower than those derived towards low-mass star-forming regions and slightly lower than those derived for the high-mass star-forming regions in Orion and the Galactic Centre. The low ratios indicate a grain surface temperature during formation $\sim$30 K, for which the efficiency of the formation of deuterated species is significantly reduced. Therefore, astration of deuterium in the Galactic Centre cannot be the explanation for its low deuteration ratio but rather the high temperatures characterising the region. 
}

\keywords{Astrochemistry - Methods: observational - Stars: protostars - ISM: individual objects: NGC 6334\RN{1} - Submillimeter: ISM}
 
\maketitle

\section{Introduction} \label{sec:introduction}
The abundance of deuterium (D) formed in the Big Bang sets the primordial D/H ratio in the universe. As stars form and start processing D in their interiors, the deuterium abundance should drop if no other source of D exists. The best estimate of the cosmic D/H ratio is therefore obtained by observing environments with little star formation and chemical processing such as the diffuse interstellar medium (ISM) for which D/H is $\sim$(1.5--2.0)$\times$10$^{-5}$ \citep{Linsky2003, Prodanovic2010}. On the other hand, environments with high star-formation rates would be expected to show lower D/H fractionation ratios as a consequence of astration, i.e., the processing of D in stellar interiors, and a generally higher temperature. This however, contradicts observations of molecules in both high and low-mass star-forming regions that not only show D/H ratios which are orders of magnitude higher than that of the ISM, but also display multiply deuterated species. An example of a source exhibiting such high deuterium fractionation is the well-studied low-mass protostellar binary IRAS 16293--2422 (hereafter IRAS 16293) \citep{vanDishoek1995, Ceccarelli1998}. This source is especially interesting because it was the first source towards which both doubly as well as triply-deuterated methanol was detected \citep{Parise2002, Parise2004}. %(at levels of 0.2 and 0.014 respectively)
More recently, IRAS 16293 has been studied by \cite{Jorgensen2017} who have characterised the isotope composition of a number of complex organic molecules. They derived D/H ratios, i.e., the column density ratio of isotopologues with respect to their hydrogenated counterparts including the statistical correction for the location of the substituted deuterium, for all detected species in the range 2--8$\%$. Specifically, the D/H ratio of methanol (CH$_3$OH) is found to be $\sim$2$\%$. In addition, the detection of singly and doubly-deuterated formaldehyde (H$_2$CO) and methanol towards seven other low-mass protostars, with deuteration fractions similar to those observed towards IRAS 16293, is reported by \cite{Parise2006}. For high-mass star-forming regions the most thorough studies of deuterated species have been carried out towards the hot cores located in the Galactic Centre and the Orion Complex. For Sagittarius B2 (Sgr B2), \cite{Belloche2016} report D/H ratios of 0.38$\%$ for  acetonitrile (CH$_3$CN) and (0.05-0.12)$\%$ for the tentative detections of CH$_3$CH$_2$CN, HC$_3$N and CH$_3$OH. Approximately the same low levels of deuteration were found by \cite{Neill2013}, who studied ammonia (NH$_3$), formaldehyde and methanol towards Orion KL and find D/H $\sim$(0.2--0.8)$\%$ towards the Compact Ridge and Hot Core regions. Similar levels of deuterium in methanol are found by \cite{Peng2012}. 

While these observations clearly illustrate that deuterated species, including CH$_3$OH, are enhanced in star-forming regions, the explanation of how the various ratios came to be remains incomplete. For the low-mass objects, the high deuteration ratios can, for the most part, be explained by gas-grain astrochemical models where high densities and low dust temperatures allow simple deuterated species to build up rapidly in precursor dark cores \citep{Taquet2012}. In the case of Sgr B2 however, the low deuteration fractions are attributed to the combined effects of astration and a less efficient deuteration process at the high temperatures characterising the Galactic Centre. In order to quantify which of these processes influence the deuteration fractionation more, observations of other high-mass star-forming regions, located away from the Galactic Centre, are essential. 

Deuterium fractionation of simple interstellar molecular species was first studied in detail by \cite{Watson1976} who argued that the large D/H ratios, especially present in DCN and DCO$^+$, can be understood as a result of ion-molecule exchange reactions in the gas and the difference in zero-point vibrational energies of the hydrogen versus deuterium-containing molecules. The observed deuterated species are then the end product of a chain of reactions starting with the formation of H$_2$D$^+$ from H$_3^+$ and HD, reacting with neutral molecules to form deuterated ions and subsequently recombining with electrons. This recombination also results in enhanced atomic D/H in the gas at low temperatures, which increases even more when CO, the main destroyer of H$_3^+$ and H$_2$D$^+$, is frozen out on grains \citep{Roberts2003}. As pointed out by \cite{Tielens1983}, this enhanced atomic D/H in the gas can be transferred to those molecules that are primarily formed by hydrogenation on grain surfaces. Key examples are H$_2$CO and CH$_3$OH, which both result from the hydrogenation of CO. Thus, the abundances of these deuterated molecules are good tracers of the physical conditions in the gas. 

The process of deuteration at low temperatures has also been studied in the laboratory. In particular, \cite{Nagaoka2005} have shown that grain-surface H-D abstraction-addition reactions in solid methanol can account for the methanol D/H ratios derived from observations. They also note that if the gaseous atomic D/H ratio is higher than 0.1, deuterated methanol may be formed directly through successive hydrogenation/deuteration of CO.

The focus of this work is on methanol, especially its isotopologues $^{13}$CH$_3$OH, CH$_3$$^{18}$OH and the two single deuterated species CH$_2$DOH and CH$_3$OD. As described above, methanol is formed primarily on the surface of dust grains \citep{Watanabe2002,Geppert2005,Fuchs2009} and therefore presents a good tracer of the chemistry of interstellar ice. In addition, if formed in environments with high gaseous atomic D/H ratios, methanol will be deuterated. Deuteration levels can therefore be considered a fossil record of the chemical composition not only of the ice, but also of the gas, characterising the region in which it was formed, with the highest ratios associated with the lowest temperatures. However, in order to deduce accurate deuteration ratios, it is critical that the transitions used to derive column densities are optically thin. If this is not the case, deuterium fractions are likely to be overestimated since the lines of the more abundant hydrogenated species are generally optically thick, resulting in underestimated values. To ensure transitions are optically thin, one generally needs to target lines that are weak, i.e., have low Einstein $A_{\textrm{ij}}$ values. Observations of such weak lines have however presented a challenge since line surveys have, for the most part, been conducted using single dish telescopes which are, by and large, less sensitive when compared with interferometric observations. These observations have therefore mostly targeted the brightest lines which may in many cases be optically thick and therefore result in deuteration ratios which are higher than is actually the case. With the unique sensitivity and resolving power offered by the Atacama Large Millimeter/submillimeter Array (ALMA), this is changing. With ALMA it has become possible to probe molecular transitions that are both weaker and are emitted from objects less bright than ever before. Consequently, the field of molecular line surveys and analysis has entered a new epoch where column density determinations no longer, or to a much lesser extent, suffer from misinterpretation due to the increased access to transitions in the optically thin regime. 

In this work we focus on NGC 6334\RN{1}, a region of active high-mass star-formation in the giant molecular cloud complex NGC 6334, also referred to as the "Cat's Paw Nebula". The complex is located in the Scorpius constellation in the southern hemisphere. The region constitutes six main dense cores, identified as discrete continuum sources in the far-infrared, labelled with Roman numerals \RN{1} - \RN{6} \citep{McBreen1979}. Later, an additional, very low temperature source, NGC 6334\RN{1}(N) was identified $\sim$2$'$ north of NGC 6334\RN{1} \citep{Gezari1982}. The distance to the NGC 6334 site has commonly been cited as (1.7 $\pm$ 0.3) kpc \citep{Russeil2012} but recent work on H$_2$O and CH$_3$OH masers associated with the star-forming complex, carried out by \cite{Reid2014} and \cite{Chibueze2014}, place the region closer, at distances of 1.34 and 1.26 kpc respectively. Here we will assume a mean distance of 1.3 kpc, corresponding to a galactocentric distance, D$_{\textrm{GC}}$, of $\sim$7.02 kpc. With a synthesised beam of the observations of 1$\overset{\second}{.}$00$\times$0$\overset{\second}{.}$74, this allows us to probe  NGC 6334\RN{1} at scales of $\sim$1300 AU. NGC 6334\RN{1} is complex both in structure and composition and very rich in molecular lines \citep{McCutcheon2000}, but has the great advantage of lines with widths of only $\sim$3 km s$^{-1}$ (as compared to $\sim$20 km s$^{-1}$ for Sgr B2), reducing the problem of line confusion considerably. The molecular diversity of the region is demonstrated by \cite{Zernickel2012} who use the \textit{Herschel} Space Observatory and Submillimeter Array (SMA) to investigate $\sim$4300 emission and absorption lines belonging to 46 different molecules and their isotopologues.

NGC 6334\RN{1} is comprised of multiple hot cores, some of which are themselves multiples. This multiplicity was shown by \cite{Brogan2016} who recently presented an in-depth, high-resolution continuum study of the morphology of the site. In addition to the four previously known millimetre sources associated with the region, labelled MM1-4 and identified by \cite{Hunter2006}, \cite{Brogan2016} identify six new sources: five at millimetre wavelengths, MM5-9, and one at centimetre wavelengths, CM2, all within a radius of $\sim$10$^{\second}$ corresponding to 0.06 pc at the distance of the site. Another centimetre source, CM1, is identified 18$^{\second}$ ($\sim$0.1 pc) west of the main cluster. In addition to molecular lines and the far-IR and radio continuum sources, indicative of cool, dense and dusty cores, NGC 6334\RN{1} displays a variety of other star-formation tracers including ultra-compact H\RN{2} regions and outflows \citep[see e.g.][and references therein]{Persi1996, Leurini2006, Qiu2011}.

The paper is structured in the following way: Section \ref{sec:observations} introduces the observations, the data reduction process and the analysis method. In Section \ref{sec:results} the process of identifying and fitting individual lines and estimating molecular column densities is presented. Section \ref{sec:ratios} discusses the deuteration levels derived and compare these to those obtained for other objects as well as the predictions from astrochemical models. Finally, Section \ref{sec:conclusion} summarises our findings. 

\section{Observations and analysis method} \label{sec:observations}
\subsection{Observations}
The observations have been described in detail in \cite{McGuire2017}, while the analysis and reduction procedures are described in detail in \cite{Brogan2016} and \cite{Hunter2017}, and as such only a brief discussion is presented here. Two sets of observations are used. The primary dataset (ALMA Project 2015.1.00150.S) was taken in Cycle 3 at $\sim$1$\arcsec$ resolution ($\sim$1300 AU at the distance of NGC 6334\RN{1}), centered on 302 GHz, and covering $\sim$3 GHz of total bandwidth with an rms noise of $\sim$500 mK. The second dataset (ALMA Project 2015.A.00022.T) was also observed in Cycle 3 but at $\sim$0$\overset{\second}{.}$2 (260 AU) resolution, centered around four frequency windows at approximately 281, 293, 338, and 350 GHz each $\sim$3.75 GHz wide, and with an rms noise of $\sim$620 mK and $\sim$900 mK in the lower and higher frequency ranges, respectively.  In each case, the most line-free continuum channels were selected and used for both self calibration and continuum subtraction

%The baseline is corrected after baseline subtracting using a linear extrapolation. 

\subsection{Analysis method} \label{subsec:method}
Nine spectra are extracted from the data cube. Each spectrum represents the average of a 1$\overset{\second}{.}$00$\times$0$\overset{\second}{.}$74 region, equivalent to the area of the synthesised beam. The coordinates of the central pixel of each of the regions are summarised in Table \ref{tab:regions} and the locations shown in Fig. \ref{fig:Map}. The positions are chosen so they present a fairly homogeneous sampling of the continuum sources MM1, located in the northern part of NGC 6334\RN{1}, MM2, located in the western part of the region, and MM3, located in the southern part of the region, as well as their surroundings.

%--------BEGIN TABLE: REGIONS ---------------
\begin{sidewaystable}[]%[htbp] %[here,top,bottom,page]
	\centering
	\caption{Summary of regions and column densities}
	\label{tab:regions}
	\begin{tabular}{lcccccclll}
		\toprule
		Region & \multicolumn{2}{c}{Location (J2000)} & $v_{\textrm{LSR}}$ & FWHM & $T_{\textrm{ex}}$\tablefootmark{a} & \multicolumn{4}{c}{$N_{\textrm{s}}$} \\
		\cline{2-3} 
		\cline{7-10} 
		& R.A. & Decl. & & & & $^{13}$CH$_3$OH & CH$_3^{18}$OH & CH$_2$DOH\tablefootmark{b} & CH$_3$OD \\
		& & & [km s$^{-1}$] & [km s$^{-1}$] & [K] & [$\times$10$^{17}$ cm$^{-2}$] & [$\times$10$^{17}$ cm$^{-2}$] & [$\times$10$^{17}$ cm$^{-2}$] & [$\times$10$^{17}$ cm$^{-2}$] \\
		\midrule
		MM1 \RN{1} & 17:20:53.437 & -35:46:57.902 & -5.5 & 4.0 & 336 & 8.3 & 2.0--3.4 & 0.56--1.63 & 5.5 \\
		MM1 \RN{2} & 17:20:53.371 & -35:46:57.013 & -6.7 & 3.0 & 215 & 7.4 & 2.0--2.4 & 0.25--0.79 & 3.8 \\		
		MM1 \RN{3} & 17:20:53.397 & -35:46:59.209 & -8.3 & 3.0 & 157 & 8.3 & 1.7 & 0.38--1.10 & 2.3 \\
		MM1 \RN{4} & 17:20:53.381 & -35:46:56.315 & -6.7 & 3.0 & 195 & 5.2 & 1.5 & 0.13--0.50 & 1.7 \\
		MM1 \RN{5} & 17:20:53.552 & -35:46:57.415 & -5.2 & 2.8 & 174 & 1.3 & 0.41 & 0.08--0.17 & 0.43 \\
		MM2 \RN{1} & 17:20:53.165 & -35:46:59.231 & -9.0 & 3.5 & 164 & 6.6 & 0.8--1.5 & 0.45--1.38 & 1.8 \\
		MM2 \RN{2} & 17:20:53.202 & -35:46:57.613 & -7.0 & 2.8 & 143 & 1.8 & 0.32 & 0.04--0.09 & 0.4 \\
		MM3 \RN{1} & 17:20:53.417 & -35:47:00.697 & -9.0 & 3.0 & 122 & 0.9 & 0.14 & 0.03--0.06 & 0.15 \\
		MM3 \RN{2} & 17:20:53.365 & -35:47:01.541 & -9.8 & 2.5 & 122 & 0.8 & 0.13 & 0.03--0.06 & 0.16 \\
		\bottomrule
	\end{tabular}
	\tablefoot{Range of $N_{\textrm{s}}$ corresponds to the range in column density, within the synthesised beam of 1$\overset{\second}{.}$00$\times$0$\overset{\second}{.}$74, with and without blending. \tablefoottext{a}{Excitation temperature of the best-fit $^{13}$CH$_3$OH model.} \tablefoottext{b}{Numbers include the vibrational correction factor of 1.25 as well as the uncertainty on the line strength (see Section \ref{subsec:method}).}}
\end{sidewaystable}
%-------------------END TABLE---------------------------

%--------------BEGIN FIGURE: MAP OF REGION -------------------------------
\begin{figure*}
	\centering
	\includegraphics[width=0.68\textwidth]{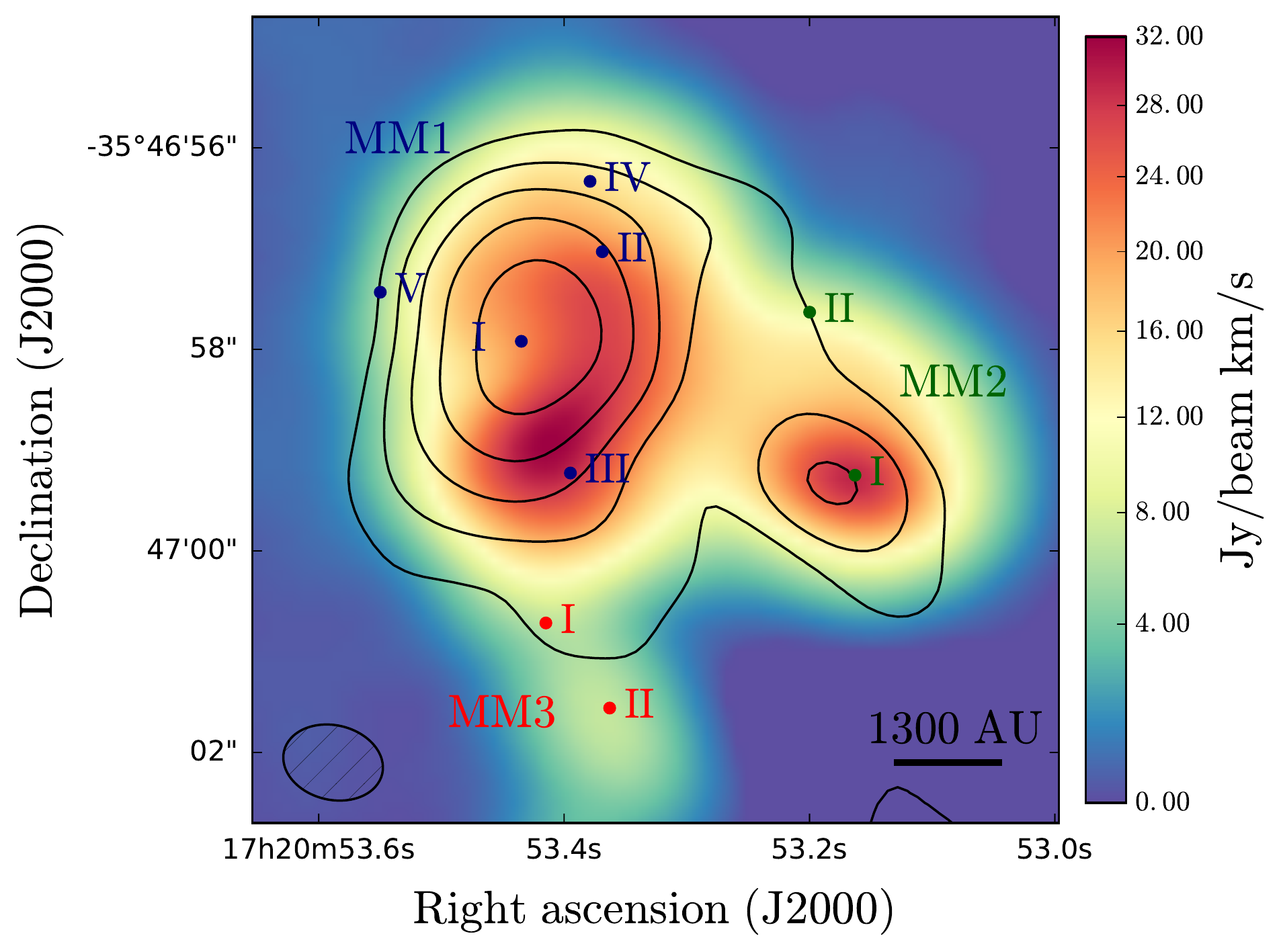}
	\caption[]{Velocity integrated intensity map of the $^{13}$CH$_3$OH transition at 303.692 GHz with the 1 mm continuum image overlaid in black contours (levels are [10, 20, 40, 80, 160]$\sigma$ with $\sigma$=0.02 Jy/beam). The locations at which spectra have been extracted are marked in blue for MM1, green for MM2 and red for MM3. The synthesized beam ($\sim$1300$\times$962 AU) is shown in the bottom left corner.}
	\label{fig:Map}
	\addtocounter{figure}{1}
\end{figure*}
%--------------END FIGURE--------------------------------------

For our line analysis, the software package CASSIS\footnote{http://cassis.irap.omp.eu} is used. CASSIS is a tool for spectral line identification and analysis which reads a list of line data, including rest frequency, $\nu$, upper state energy, $E_{\textrm{up}}$, and Einstein $A_{ij}$ coefficients. Once a line has been identified, the line analysis tool is used to produce a synthetic spectrum by providing CASSIS with a number of parameters: excitation temperature, $T_{\textrm{ex}}$ [K], column density of the species, $N_{\textrm{s}}$ [cm$^{-2}$], source velocity, $v_{\textrm{LSR}}$ [km s$^{-1}$], line width, FWHM [km s$^{-1}$], and angular size of the emitting region (which is assumed to be equal to the area of the synthesised beam), $\theta_{\textrm{s}}$ [$^{\second}$]. For line input the Jet Propulsion Laboratory (JPL\footnote{http://spec.jpl.nasa.gov}) and Cologne Database for Molecular Spectroscopy (CDMS\footnote{http://www.ph1.uni-koeln.de/cdms/}) databases are used. 

For the analysis of $^{13}$CH$_3$OH and CH$_3^{18}$OH, the CDMS database is used. These entries include the ground and first vibrational states for $^{13}$CH$_3$OH and the ground, first and second vibrational states for CH$_3^{18}$OH. For the analysis of CH$_2$DOH the JPL datatbase is used. The line intensities in this list are based on the method described by \cite{Pearson2012}. The JPL catalog warns that extreme caution should be taken in determining columns (or concentrations) of CH$_2$DOH directly from b-type and c-type transitions as significant errors can occur, while the a-type transitions are more reliable\footnote{https://spec.jpl.nasa.gov/ftp/pub/catalog/doc/d033004.pdf}. The symmetry of the transitions are defined as follows: $\Delta$K$_{\textrm{a}}$ = even, $\Delta$K$_{\textrm{c}}$ = odd for a-type, $\Delta$K$_{\textrm{a}}$ = odd, $\Delta$K$_{\textrm{c}}$ = odd for b-type and $\Delta$K$_{\textrm{a}}$ = odd, $\Delta$K$_{\textrm{c}}$ = even for c-type \citep{Pearson2012}. We optimise our CH$_2$DOH fit to the 12$_{(2,11)}$$\rightarrow$12$_{(1,12)}$ b-type transition at 301.514 GHz (see Sec. \ref{subsec:D}) and use an updated value for the base 10 logarithm of the integrated line intensity at 300 K for this transition of -3.84$\pm$0.2 nm$^2$ MHz (John Pearson, private communication). Also, since the CH$_2$DOH entry only includes the ground vibrational state, a vibrational correction factor of 1.25 has been applied to all listed values (Holger M{\"u}ller, private communication). To model CH$_3$OD we use the summary of CH$_3$OD frequencies reported by \cite{Walsh2000}, the line strengths reported by \cite{Anderson1988} and \cite{Anderson1993} and the partition function of CH$_3^{18}$OH (from the CDMS database) which includes the vibrational levels. 

Because CH$_3$OH is a very abundant species in star-forming regions, the lines of the primary $^{12}$C-isotope are optically thick and cannot be used to derive column densities and excitation temperatures. Fortunately, many lines of the isotopologues $^{13}$CH$_3$OH, CH$_3^{18}$OH, CH$_2$DOH and CH$_3$OD remain in the optically thin regime, making column density determination of these species possible. We identify a total of fifteen transitions belonging to the methanol isotopologues in the covered frequency range: eight of these belong to $^{13}$CH$_3$OH, four to CH$_3^{18}$OH, one to CH$_2$DOH and two to CH$_3$OD. The transitions are summarised in Table \ref{tab:line_summary}. Since the frequency range covered is limited to $\sim$3 GHz (301.180--304.165 GHz) only a few transitions of each species are covered. This complicates the definite identification of lines, especially in the case of CH$_3^{18}$OH, CH$_2$DOH and CH$_3$OD for which many lines are either very weak or blended to an extent that it is impossible to distinguish the contribution from individual molecular species. 

In order to verify the presence of these species in NGC 6334\RN{1}, a second set of data, covering a larger range of frequencies, is also consulted (see Section \ref{sec:observations}). Based on transitions in these data the presence of the less abundant CH$_3$OH-isotopologues is confirmed. However, since these data are of higher angular resolution ($\sim$0$\overset{\second}{.}$2) emission is resolved out and we are not able to constrain the column density of either of the deuterated species better than from our primary data set. The upper limits for CH$_2$DOH (which are similar for a- and b-type transitions covered) and CH$_3$OD as well as the column densities for $^{13}$CH$_3$OH and CH$_3^{18}$OH based on these data are consistent with the values derived based on our primary data set. The data are presented in Appendix \ref{app:Brogan} but will not be discussed further in the main manuscript.

For each methanol line candidate we carefully examine transitions belonging to other species (for which the spectroscopy is known and listed in either of the databases mentioned above) at similar frequencies to ensure lines are not incorrectly assigned. In the first step, all species which have not previously been detected in space are excluded. Secondly, species are excluded based on Einstein $A_{ij}$ coefficient and upper state energy. For the remaining blending candidates we investigate if any additional transitions are covered in the data range and if so, whether these can provide additional constraints. If the blending candidates are isotopologues, we search for transitions belonging to the parent species and ensure that column densities are consistent. The process of line assignment and analysis of potential blended transitions will be discussed in detail in Sections \ref{subsec:13C} - \ref{subsec:CH3OD}.

Assuming local thermodynamic equilibrium (LTE) and optically thin lines synthetic spectra are created for each methanol isotopologue. Firstly, the excitation temperatures and $^{13}$CH$_3$OH column densities for each region are derived by creating a grid of models, with $T_{\textrm{ex}}$ ranging between 50 and 350 K and $N$ ranging between 5$\times$10$^{16}$ and 5$\times$10$^{19}$ cm$^{-2}$, and selecting the model with the minimal $\chi^2$ as the best fit. The $^{13}$CH$_3$OH lines are fitted first because these lines are the most numerous and span the largest range of upper state energies. Secondly, the column densities of the remaining methanol isotopologues are optimised keeping the excitation temperature fixed at the value of the best-fit $^{13}$C-methanol model. The transitions of each methanol isotopologue, as well as the blending species, are modelled separately, and summed to obtain a full spectrum for each of the nine regions. These spectra are shown in Fig. \ref{fig:fullMM1} - \ref{fig:fullMM3}. 

Finally, to estimate the $^{12}$CH$_3^{16}$OH column density used to calculate the deuterium fraction for each spectral region, a $^{12}$C/$^{13}$C ratio of 62 and a $^{16}$O/$^{18}$O ratio of 450 is adopted, both derived assuming D$_{\textrm{GC}}$ $\sim$7.02 kpc and the relations for $^{12}$C/$^{13}$C and $^{16}$O/$^{18}$O reported by \cite{Milam2005} and \cite{Wilson1999} respectively.

%--------BEGIN TABLE: LINE SUMMARY---------------
\begin{table*}[]%[htbp] %[here,top,bottom,page]
	\centering
	\caption{Summary of detected lines}
	\label{tab:line_summary}
	\begin{tabular}{lllcccc}
		\toprule
		Species & \multicolumn{2}{c}{Transition} & Frequency & $E_{\textrm{up}}$ & $A_{\textrm{ij}}$ & Database\\
		\cline{2-3}
		& [QN]$_{\textrm{up}}$\tablefootmark{a} & [QN]$_{\textrm{low}}$\tablefootmark{a} & [MHz]  & [K] & $\times$10$^{-5}$ [s$^{-1}$] & \\  
		\midrule
		% CH$_3$OH & & & &  \\
		$^{13}$CH$_3$OH  
		& 18 2 16 1 & 17 1 17 1 & 301 238.558 & 684.91 & 8.96 & CDMS \\
		& 20 3 18 0 & 19 4 15 0 & 301 272.475 & 525.44 & 4.89 & \\
		& 14 -1 14 0 & 13 2 11 0 & 302 166.269 & 243.11 & 0.12 & \\
		& 10 0 10 0 & 9 1 8 0 & 302 590.285 & 137.53 & 6.91 & \\
		& 20 3 17 0 & 19 4 16 0 & 302 882.003 & 525.52 & 4.99 & \\
		& 7 1 6 0 & 6 2 4 0 & 303 319.623 & 84.49 & 4.27 & \\
		& 1 1 0 0 & 1 0 1 0 & 303 692.682 & 16.84 & 32.2 & \\
		& 15 -3 13 0 & 16 0 16 0 & 303 865.391 & 334.68 & 0.10 & \\
		\midrule
		CH$_3^{18}$OH 
		& 3 1 2 0 & 3 0 3 0 & 301 279.428 & 27.81 & 30.6 & CDMS \\
		& 4 1 3 0 & 4 0 4 0  &  302 848.743 & 36.78 &  30.9 &  \\
		& 16 1 16 1 & 15 2 14 1 & 303 016.300 & 307.63 & 4.14 & \\
		& 3 1 2 2 & 2 0 2 1 &  303 855.874 & 34.10 & 8.09 & \\
		\midrule
		CH$_2$DOH 
		& 12 2 11 0 & 12 1 12 0 & 301 514.152 & 183.10 & 8.31 & JPL \\
		\midrule
		CH$_3$OD 
		& 7 2 + 0 & 7 1 - 0 & 303 296.120 & 82.73 & 18.2 & \tablefootmark{b} \\
		& 7 4 -- 0 & 8 3 -- 0 & 303 904.827 & 130.79 & 2.82 & \\
		\bottomrule
	\end{tabular}
	\tablefoot{\tablefoottext{a}{Quantum numbers for $^{13}$CH$_3$OH, CH$_3^{18}$OH and CH$_2$DOH are (J K$_\textrm{a}$ K$_\textrm{c}$ v) and quantum numbers for CH$_3$OD are (J K P v) where v=0, 1, 2 refers to the three sub-states $e_0$, $e_1$ and $o_1$ of the ground state respectively.} \tablefoottext{b}{\cite{Walsh2000}, and references therein.}}
\end{table*}
%-------------------END TABLE---------------------------

\section{Results} \label{sec:results}
\subsection{$^{13}$CH$_3$OH} \label{subsec:13C}
Eight transitions of $^{13}$CH$_3$OH are detected towards NGC 6334\RN{1}. For each of these, a synthetic spectrum is created and optimised simultaneously to obtain the best-fit values for $T_{\textrm{ex}}$ and $N$. However, the same range of frequencies which host the $^{13}$C-methanol lines are also known to be occupied by a number of methyl formate (CH$_3$OCHO) transitions. Especially the $^{13}$CH$_3$OH transitions at 302.590, 302.882, 303.319 and 303.865 GHz are overlapping with transitions of CH$_3$OCHO. Fortunately, since many transitions of this species are covered in the spectral range, the column density of CH$_3$OCHO can be constrained to a range of (0.6--2.5)$\times$10$^{17}$ cm$^{-2}$ in our beam for all regions. While the contribution from CH$_3$OCHO to the $^{13}$CH$_3$OH line fits does not change the value of the best-fit $^{13}$CH$_3$OH column density, likely because the upper state energy of these transitions are high, $\sim$700 K, as compared with those of the affected $^{13}$CH$_3$OH lines, it is included for completeness. Figure \ref{fig:13CAll} shows the synthetic spectra of the best fit models to $^{13}$CH$_3$OH, CH$_3$OCHO and the combination of the two for the transition at 302.590 GHz detected towards each of the regions. Despite this transition having the largest contribution from CH$_3$OCHO, it is evident that at the abundances of CH$_3$OCHO present in NCG 6334\RN{1}, its effect on the $^{13}$CH$_3$OH line is small and it is therefore reasonable to assume that the observed peak in the data at 302.590 GHz is due mainly to $^{13}$CH$_3$OH. 

The best-fit column density and excitation temperature for $^{13}$CH$_3$OH for each of the regions are listed in Table \ref{tab:regions}. The values range about an order of magnitude, with the lowest values associated with MM3 \RN{1} and \RN{2}, (0.8--0.9)$\times$10$^{17}$ cm$^{-2}$, while the highest, $\sim$8.3$\times$10$^{18}$ cm$^{-2}$, are found in MM1 \RN{1} and \RN{3}. For the remaining regions the values span a range of (1.3--7.4)$\times$10$^{18}$ cm$^{-2}$.

Zoom-ins of the spectra showing the remaining $^{13}$CH$_3$OH lines can be found in Appendix \ref{app:13C}. Generally the single-$T_{\textrm{ex}}$, single-$N$ models reproduce the data well. It should be noted however, that the transition with the lowest $A_{\textrm{ij}}$ $\sim$10$^{-6}$ s$^{-1}$ at 303.865 GHz is consistently under produced by the models as compared with the data by factors $\sim$3--6. The same is true for the line at 302.166 GHz towards the regions around MM1. The reason for this could either be uncertainties in the spectroscopic values for these particular transitions, or that the lines are blended with some unknown species, the spectroscopy of which is not included in either of the JPL or CDMS databases. Alternatively, the explanation could be that some of the transitions which we assume to be optically thin are in fact slightly thick, resulting in underestimated column densities. To test this, we optimise the column density and excitation temperature to the transitions at 302.166 and 303.865 GH. Doing so, we find that for the modelled spectra to reproduce the data at these frequencies, the column densities need to be higher by factors of 5--10, as compared to the best-fit column density when fitting to all lines, resulting in the remaining transitions being largely saturated. To improve the fits the individual sources would need to be modelled using an excitation model taking both density and temperature gradients into account. Such a full excitation model would potentially make the saturated lines appear more gaussian.

\subsection{CH$_3^{18}$OH} \label{subsec:18O}
Assuming a $^{12}$C/$^{13}$C ratio of 62 and a $^{16}$O/$^{18}$O ratio of 450 implies that the $^{13}$CH$_3$OH/CH$_3^{18}$OH column density ratio is a factor $\sim$7. Adopting this ratio and using the best-fit column density values for $^{13}$CH$_3$OH, the majority of the modelled CH$_3^{18}$OH lines appear weaker as compared with the data by about a factor of two. Indeed, when optimising the CH$_3^{18}$OH column density, assuming $T_{\textrm{ex}}$ to be the same as for $^{13}$CH$_3$OH, the values are only factors of 2--3 lower than those of $^{13}$CH$_3$OH. This result suggests that the $^{13}$CH$_3$OH lines, which are assumed to be optically thin, are in fact partially optically thick, and therefore the derived column densities may be slightly underestimated, as discussed above.      

As in the case of $^{13}$CH$_3$OH, the CH$_3^{18}$OH lines are also partly blended. Especially the line at 303.016 GHz overlaps with the transition from another molecular species: O$^{13}$CS, although the slight shift in frequency of the O$^{13}$CS line with respect to the line of CH$_3^{18}$OH means that the observed peak in the data cannot be due purely to O$^{13}$CS. Unfortunately no other lines of O$^{13}$CS are covered in our frequency range and likewise only one line of the parent species, OCS, is in the data range. It is therefore difficult to constrain the contribution of O$^{13}$CS to the blend from the data itself. Instead the column density for OCS derived by \cite{Zernickel2012} of 1.2$\times$10$^{18}$ cm$^{-2}$ (assuming a source size of 2.5${^{\second}}$) at $T_{\textrm{ex}}$ = 100 K and the $^{12}$C/$^{13}$C ratio of 62 is used to estimate a column density of O$^{13}$CS in NGC 6334\RN{1} of 1.9$\times$10$^{16}$ cm$^{-2}$. The modelled spectra of this transition as well as the data are shown in Fig. \ref{fig:18OAll}. A column density of O$^{13}$CS of 1.9$\times$10$^{16}$ cm$^{-2}$ is also consistent with the data in Appendix \ref{app:Brogan}; however, while the O$^{13}$CS lines covered here are not overproduced at this column density, they cannot be better constrained due to blends. To derive the column density of CH$_3^{18}$OH two sets of fits are preformed: the first excludes the O$^{13}$CS-blended line and optimises the column density to the remaining three transitions (for which no known species listed in the databases contribute significantly to the data peaks), this value is used to set an upper limit for the column density. In the second fit, a contribution from O$^{13}$CS is included and the CH$_3^{18}$OH column density is optimised to all lines. The O$^{13}$CS contribution is kept constant for all regions.

The column densities derived for CH$_3^{18}$OH are listed in Table \ref{tab:regions}. For most regions, the difference between the fit purely considering CH$_3^{18}$OH and the fit of CH$_3^{18}$OH with a contribution from O$^{13}$CS, is less than a factor of two. For regions MM1 \RN{3}-\RN{5}, MM2 \RN{2} and MM3 however, the O$^{13}$CS column density derived from \cite{Zernickel2012} results in modelled spectra that overshoot the data at the specific frequency. For these regions only the value for the pure-CH$_3^{18}$OH fit is reported. As is the case of $^{13}$CH$_3$OH, the derived CH$_3^{18}$OH column densities are lowest in regions MM3 \RN{1} and \RN{2}, with values between (1.3--1.4)$\times$10$^{16}$ cm$^{-2}$. Slightly higher values, (0.32--1.5)$\times$10$^{17}$ cm$^{-2}$ and (0.41--2.4)$\times$10$^{17}$ cm$^{-2}$, are derived for regions MM2 \RN{1} and \RN{2} and MM1 \RN{2}-\RN{5} respectively. The highest value is again associated with region MM1 \RN{1}, (2.0--3.4)$\times$10$^{17}$ cm$^{-2}$. Zoom-ins of all detected CH$_3^{18}$OH transitions can be seen in Appendix \ref{app:18O}. Again, the single density and temperature models reproduce the data well, keeping in mind that the main contributor to the peak at 303.016 GHz is O$^{13}$CS.  

\subsection{CH$_2$DOH} \label{subsec:D} 
For CH$_2$DOH, the 12$_{(2,11)}$$\rightarrow$12$_{(1,12)}$ transition at 301.514 GHz is detected. Unfortunately, this line too is blended. We constrain the column density of the blending species, CH$_3$NC, based on its characteristic double feature around 301.53 GHz, and derive values in the range (1.1-2.0)$\times$10$^{14}$ cm$^{-2}$ for the regions MM1 \RN{1}-\RN{4} and MM2 \RN{1}. As in the case of O$^{13}$CS, these column densities are consistent with the data in Appendix \ref{app:Brogan} but cannot be further constrained. For the remaining regions, MM1 \RN{5}, MM2 \RN{2} and MM3 \RN{1}-\RN{2}, the data do not display any signs of CH$_3$NC. Figure \ref{fig:DAll} shows the modelled spectra of CH$_2$DOH, both with and without the contribution from the blending species, CH$_3$NC, as well as the sum of the spectra of CH$_2$DOH with blending and CH$_3$NC. For the regions where CH$_3$NC is detected, the difference between the values of the column density of CH$_2$DOH fitted with and without the contribution from CH$_3$NC is a factor $\sim$3. No transitions of the more common isomer CH$_3$CN are covered by the data, so the ratio of -CN/-NC cannot be constrained.

The column densities derived for CH$_2$DOH are listed in Table \ref{tab:regions}. Note that these include the vibrational correction and uncertainty on the line strength as discussed in Section \ref{subsec:method}. As regions MM1 \RN{5}, MM2 \RN{2} and MM3 \RN{1}-\RN{2} show no clear CH$_3$NC feature, only the value for the pure CH$_2$DOH fit is reported. As in the case of $^{13}$CH$_3$OH and CH$_3^{18}$OH, the lowest CH$_2$DOH column densities are detected towards regions MM3 \RN{1} and \RN{2}, both with values in the range of (3.0--6.0)$\times$10$^{15}$ cm$^{-2}$. Region MM2 \RN{2} also displays low values, (4.0--9.0)$\times$10$^{15}$ cm$^{-2}$, while the values detected towards MM2 \RN{1} are fairly high, (0.45--1.38)$\times$10$^{17}$ cm$^{-2}$. The highest column density is again detected towards MM1 \RN{1}, (056--1.63)$\times$10$^{17}$ cm$^{-2}$, with the remain regions extracted from the area around MM1 show column densities spanning a range of about an order of magnitude, (0.08--1.10)$\times$10$^{17}$ cm$^{-2}$.

\subsection{CH$_3$OD}\label{subsec:CH3OD}
In addition to the transitions of CH$_2$DOH, we have searched for lines belonging to CH$_3$OD, for which two transitions, one at 303.296 GHz and one at 303.904 GHz \citep[see summary by][]{Walsh2000}, are covered. To derive a column density for this isotopologue, our fits were optimised to the transition at 303.296 GHz. No other transitions in either the JPL nor the CDMS databases are listed at this frequency so the line in the data is considered to be purely due to CH$_3$OD. The modelled spectra and data are shown in Fig. \ref{fig:ODAll} and Appendix \ref{app:OD} and the derived column densities are listed in Table \ref{tab:regions}. 

As in the case of CH$_2$DOH, the lowest column densities of CH$_3$OD are detected towards the regions MM3 \RN{1} and \RN{2}, with values in the range of (1.5--1.6)$\times$10$^{16}$ cm$^{-2}$. Regions MM2 \RN{2} and MM1 \RN{5} have similar column densities of 4.0 and 4.3$\times$10$^{16}$ cm$^{-2}$ respectively. For the remaining regions, column densities in the range of (1.7--5.5)$\times$10$^{17}$ cm$^{-2}$ are derived. The highest value is again associated with region MM1 \RN{1}. It is very interesting to note that the column densities derived for CH$_3$OD are consistently higher than those derived for CH$_2$DOH in all regions. 

%--------------BEGIN FIGURE: 18O - ALL REGIONS - OPTIMISED LINE -------------------------------
\begin{figure*}[]
	\rotatebox{90}{\begin{minipage}{0.48\textheight}
			\includegraphics[width=1.\textwidth, trim={0 1cm 0 1cm}, clip]{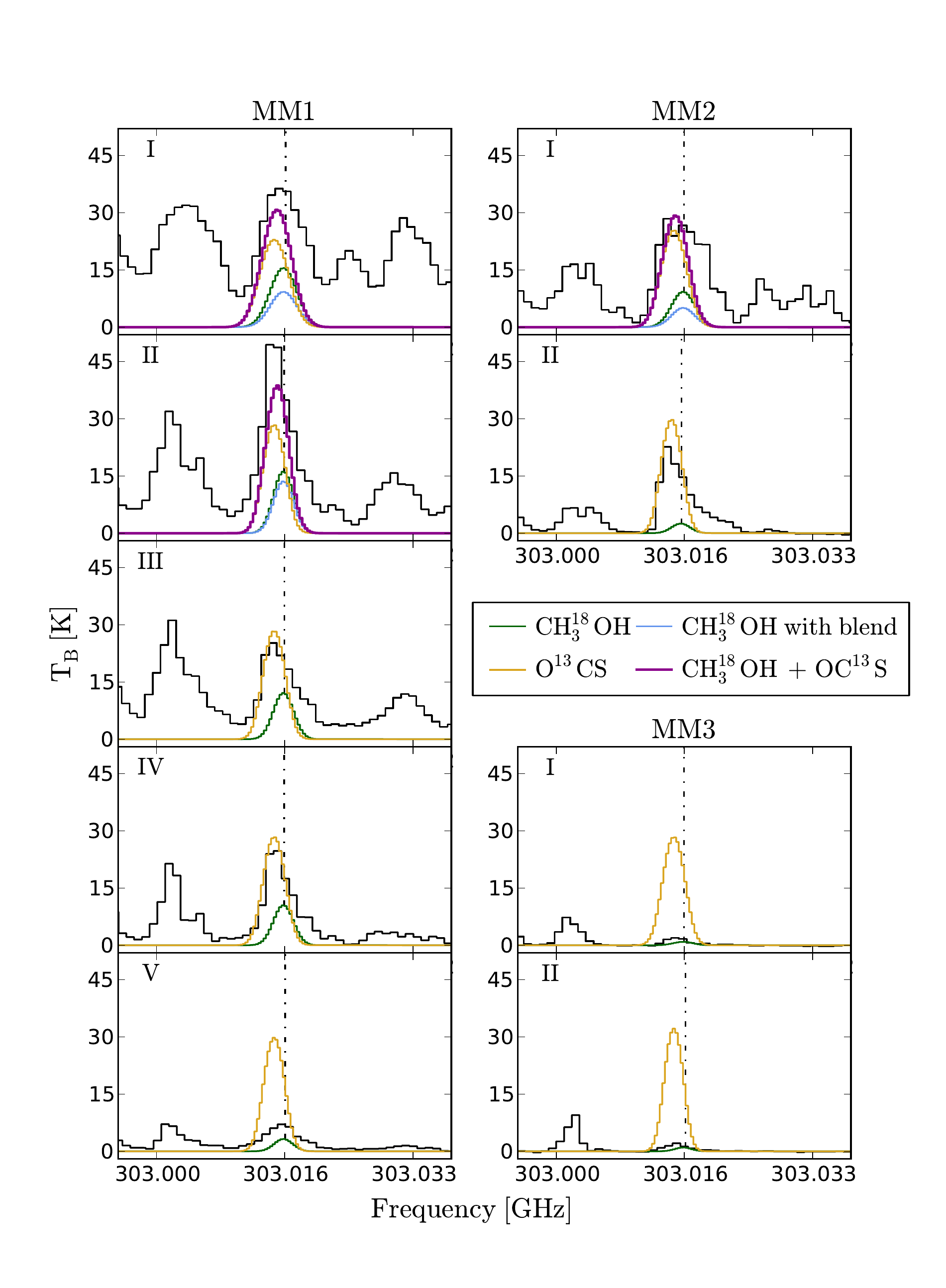} % trim={<left> <lower> <right> <upper>}
			\captionof{figure}{CH$_3^{18}$OH 16$_{(1,16)}$$\rightarrow$15$_{(2,14)}$ transition at 303.016 GHz (indicated by dash-dotted lines) detected towards each region. Frequencies are shifted to the rest frame of the individual regions. Blue and green lines represent the modelled spectra of CH$_3^{18}$OH with and without blending, i.e., including and excluding the contribution from O$^{13}$CS, respectively. Yellow lines represent the modelled spectra of O$^{13}$CS, assuming a fixed column density of 1.9$\times$10$^{16}$ cm$^{-2}$ for all regions, and magenta lines represent the sum of the spectra of CH$_3^{18}$OH with blending and O$^{13}$CS.}
			\label{fig:18OAll}
			\addtocounter{figure}{-2}
	\end{minipage}}	
\end{figure*}
%--------------END FIGURE--------------------------------------

%--------------BEGIN FIGURE: 13C - ALL REGIONS - OPTIMISED LINE -------------------------------
\begin{figure*}[]
	\rotatebox{90}{\begin{minipage}{0.48\textheight}
			\includegraphics[width=1.\textwidth, trim={0 1cm 0 1cm}, clip]{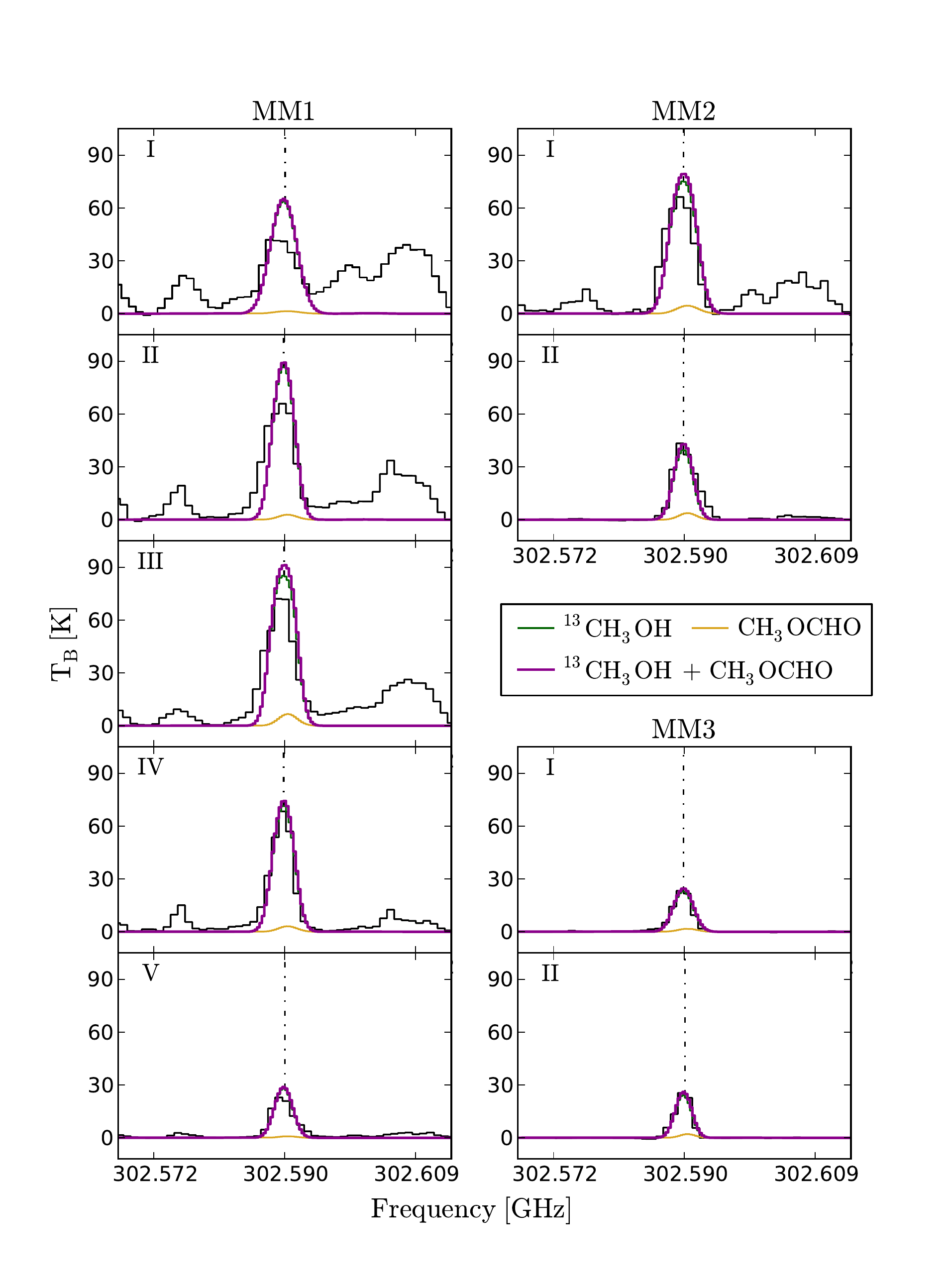} % trim={<left> <lower> <right> <upper>}
			\captionof{figure}{$^{13}$CH$_3$OH 10$_{(0,10)}$$\rightarrow$9$_{(1,8)}$ transition at 302.590 GHz (indicated by dash-dotted lines) detected towards each region. Frequencies are shifted to the rest frame of the individual regions. Green lines represent the modelled spectra of $^{13}$CH$_3$OH without blending, i.e., excluding the contribution from CH$_3$OCHO (including the contribution from CH$_3$OCHO does not change the $^{13}$CH$_3$OH column density of the best-fit model). Yellow lines represent the modelled spectra of CH$_3$OCHO and magenta lines represent the sum of the $^{13}$CH$_3$OH and CH$_3$OCHO spectra.}
			\label{fig:13CAll}
			\addtocounter{figure}{2}
	\end{minipage}}
\end{figure*}
%--------------END FIGURE--------------------------------------

%--------------BEGIN FIGURE: OD - ALL REGIONS - OPTIMISED LINE -------------------------------
\begin{figure*}
	\rotatebox{90}{\begin{minipage}{0.48\textheight}
			\includegraphics[width=1.\textwidth, trim={0 1cm 0 1cm}, clip]{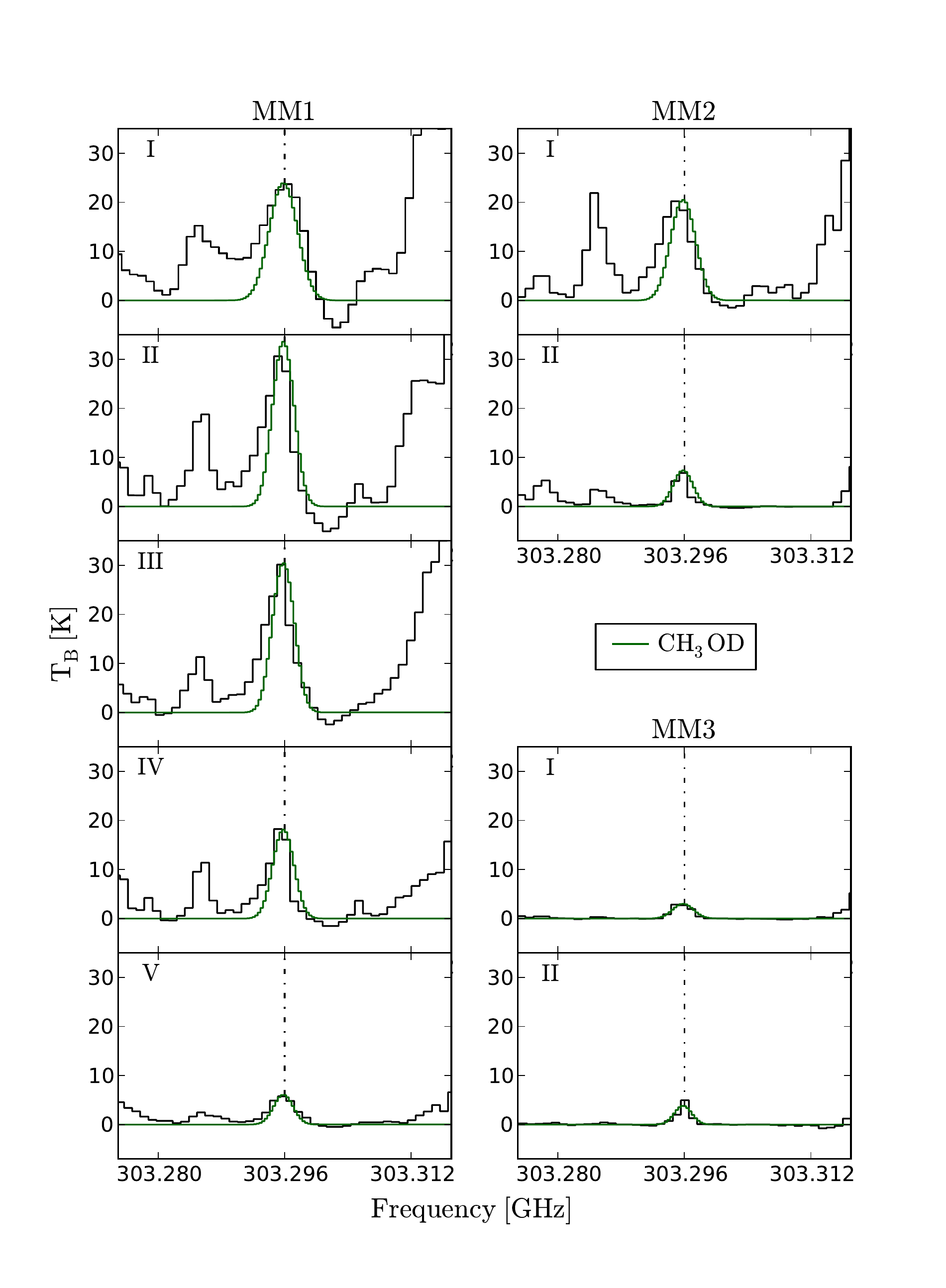} % trim={<left> <lower> <right> <upper>}
			\caption{CH$_3$OD 7$_{(2+)}$$\rightarrow$7$_{(2,-)}$ transition at 303.296 GHz (indicated by dash-dotted lines) detected towards each region. Frequencies are shifted to the rest frame of the individual regions. Green lines represent the modelled spectra of CH$_3$OD.}
			\label{fig:ODAll}
			\addtocounter{figure}{-2}
	\end{minipage}}
\end{figure*}
%--------------END FIGURE--------------------------------------

%--------------BEGIN FIGURE: DOH - ALL REGIONS - OPTIMISED LINE -------------------------------
\begin{figure*}
	\rotatebox{90}{\begin{minipage}{0.48\textheight}
			\includegraphics[width=1.\textwidth, trim={0 1cm 0 1cm}, clip]{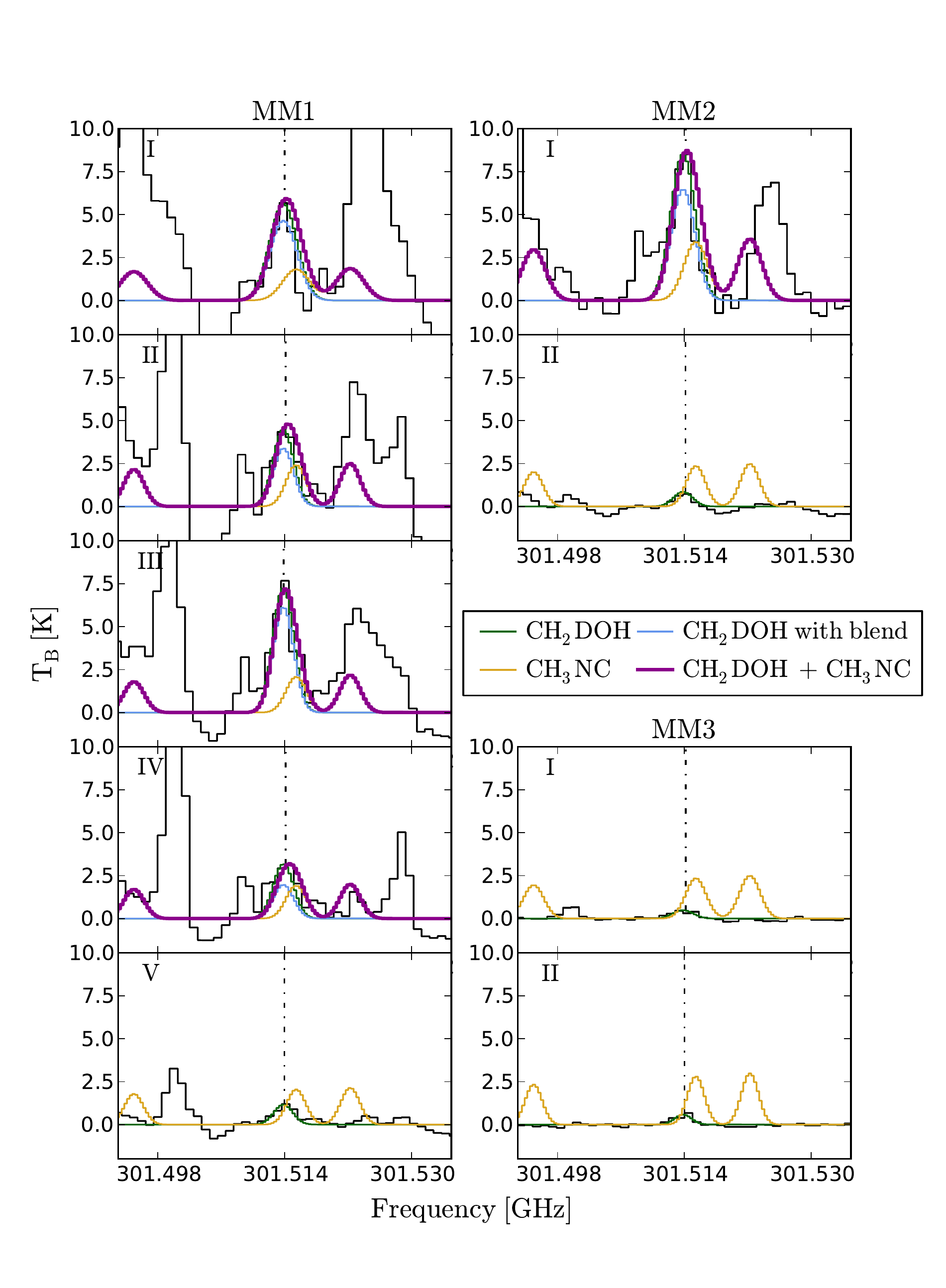} % trim={<left> <lower> <right> <upper>}
			\caption{CH$_2$DOH 12$_{(2,11)}$$\rightarrow$12$_{(1,12)}$ transition at 301.514 GHz (indicated by dash-dotted lines) detected towards each region. Frequencies are shifted to the rest frame of the individual regions. Blue and green lines represent the modelled spectra of CH$_2$DOH with and without blending, i.e., including and excluding the contribution from CH$_3$NC, respectively. Yellow lines represent the modelled spectra of CH$_3$NC and magenta lines represent the sum of the spectra of CH$_2$DOH with blending and CH$_3$NC.}
			\label{fig:DAll}
			\addtocounter{figure}{1}
	\end{minipage}}
\end{figure*}
%--------------END FIGURE--------------------------------------

\section{Methanol deuteration fractions} \label{sec:ratios}
\begin{figure}[]
	\centering
	\includegraphics[width=0.48\textwidth]{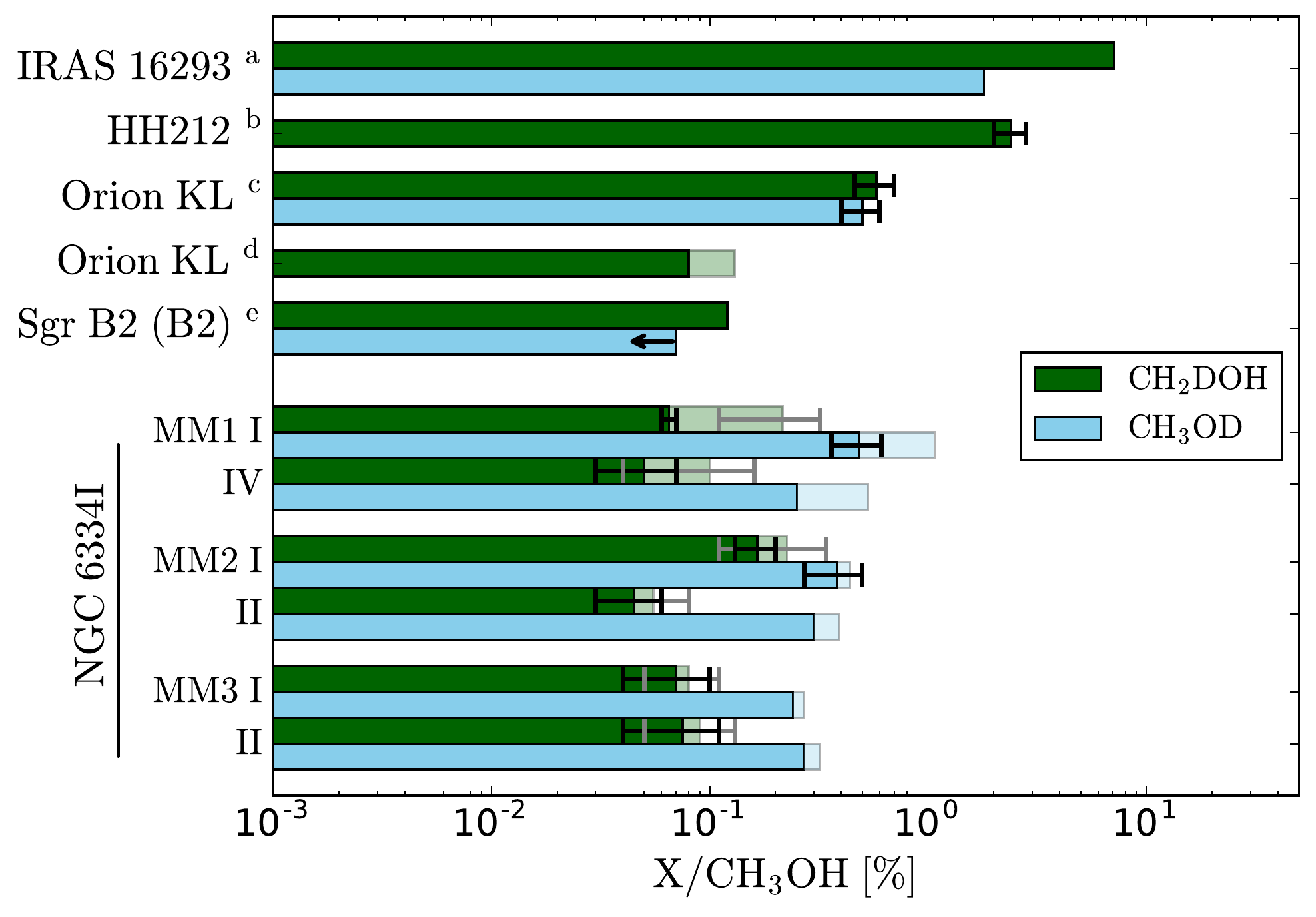} % trim={<left> <lower> <right> <upper>}
	\caption{Column density ratio of CH$_2$DOH/CH$_3$OH (green) and CH$_3$OD/CH$_3$OH (blue) for NGC 6334\RN{1} and other objects. For Orion KL the shaded bar indicates the range of ratios derived by \cite{Peng2012}. For NGC 6334\RN{1} shaded and filled bars indicate the ratios derived using the $^{13}$C and $^{18}$O isotopologues as base respectively. Error bars indicate the range of ratios derived with and without blending. For NGC 6334\RN{1} MM1 only the region with the lowest (MM1 \RN{4}) and highest (MM1 \RN{1}) ratios are plotted. $^{(a)}$ \cite{Jorgensen2017}. $^{(b)}$ \cite{Bianchi2017}. $^{(c)}$ \cite{Neill2013}. $^{(d)}$ \cite{Peng2012}. $^{(e)}$ \cite{Belloche2016}.}
	\label{fig:histogram}
\end{figure}

In the following sections the CH$_2$DOH/CH$_3$OH and CH$_2$DOH/CH$_3$OD ratios derived for the nine regions in NGC 6334\RN{1}, as well as those derived for a number of other sources, will be discussed. These are summarised in Fig. \ref{fig:histogram} and listed in Table \ref{tab:isotope_ratios}. 

%--------------BEGIN FIGURE:Histrogram -------------------------------
%--------------END FIGURE--------------------------------------

\subsection{NGC 6334\RN{1}} \label{subsec:ratio_vs_temp}
Using the column densities derived from the $^{13}$CH$_3$OH, CH$_3^{18}$OH, CH$_2$DOH and CH$_3$OD transitions, we calculate the CH$_2$DOH/CH$_3$OH and CH$_3$OD/CH$_3$OH ratios for each of the spectral regions. These are listed in Table \ref{tab:isotope_ratios}. The CH$_2$DOH/CH$_3$OH ratios range between (0.03--0.34)$\%$ and (0.03--0.20)$\%$ and the CH$_3$OD/CH$_3$OH ratios range between (0.27--1.07)$\%$ and (0.22--0.61)$\%$, derived from $^{12}$C/$^{13}$C and $^{16}$O/$^{18}$O respectively. In the case of the CH$_2$DOH/CH$_3$OH, both the lowest and the highest ratios are associated with region MM2, the lowest detected towards region MM1 \RN{2} and the highest towards MM2 \RN{1}. For the CH$_3$OD/CH$_3$OH ratio the lowest values are detected towards MM3 \RN{1} and MM1 \RN{5} while the highest is detected towards MM1 \RN{1}. The mean values over all regions (including the vibrational correction and uncertainty on the line strength of CH$_2$DOH) are 0.13$\%$$\pm$0.06$\%$ and 0.53$\%$$\pm$0.27$\%$ based on the $^{13}$C isotopologue and 0.08$\%$$\pm$0.04$\%$ and 0.32$\%$$\pm$0.09$\%$ based on the $^{18}$O isotopologue, for the CH$_2$DOH/CH$_3$OH and CH$_3$OD/CH$_3$OH ratios respectively. Based on these means the average CH$_3$OD/CH$_3$OH ratio is twice that of the average CH$_2$DOH/CH$_3$OH ratio. The CH$_2$DOH/CH$_3$OH and CH$_3$OD/CH$_3$OH ratios derived based on the $^{13}$C and $^{18}$O isotopologues agree within factors of $\sim$2 for all regions.      

If the exchange of D into the CH$_3$ and OH functional groups of methanol is equally efficient, the column density ratio of CH$_2$DOH/CH$_3$OD is expected to be 3. Interestingly, this is not the case for the regions presented here. Instead, we derive CH$_2$DOH/CH$_3$OD column density ratios of the order $\sim$0.3 (not including the statistical correction factor which would further decrease the ratio by a factor of 3), lower by a factor of 6 as compared with the lower limit derived by \cite{Belloche2016} for Sgr B2 and factors of 4 and 2 as compared with the values derived for Orion KL by \cite{Neill2013} and \cite{Peng2012} respectively. This very low CH$_2$DOH/CH$_3$OD ratio is just opposite to values found for pre-stellar cores and low-mass protostars, where CH$_2$DOH/CH$_3$OD ratios up to 10 are found \citep{Bizzocchi2014, Parise2006}, although consistent with the trend of lower ratios inferred for high mass protostars \citep{Ratajczak2011}. The low CH$_2$DOH/CH$_3$OD ratio implies that the deuteration of the OH functional group is more efficient than that of the CH$_3$ group, or that D is more easily abstracted from the CH$_3$ group rather than from the OH group. 

Various chemical processes in the gas and ice that can cause the CH$_2$DOH/CH$_3$OD ratio to deviate from the statistical ratio of 3 are described in \cite{Ratajczak2011} and in \cite{Faure2015}. In the gas phase these processes include protonation and dissociative recombination reactions which destroy CH$_3$OD more efficiently than CH$_2$DOH since all recombinations of CH$_2$DOH$_2^+$ lead to CH$_2$DOH while CH$_3$OHD$^+$ can recombine to either CH$_3$OH or CH$_3$OD with an equal branching ratio \citep{Charnley1997}. However the timescale needed to significantly alter the CH$_2$DOH/CH$_3$OD ratio through these reactions is likely too long, i.e., more than 10$^5$ years, when compared with the typical lifetime of a hot core. In the solid state, experiments carried out by \cite{Nagaoka2005} have shown that H/D substitution in solid CH$_3$OH forms CH$_2$DOH but no CH$_3$OD. Also in the solid state, experiments by \cite{Ratajczak2009} and \cite{Faure2015} have shown that H/D exchanges can occur between water and methanol in warm ices (T $\sim$120 K), but only on the OH functional group of methanol and not on the CH$_3$ group. The processes mentioned above favour the formation of CH$_2$DOH over that of CH$_3$OD and consequently lead to an increase in the CH$_2$DOH/CH$_3$OD ratio rather than decrease. The low observed ratio therefore remains unexplained. 

It should be noted however, that because the spectroscopy of CH$_2$DOH is not well understood, the CH$_2$DOH/CH$_3$OD ratios derived may be higher. To better constrain the ratio, future studies are well-advised to target only a-type CH$_2$DOH-transitions rather than b- or c-type transitions, for which the uncertainties on the line strength are not well-constrained, in addition to preferring weak, low $A_{\textrm{ij}}$ transitions, to ensure that lines are optically thin. A refinement of the accuracy of the column density of CH$_2$DOH, will make it possible to investigate the coupling of CH$_2$DOH/CH$_3$OD ratios with environment and chemical evolution. Therefore it may be advantageous to focus future observations on narrow bands with high spectroscopic resolution covering a few well-chosen transitions, rather than broader bands which, albeit potentially covering more lines, might make identification and analysis difficult due to uncertainties in spectroscopic values or blending with features of other species which may not be resolved.

%--------BEGIN TABLE:ISOTOPE RATIOS---------------
\begin{sidewaystable*}[]%[htbp] %[here,top,bottom,page]
%\begin{small}
	\centering
	\caption{CH$_2$DOH/CH$_3$OH and CH$_3$OD/CH$_3$OH ratios derived from $^{13}$CH$_3$OH or CH$_3^{18}$OH}
	\label{tab:isotope_ratios}
	\begin{tabular}{!{\extracolsep{4pt}}llccccccc!{}}
		\toprule
		Source & Region & \multicolumn{2}{c}{CH$_2$DOH/CH$_3$OH} & \multicolumn{2}{c}{(CH$_2$DOH/CH$_3$OH)$_{corr}$\tablefootmark{a}} & \multicolumn{2}{c}{CH$_3$OD/CH$_3$OH} & CH$_2$DOH/CH$_3$OD\tablefootmark{b} \\
		\cline{3-4} 
		\cline{5-6} 
		\cline{7-8}
		& & $^{12}$C/$^{13}$C & $^{16}$O/$^{18}$O &  $^{12}$C/$^{13}$C  & $^{16}$O/$^{18}$O & $^{12}$C/$^{13}$C & $^{16}$O/$^{18}$O & \\
		& & [$\%$] & [$\%$] & [$\%$] & [$\%$] & [$\%$] & [$\%$] & \\
		\midrule
		NGC 6334 \RN{1} & MM1 \RN{1} & 0.11--0.32 & 0.06--0.07 & 0.04--0.11 & 0.02 & 1.07 & 0.36--0.61 & 0.16--0.19 \\
		& MM1 \RN{2} & 0.05--0.17 & 0.03--0.07 & 0.02--0.06 & 0.01--0.02 & 0.83 & 0.35--0.42 & 0.10--0.13 \\
		& MM1 \RN{3} & 0.07--0.21 & 0.06--0.14 & 0.02--0.07 & 0.02--0.05 & 0.45 & 0.30 & 0.23--0.29 \\
		& MM1 \RN{4} & 0.04--0.16 & 0.03--0.07 & 0.01--0.05 & 0.01--0.03 & 0.53 & 0.25 & 0.12--0.19 \\
		& MM1 \RN{5} & 0.09--0.20 & 0.04--0.09 & 0.03--0.07 & 0.01--0.03 & 0.50 & 0.22 & 0.25 \\
		& MM2 \RN{1} & 0.11--0.34 & 0.13--0.20 & 0.04--0.11 & 0.04--0.07 & 0.44 & 0.27--0.50 & 0.38--0.50 \\
		& MM2 \RN{2} & 0.03--0.08 & 0.03--0.06 & 0.01--0.03 & 0.01--0.02 & 0.39 & 0.30 & 0.15 \\
		& MM3 \RN{1} & 0.05--0.11 & 0.04--0.10 & 0.08--0.04 & 0.01--0.03 & 0.27 & 0.24 & 0.25 \\
		& MM3 \RN{2} & 0.05--0.13 & 0.04--0.11 & 0.02--0.04 & 0.01--0.04 & 0.32 & 0.27 & 0.23 \\
		\midrule
		Sgr B2\tablefootmark{c} & N2 & 0.12 & -- & 0.04 & -- & $<$0.07 & -- & $>$1.8 \\
		Orion KL\tablefootmark{d,e} & Compact Ridge & 0.58$\pm$0.12  & -- & 0.19 & -- & 0.5$\pm$0.1 & -- & 1.2$\pm$0.3\\
		&  & (0.08--0.13) & -- & 0.03--0.04 & -- &   & -- & 0.7$\pm$0.3\tablefootmark{f} \\
		Orion B\tablefootmark{g} & HH212 & 2.4$\pm$0.4 & -- & 0.8$\pm$0.1 & -- & -- & -- & -- \\
		IRAS 16293--2422\tablefootmark{h} & B & -- & 7.1\tablefootmark{i} & -- & 2.4\tablefootmark{i} & -- & 1.8 & 3.9 \\
		\bottomrule
	\end{tabular}
	\tablefoot{Ranges of CH$_2$DOH/CH$_3$OH and CH$_3$OD/CH$_3$OH correspond to the range in column density of $^{13}$CH$_3$OH, CH$_3^{18}$OH and CH$_2$DOH with and without blending. All CH$_2$DOH/CH$_3$OH ratios include the vibrational correction of 1.25 as well as the uncertainy in the line strength.  \tablefoottext{a}{Corrected for statistical weight of the location of the substituted deuterium. For CH$_2$DOH this value is 3, for CH$_3$OD it is 1.} \tablefoottext{b}{Ratios do not include statistical correction factors.} \tablefoottext{c}{\cite{Belloche2016}.} \tablefoottext{d}{\cite{Neill2013}.} \tablefoottext{e}{\cite{Peng2012}.} \tablefoottext{f}{Mean ratio over central region.} \tablefoottext{g}{\cite{Bianchi2017}.} \tablefoottext{h}{\cite{Jorgensen2017}.} \tablefoottext{i}{Includes vibration correction factor for CH$_2$DOH of 1.457 at 300 K (see Section 2.2 of \cite{Jorgensen2017}).}}
%\end{small}
\end{sidewaystable*}
%-------------------END TABLE--------------------------- 

\begin{figure}[]
	\centering
	\includegraphics[width=0.48\textwidth, trim={0 0 1.5cm 1cm}, clip]{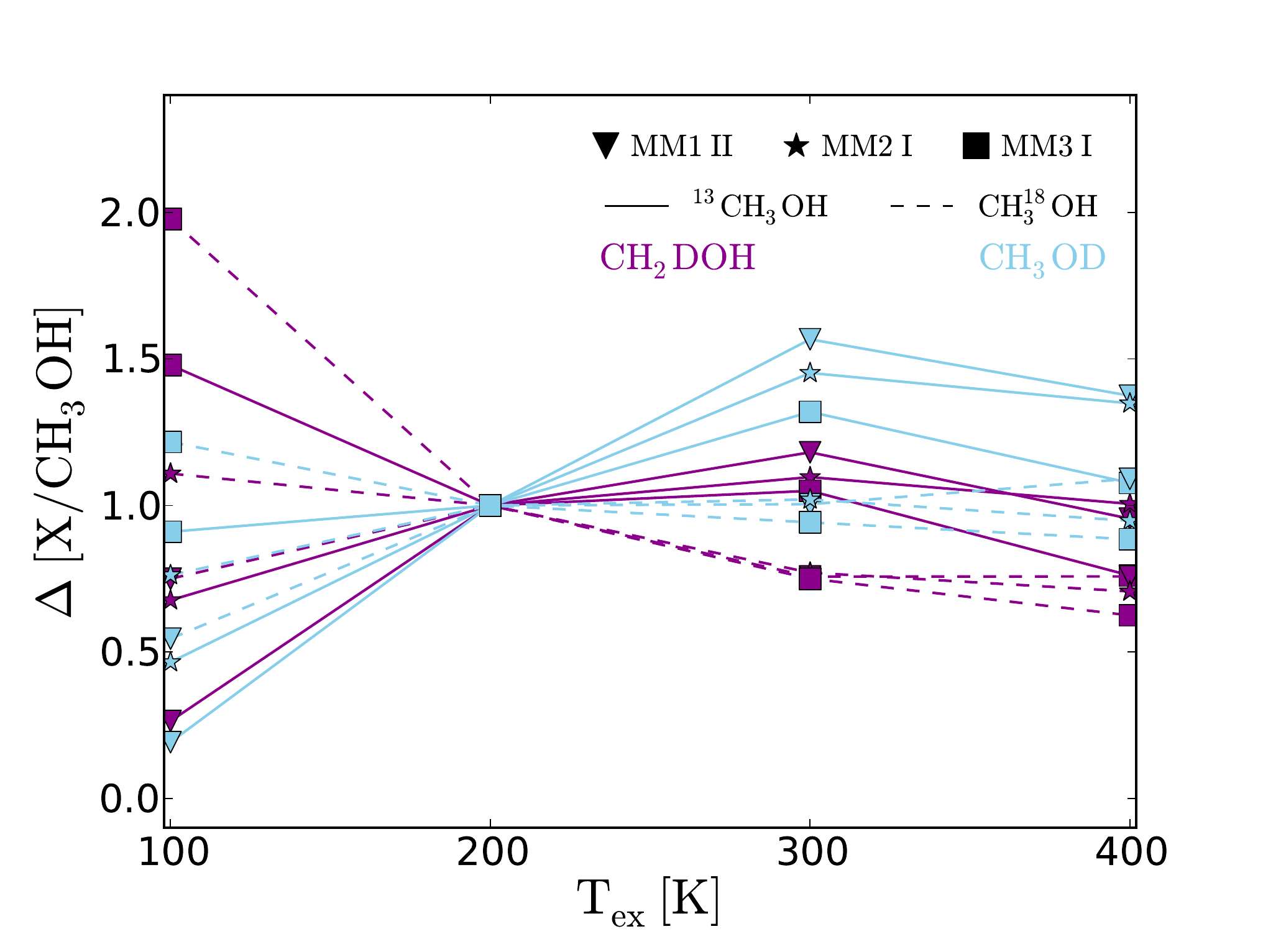} % trim={<left> <lower> <right> <upper>}
	\caption{Column density ratio as function of excitation temperature. Ratios are normalised to the value derived for $T_{\textrm{ex}}$ = 200 K. Magenta and blue lines represent the CH$_2$DOH/CH$_3$OH and CH$_3$OD/CH$_3$OH column density ratios respectively, while sold and dashed lines indicate the values derived assuming $^{12}$C/$^{13}$C = 62 and $^{16}$O/$^{18}$O = 450 respectively. Different regions are indicated by different markers.} 
	\label{fig:ratio_vs_temp}
\end{figure}

As discussed in Section \ref{subsec:method}, the excitation temperature for each region is determined based on a minimal $\chi^2$ technique comparing a grid of models with the spectra. The excitation temperatures of the best-fit models range between $\sim$120 and 330 K, the warmest regions associated with the MM1 core and the coldest with the MM3 core. We investigate the effect of varying the excitation temperature on the derived column densities and plot the ratios of CH$_2$DOH/CH$_3$OH and CH$_3$OD/CH$_3$OH as function of $T_{\textrm{ex}}$ in Fig. \ref{fig:ratio_vs_temp}. From this analysis, it is evident that the spread in CH$_2$DOH/CH$_3$OH and CH$_3$OD/CH$_3$OH ascribed to different excitation temperatures (over all regions) is within a factor 2 of the value derived for $T_{\textrm{ex}}$= 200 K. For the individual regions, the variations in CH$_2$DOH/CH$_3$OH and CH$_3$OD/CH$_3$OH over the full range of excitation temperatures are within factors of 2--4. Therefore, it is reasonable to assume that the lines used to optimise our fits are optically thin at excitation temperatures of (100--400) K and consequently that our derived CH$_2$DOH/CH$_3$OH and CH$_3$OD/CH$_3$OH ratios do not depend critically on an exact determination of $T_{\textrm{ex}}$. For each value of $T_{\textrm{ex}}$ we also calculate the ratio of the $^{13}$C- and $^{18}$O-methanol column densities. These ratios differ by less than a factor two for MM2 and MM3 and a factor 4 for MM1, verifying that the transitions are excited under similar conditions. The vibrational correction factor applied to CH$_2$DOH has only little effect on this result. 

%--------------BEGIN FIGURE: RATIO VS TEMP -------------------------------
%--------------END FIGURE-------------------------------------- 

\subsection{Comparison with other sources}
Using ALMA, \cite{Belloche2016} investigated the deuterium fractionation of complex organic molecules towards the high mass star-forming region Sgr B2 in the Galactic central region. With the high spatial resolution observations, $\sim$1.4$^{\second}$, they probe scales down to 0.06 pc ($\sim$11620 AU assuming a distance $\sim$8.3 kpc). They report a tentative detection of CH$_2$DOH and derive a deuteration fraction for CH$_2$DOH/CH$_3$OH of 0.12\%. Including the statistical weight of the location of the substituted deuterium this value becomes 0.04\%. CH$_3$OD is not detected, but an upper limit of 0.07\% for the CH$_3$OD/CH$_3$OH ratio is reported. This is translated into a lower limit on the CH$_2$DOH/CH$_3$OD ratio of 1.8. Their CH$_3$OH column density is based on LTE modelling of both the $^{13}$C and $^{18}$O methanol isotopologues. The authors note that the CH$_2$DOH/CH$_3$OH ratio they derive is lower than what is predicted by the chemical models of \cite{Taquet2014} and \cite{Aikawa2012} but may be explained by the high temperature that characterises the Galactic Centre or result from an overall low abundance of deuterium in this region due to the high star formation rates. 

For the high-mass star-forming regions in Orion, CH$_2$DOH/CH$_3$OH ratios are of the same order as for Sgr B2, ranging between (0.08--0.58)\%, equivalent to (0.03--0.19)\% when the statistical weights are accounted for. These values are derived by \cite{Peng2012}, using observations from the IRAM Plateau de Bure Interferometer, and \cite{Neill2013}, using data from \textit{Herschel}/HIFI. In addition, \cite{Neill2013} report a CH$_3$OD/CH$_3$OH ratio of 0.5$\pm$0.1\% and a CH$_2$DOH/CH$_3$OD ratio of 1.2$\pm$0.3. A slightly lower CH$_2$DOH/CH$_3$OD ratio of 0.7$\pm$0.3 is reported by \cite{Peng2012}. To derive CH$_3$OH column densities \cite{Neill2013} and \cite{Peng2012} use slightly different approaches: \cite{Neill2013} assume a $^{12}$C/$^{13}$C ratio of 60 and derive the CH$_3$OH density based on transitions of $^{13}$CH$_3$OH, while \cite{Peng2012} detect a number of E-type methanol transitions and derive the total CH$_3$OH density assuming an A/E-type abundance of 1.2. Both studies of Orion KL probe scales which are smaller than those studied in Sgr B2: $\sim$10$^{\second}$ and $\sim$2$^{\second}$, corresponding to $\sim$4140 AU and $\sim$830 AU at the distance of Orion KL ($\sim$414 pc), for \cite{Neill2013} and \cite{Peng2012} respectively. However, since the beam dilution factor is higher in these studies, i.e., the area over which the emission is averaged is larger, the column density derived, assuming the total emission to be the same, is lower. Also, observations with larger beam sizes, which are more sensitive to large scale structures, generally probe regions of lower temperature, meaning that some molecules may be locked up in icy grain mantles, resulting in lower gas phase abundances. This combination of effects means that the derived column densities, as well as the inferred deuteration ratios, for Orion KL may in fact be higher, if derived from observations with higher angular resolution.  

An example of such high-resolution observations are presented by \cite{Bianchi2017} who use 0.15$^{\second}$-resolution ALMA observations to study the Sun-like class 0 protostar HH212, located in the Orion B cloud, on scales of $\sim$70 AU. From transitions of $^{13}$CH$_3$OH and CH$_2$DOH they derive a CH$_2$DOH/CH$_3$OH ratio of (2.4$\pm$0.4)\%, equivalent to (0.8$\pm$0.1)\% after accounting for statistical weights, assuming a $^{12}$C/$^{13}$C ratio of 70. This deuteration ratio is higher than what has been derived for high-mass star-forming regions but lower by an order of magnitude as compared with observations (carried out with single dish telescopes) towards protostars in Perseus. \cite{Bianchi2017} argue that the lower deuteration ratio they find is consistent with the dust temperature of the Orion region being higher than that of the Perseus cloud. 

Similarly to HH212, the low-mass protostellar binary system IRAS 16293, located in the $\rho$ Ophiuchi cloud complex, exhibits a CH$_2$DOH/CH$_3$OH ratio which is much higher than that of the high-mass star-forming regions. With the ALMA-PILS survey (see \citealt{Jorgensen2016} for full PILS overview), sampling spatial scales of the order 0.5$^{\second}$, corresponding to 60 AU at the distance of the source ($\sim$120 pc), \cite{Jorgensen2017} derive a CH$_2$DOH/CH$_3$OH ratio of 7.1$\%$, equivalent to 2.4\% after corrections for statistical weights, assuming a $^{16}$O/$^{18}$O isotope ratio of 560 to estimate the CH$_3$OH abundance, and a CH$_3$OD/CH$_3$OH ratio of 1.8\% resulting in a CH$_2$DOH/CH$_3$OD ratio of 3.9. 

Because of the comparable resolution and methods used, the methanol deuteration ratios derived for IRAS 16293, HH212 and Sgr B2, may be directly compared to the ratios derived in this study. When doing so, the low levels of deuterium fractionation associated with the regions in NGC 6334\RN{1}, as compared with the low-mass star-forming regions IRAS 16293 and HH212, become very apparent (Table \ref{tab:isotope_ratios}). It is also interesting to note that while our inferred CH$_2$DOH/CH$_3$OH ratios for NGC 6334\RN{1} are similar to what has been derived by \cite{Belloche2016} for Sgr B2, the CH$_3$OD/CH$_3$OH ratios are higher by up to an order of magnitude as compared with the lower limit reported for Sgr B2.
 
\subsection{Comparison with models}
In an effort to better understand the observed variety of deuterated species and their column density ratios, a number of astrochemical models have been put forward, among these the GRAINOBLE model. This code has been described in detail in previous studies \citep{Taquet2012,Taquet2013,Taquet2014} and here we give only a brief presentation before discussing how the model compares with the levels of deuteration we find in the regions of NGC 6334\RN{1}. 

GRAINOBLE follows the gas-ice interstellar chemistry through a 3-phase (gas, ice surface, ice bulk) rate equations approach initially developed by \cite{Hasegawa1993}. The model takes into account a number of gas-grain processes including the accretion of species onto the grains, desorption back into the gas-phase and reactions between particles at the ice surface and in the ice mantle. The gas-phase chemical network is described in \cite{Taquet2014} and includes the spin states of H$_2$, H$_2^+$, and H$_3^+$, as well as the deuterated isotopologues of hydrogenated species with four or less atoms and molecules involved in the chemistry of water, methanol, ammonia, and formaldehyde. The ice chemical network follows the formation and the deuteration of the main ice species, following a series of laboratory experiments, and complex organic molecules through radical-radical recombination reactions. Specifically, the pathways for formation and deuteration of methanol include both addition reactions involving atomic D and hydrogenation of solid CO and H$_2$CO at cold ($T$ = 10--15 K) temperatures as well as abstraction reactions, as shown by \cite{Hidaka2009}. 

\begin{figure*}[]
	\centering
	\begin{subfigure}[]{0.45\textwidth} 
		\includegraphics[width=0.9\textwidth, trim={0 0 0cm 0cm}, clip]{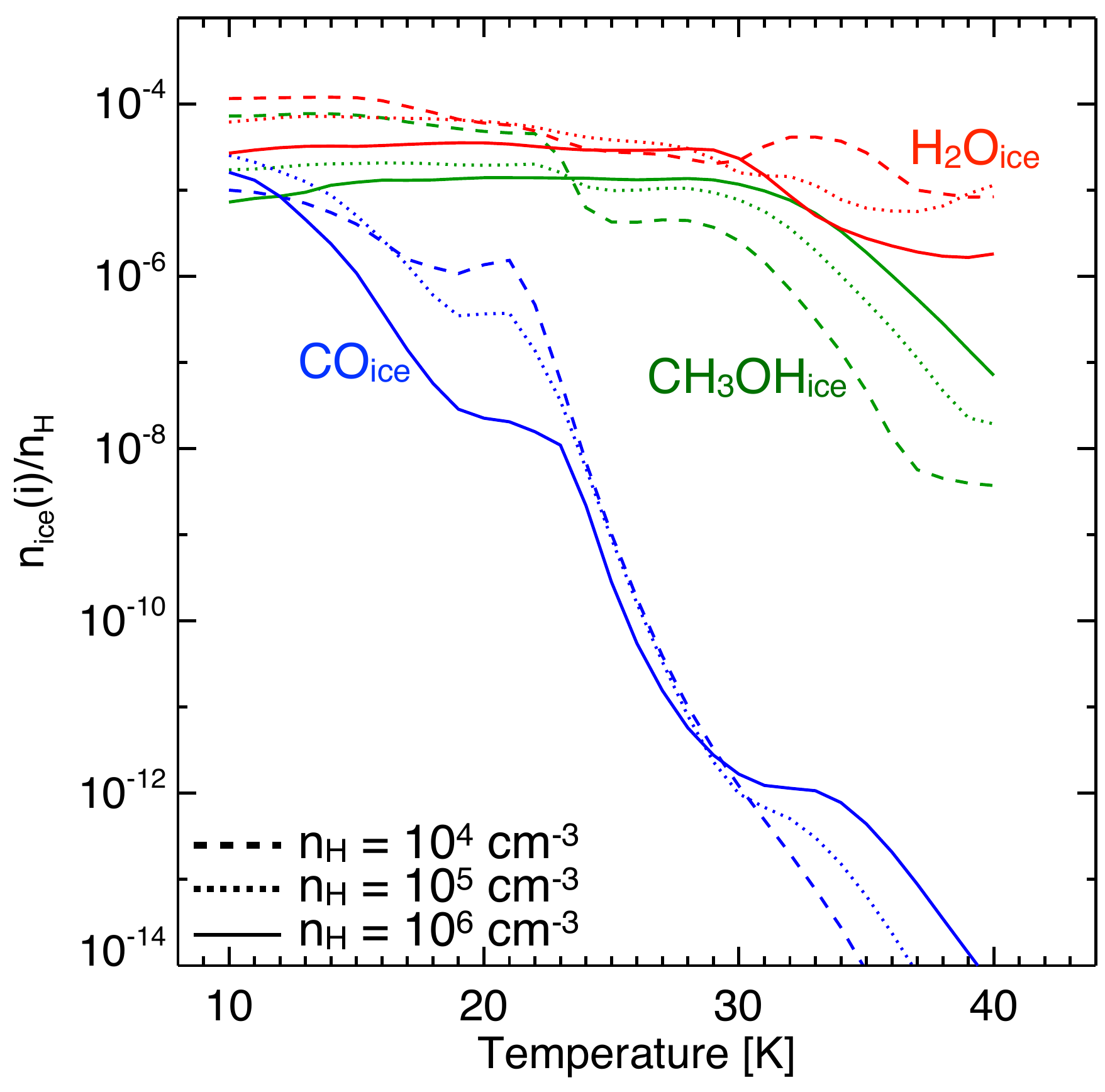} %trim={<left> <lower> <right> <upper>}
		\caption*{}
		\label{fig:Xice}
	\end{subfigure}
	\begin{subfigure}{0.45\textwidth}
		\includegraphics[width=0.88\textwidth, trim={0 0 0cm 0cm}, clip]{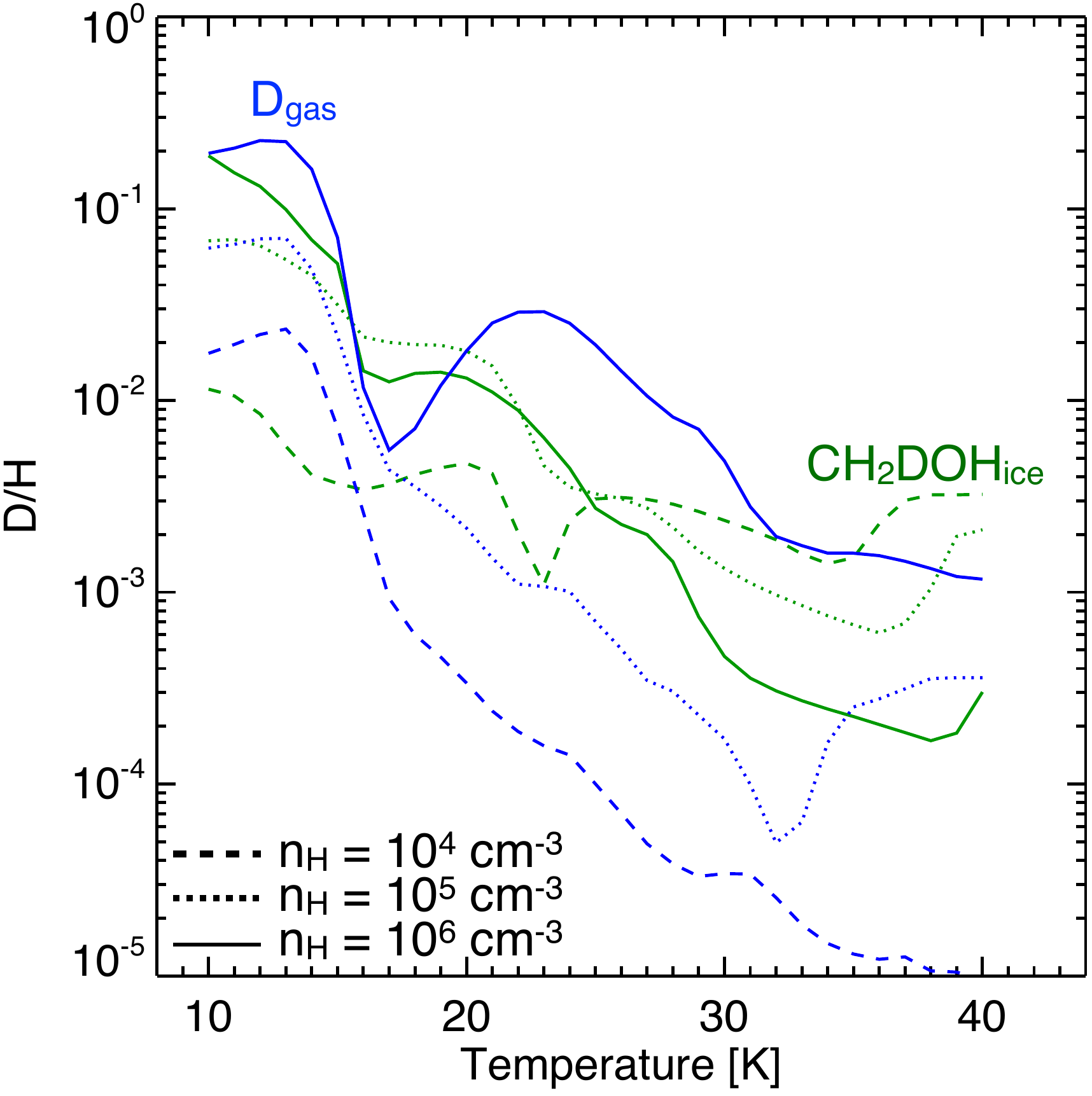}
		\caption*{}
		\label{fig:Dratio}
	\end{subfigure}
	\begin{subfigure}{0.45\textwidth}
		\includegraphics[width=1\textwidth, trim={0 0cm 0 0cm}, clip]{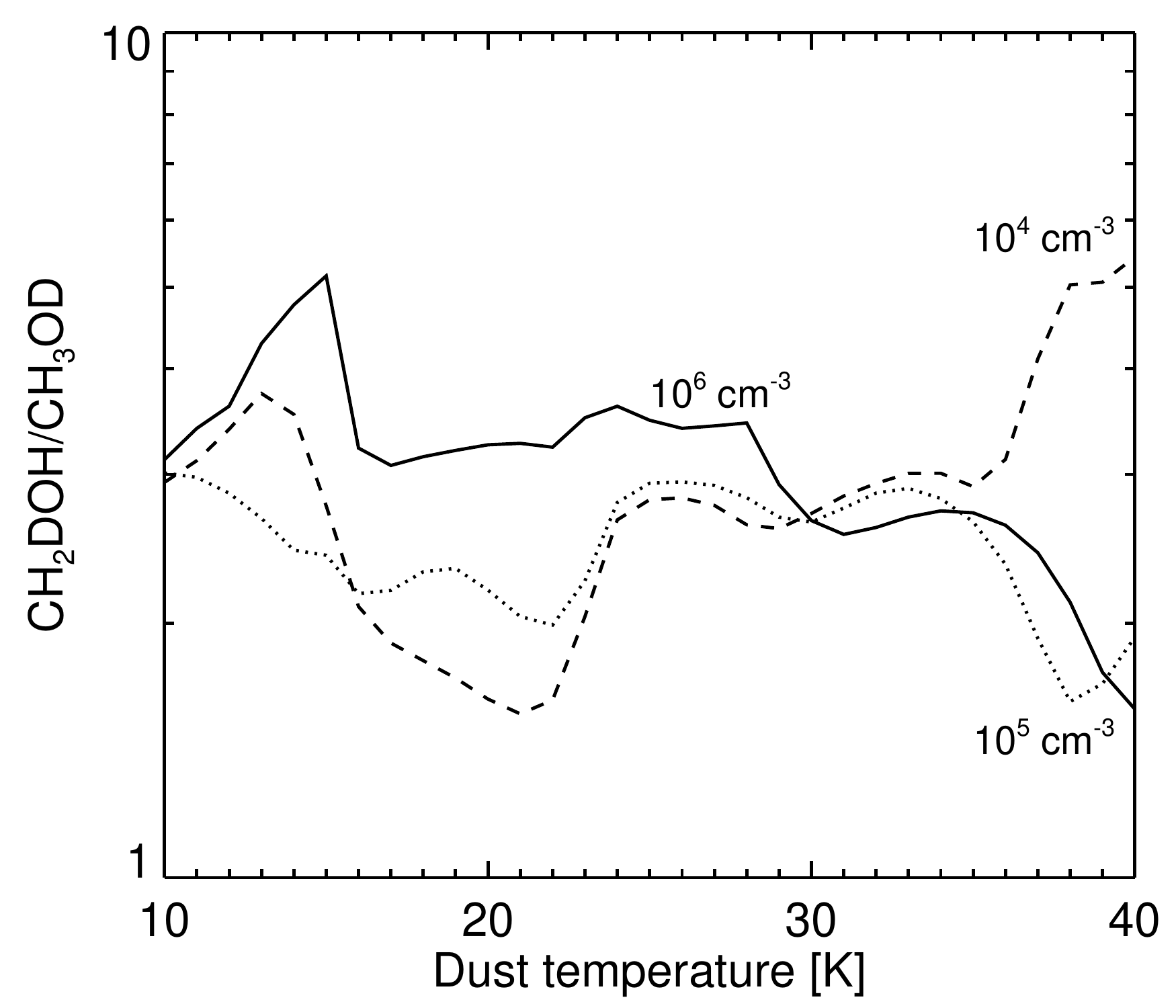}
		\caption*{}
		\label{fig:DOH_ODratio}
	\end{subfigure}
	\caption{Top left panel: Final abundances of H$_2$O (red lines), CO (blue lines), and CH$_3$OH (green lines) in ices as a function of temperature. Top right panel: Final CH$_2$DOH/CH$_3$OH abundance ratio in ices (green lines) and final atomic D/H in the gas phase (blue lines) as a function of temperature. Bottom panel: Final CH$_2$DOH/CH$_3$OD abundance ratio as function of temperature. In all panels three densities are considered: $n_{\rm H} = 10^4$ (dashed), $10^5$ (dotted), and $10^6$ (solid) cm$^{-3}$.}
	\label{fig:model}
\end{figure*}

We investigated the effect of the dust and gas temperatures, $T$, assumed to be equal, and the total density, $n_{\rm H}$, on the methanol deuteration. For this purpose, a series of pseudo-time dependent simulations in which the chemistry evolves over time whilst the temperature and density remain constant were run. Figure \ref{fig:model} shows the final ice abundances of CO, H$_2$O, and CH$_3$OH, the deuteration in ices of methanol and the gas phase atomic D/H ratio and the CH$_2$DOH/CH$_3$OD ratio as a function of temperature. Models are shown for three different densities, $n_{\rm H}$ = 10$^4$, 10$^5$ and 10$^6$ cm$^{-3}$, with respective ages of 2$\times$10$^7$, 2$\times$10$^6$, and 2$\times$10$^5$ years.

It is evident that both the ice composition and the methanol deuteration are highly dependent on the considered temperature and total density. The barrierless reactions which form water from atomic oxygen mean that this species is the least dependent on $T$ and $n_{\rm H}$. Methanol and its deuterated isotopologues are, on the other hand, formed through hydrogenation and/or deuteration of CO and H$_2$CO, which are thought to have high activation barriers. As a consequence, methanol formation can be inhibited at high densities, decreasing the CO hydrogenation, or at high temperatures where CO no longer efficiently freezes-out onto grains. Despite the lower residence time of H and CO on the grains at temperatures above $\sim$20 K, limiting the formation of H$_2$CO and later CH$_3$OH, significant deuteration of methanol occur in the model. This is due to the long timescales considered which enhance the probability that the CO and H or D particles meet on the dust grain and recombine. Also, since it is formed through hydrogenation and/or deuteration processes, the methanol deuteration is governed by the atomic D/H in the gas phase.

Specifically, the strong decrease of the methanol deuteration with temperature, up to three orders of magnitudes with an increase of $T$ from 10 to 40 K, is due to the decreased efficiency of deuterium chemistry in the gas phase with increasing temperature. Atomic D is mostly formed via electronic recombination of H$_3^+$ isotopologues, formed through exothermic reactions between H$_3^+$ (or with its isotopologues) and HD. Because the efficiency of the backward reactions increase with an increase in temperature, the deuteration of H$_3^+$ and therefore the production of atomic D is limited at high temperatures. However, the solid CH$_2$DOH/CH$_3$OH ratio does not always scale with the final atomic gas phase D/H ratio. Figure \ref{fig:model} shows how the CH$_2$DOH/CH$_3$OH ratio decreases more efficiently with temperature at high densities but remains fairly constant for $n_{\rm H}$ = $10^4$ cm$^{-3}$.  

The model distinguishes the formation of CH$_2$DOH from that of CH$_3$OD. As seen in Fig. \ref{fig:model}, the ratio remains close to 3 throughout the temperature range. This is due to the three reaction channels that form CH$_2$DOH compared with the single channel that forms CH$_3$OD. Also, it is important to note that while the observations trace the methanol deuteration in the gas phase of the hot core surrounding the forming massive protostars, the model describes the predicted deuteration of ices formed during the precursor pre-stellar phase. Therefore, the analysis is implicitly based on the assumption that the gas phase composition around the hot cores reflects the composition of the ices in the pre-stellar phase. As discussed in section \ref{subsec:ratio_vs_temp}, a number of processes may alter the ice methanol deuteration. Some of these processes are already accounted for in the model, e.g., the formation of CH$_2$DOH via abstraction reactions, shown experimentally by \cite{Nagaoka2005}, is included in the model but contributes little to the overall abundance of CH$_2$DOH. Others, like the H/D exchanges shown to occur between water and methanol in warm ices \citep{Ratajczak2009, Faure2015} still need to be implemented and could potentially contribute significantly to the decrease of the CH$_3$OD/CH$_3$OH abundance resulting in a higher CH$_2$DOH/CH$_3$OD ratio in contrast to the low ratios derived for the regions in NGC 6334\RN{1}.
	
Based on the model predictions, our inferred deuterium fractionation of methanol in the NGC 6334\RN{1} regions, indicate a temperature during formation around 30 K. A similar dust temperature of the precursor dense core is reported by \cite{Russeil2013} who use data from \textit{Herschel} to derive a dust temperature map of the NGC 6334 star-forming region and show that the area is dominated by temperatures $\sim$25--35 K. Such a warm pre-stellar cloud could be the result of heating from a nearby earlier phase of star formation, potentially associated with NGC 6334E a region located between \RN{1} and \RN{1}(N) and hosting an O star and associated large shell-like H\RN{2} region \citep{Carral2002}. The small variation in deuterium fractions found towards the individual regions of NGC 6334\RN{1} may reflect local differences in $n_{\rm H}$.  

%--------------BEGIN FIGURE: MODEL OUTPUT -------------------------------
%--------------END FIGURE: MODEL OUTPUT -------------------------------

\section{Summary and conclusion} \label{sec:conclusion}
In this paper we analyse the level of deuterium fractionation in methanol towards the high-mass star-forming region NGC 6334\RN{1}. We use high spectral resolution observations carried out with ALMA and extract spectra from nine individual locations across the MM1, MM2 and MM3 regions. Based on these, we identify a total of 15 transitions belonging to the isotopologues of methanol. Assuming excitation temperatures of $\sim$120--330 K for the individual regions, we derived column densities for each of the species $^{13}$CH$_3$OH, CH$_3^{18}$OH, CH$_2$DOH and CH$_3$OD and calculate CH$_2$DOH/CH$_3$OH and CH$_3$OD/CH$_3$OH ratios. These range from 0.03\% to 0.34\% and 0.03\% to 0.20\% for the ratios of CH$_2$DOH/CH$_3$OH and from 0.27\% to 1.07\% and 0.22\% to 0.61\% for the ratios of CH$_3$OD/CH$_3$OH, using the $^{13}$C and $^{18}$O isotopologues to derive the column density of $^{12}$CH$_3^{16}$OH respectively. The highest levels of deuterium fractionation are associated with the regions MM1 \RN{1} and MM2 \RN{1}, which are also the brightest regions in the continuum image, while the lowest are detected towards the regions MM2 \RN{2} and MM3 \RN{1} both associated with less bright continuum regions. The derived deuterium fractionation ratios only vary by factors 2--4 with excitations temperatures between 100 and 400 K.   

The derived CH$_2$DOH/CH$_3$OH ratios are consistently low throughout NGC 6334\RN{1} with a mean CH$_2$DOH/CH$_3$OH ratio over all regions of 0.13$\pm$0.06$\%$ and 0.08$\pm$0.04$\%$ based on the $^{13}$C and $^{18}$O isotopologues respectively. This homogeneity of the CH$_2$DOH/CH$_3$OH ratios derived for the different cores indicate that they formed from the same precursor cloud. The levels are also similar to those derived towards Sgr B2 in the Galactic Centre \citep{Belloche2016} as well as those derived for the high-mass star-forming regions in Orion by \cite{Peng2012}, but lower by factors of a few when compared with the values derived by \cite{Neill2013} for the same region. Because of the similar values, the low deuterium fractionation found in the Galactic Centre is likely not due to astration but rather the high temperatures characterising the region. The methanol deuterium fractionation levels derived for the low-mass systems HH212 and IRAS 16293 are higher by up to two orders of magnitude when compared with the values inferred for NGC 6334\RN{1}, clearly illustrating the differences in physical and chemical environment prevailing in high- and low-mass star-forming regions.

Based on the chemical model of \cite{Taquet2014}, the CH$_2$DOH/CH$_3$OH ratios detected towards the regions of NGC 6334\RN{1} indicate a dust temperature at the time of the systems formation $\sim$30 K. This temperature is higher than what is predicted for low-mass star-forming regions where the levels of deuterium in simple molecular species indicate a dust temperature at the time of formation below 20 K.    

In addition to CH$_2$DOH, CH$_3$OD is unambiguously detected towards all nine regions of NGC 6334\RN{1} allowing CH$_2$DOH/CH$_3$OD ratios to be derived. These range from 0.10--0.50, the highest value associated with the region MM2 \RN{1} and the lowest with MM1 \RN{2}. Our CH$_2$DOH/CH$_3$OD ratios towards NGC 6334\RN{1} are lower than what is expected from statistics, assuming the CH$_3$ and OH functional groups of methanol to be equally likely to be deuterated. While CH$_2$DOH/CH$_3$OD ratios of the order of unity have been detected towards high-mass star-forming regions previously, the low levels derived here have not been seen before and may hint at a favoured formation of CH$_3$OD or destruction of CH$_2$DOH in this region which has yet to be explained by models or experiments.

\begin{acknowledgements} 
A special thanks to Dr. Holger M{\"u}ller for valuable discussions of the spectroscopy of deuterated methanol. We acknowledge data reduction support from Allegro, the European ALMA Regional Center node in the Netherlands, and Dr. Luke Maud in particular for expert advice. We thank the anonymous referee for constructive comments that improved our manuscript. Support for B.A.M. was provided by NASA through Hubble Fellowship grant \#HST-HF2-51396 awarded by the Space Telescope Science Institute, which is operated by the Association of Universities for Research in Astronomy, Inc., for NASA, under contract NAS5-26555. This paper makes use of the following ALMA data: ADS/JAO.ALMA\#2015.1.00150.S and \#2015.A.00022.T. ALMA is a partnership of ESO (representing its member states), NSF (USA) and NINS (Japan), together with NRC (Canada) and NSC and ASIAA (Taiwan) and KASI (Republic of Korea), in cooperation with the Republic of Chile. The Joint ALMA Observatory is operated by ESO, AUI/NRAO and NAOJ.
This work is based on analysis carried out with the CASSIS software and JPL: http://spec.jpl.nasa.gov/ and CDMS: http://www.ph1.uni-koeln.de/cdms/ spectroscopic databases. CASSIS has been developed by IRAP-UPS/CNRS (http://cassis.irap.omp.eu). 
\end{acknowledgements}

\bibliographystyle{aa} % style aa.bst
\bibliography{BibTex/NGC6334I} 

\clearpage
\appendix
\section{ADS/JAO.ALMA\#2015.A.00022.T}\label{app:Brogan}
This appendix summarises the additional data set introduced in Section \ref{sec:observations}. The data confirm the presence of CH$_3^{18}$OH, CH$_2$DOH and CH$_3$OD in NGC 6334\RN{1} but do not constrain the column densities of O$^{13}$CS and CH$_3$NC better than the primary data set. This is due to blending with other species and, in the case of O$^{13}$CS, a limited number of lines covered. The column densities of O$^{13}$CS and CH$_3$NC derived from the primary data set are consistent with the data in this appendix. 

%--------BEGIN TABLE: Coordinates and densities---------------
\begin{table}[!h]%[htbp] %[here,top,bottom,page]
	\centering
	\caption{Summary of data set ALMA\#2015.A.00022.T}
	\label{tab:Brogan_regions}
	\begin{tabular}{cccccccc}
		\toprule
		\multicolumn{2}{c}{Location (J2000)} & $v_{\textrm{LSR}}$ & FWHM & \multicolumn{4}{c}{$N_{\textrm{s}}$} \\
		\cline{1-2} 
		\cline{5-8} 
		R.A. & Decl. & & & $^{13}$CH$_3$OH & CH$_3^{18}$OH & CH$_2$DOH\tablefootmark{a} & CH$_3$OD \\
		& & [km s$^{-1}$] & [km s$^{-1}$] & [$\times$10$^{17}$ cm$^{-2}$] & [$\times$10$^{17}$ cm$^{-2}$] & [$\times$10$^{17}$ cm$^{-2}$] & [$\times$10$^{17}$ cm$^{-2}$] \\
		\midrule
		17:20:53.372 & -35:46:58.140 & -7.0 & 3.0 & 15 & 3 & $<$0.63 & $<$5.0 \\
		\bottomrule
	\end{tabular}
	\tablefoot{%Column densities within the syntesised beam of 0$\overset{\second}{.}$2$\times$0$\overset{\second}{.}$15. 
		All models assume $T_{\textrm{ex}}$ = 200 K. \tablefoottext{a}{Numbers include the vibrational correction factor of 1.25}.} 
\end{table}
%-------------------END TABLE---------------------------

%--------BEGIN TABLE: Coordinates and densities---------------
\begin{table}[!h]%[htbp] %[here,top,bottom,page]
	\centering
	\caption{CH$_2$DOH/CH$_3$OH and CH$_3$OD/CH$_3$OH upper limits from data set ALMA\#2015.A.00022.T}
	\label{tab:Brogan_dueteration_rates}
	\begin{tabular}{!{\extracolsep{4pt}}ccccccc}
		\toprule
		\multicolumn{2}{c}{CH$_2$DOH/CH$_3$OH} & \multicolumn{2}{c}{(CH$_2$DOH/CH$_3$OH)$_{corr}$}\tablefootmark{a} & \multicolumn{2}{c}{CH$_3$OD/CH$_3$OH} & CH$_2$DOH/CH$_3$OD\tablefootmark{b} \\
		\cline{1-2} 
		\cline{3-4} 
		\cline{5-6}
		$^{12}$C/$^{13}$C & $^{16}$O/$^{18}$O & $^{12}$C/$^{13}$C & $^{16}$O/$^{18}$O & $^{12}$C/$^{13}$C & $^{16}$O/$^{18}$O & \\
		$\textrm{[\%]}$ & [$\%$] & [$\%$] & [$\%$] & [$\%$] & [$\%$] & \\
		\midrule
		$<$0.07 & $<$0.05 & $<$0.02 & $<$0.02 & $<$0.54 & $<$0.37 & 0.13 \\
		\bottomrule
	\end{tabular}
	\tablefoot{All CH$_2$DOH/CH$_3$OH ratios include the vibrational correction of 1.25. \tablefoottext{a}{Corrected for statistical weight of the location of the substituted deuterium. For CH$_2$DOH this value is 3, for CH$_3$OD it is 1.} \tablefoottext{b}{Ratios do not include statistical correction factors.}} 
\end{table}
%-------------------END TABLE---------------------------

%--------BEGIN TABLE: LINE SUMMARY---------------
\begin{table*}[!h]%[htbp] %[here,top,bottom,page]
\begin{tiny}
	\centering
	\caption{Summary of covered lines with $A_{\textrm{ij}}>$ 10$^{-5}$ s$^{-1}$ and $T_{\textrm{ex}}<$ 600 K}
	\label{tab:Brogan_line_summary}
	\begin{tabular}{lllcccc}
		\toprule
		Species & \multicolumn{2}{c}{Transition} & Frequency & $E_{up}$ & $A_{\textrm{ij}}$ & Database \\
		\cline{2-3}
		& [QN]$_{\textrm{up}}$\tablefootmark{a} & [QN]$_{\textrm{low}}$\tablefootmark{a}  & [MHz]  & [K] & $\times 10^{-5}$ [s$^{-1}$] & \\  
		\midrule
		$^{13}$CH$_3$OH  
		& 6 1 6 +0 & 5 1 5 +0 & 282 790.743 & 61.73 & 9.36 & CDMS \\
		& 10 1 10 1 & 9 0 9 1 & 281 578.684 & 415.84 & 6.20 & \\
		& 6 1 6 +1  & 5 1 5 +1 & 282 167.288 & 372.77 & 9.50 & \\
		& 6 5 1 -1 & 5 5 0 -1 & 282 364.928 & 468.80 & 2.94 & \\
		& 6 5 2 +1 & 5 5 1 +1 & 282 364.928 & 468.80 & 2.94 & \\
		& 6 3 4 1 & 5 3 3 1 & 282 383.033 & 465.18 & 7.33 & \\
		& 6 4 3 1 & 5 4 2 1 & 282 395.053 & 411.38 & 5.39 & \\
		& 6 -3 3 1 & 5 -3 2 1 & 282 419.021 & 370.43 & 7.30 & \\
		& 6 -2 4 1 & 5 -2 3 1 & 282 424.212 & 412.25 & 8.75 & \\
		& 6 2 4 +1 & 5 2 3 +1 & 282 433.145 & 346.33 & 8.68 & \\
		& 6 -4 2 1 & 5 -4 1 1 & 282 434.336 & 453.35 & 5.44 & \\
		& 6 5 1 1 & 5 5 0 1 & 282 436.442 & 477.34 & 2.99 & \\
		& 6 -5 2 1 & 5 -5 1 1 & 282 438.034 & 594.31 & 2.97 &\\
		& 6 2 5 -1 & 5 2 4 -1 & 282 438.887 & 346.33 & 8.68 &\\
		& 6 1 6 1 & 5 1 5 1 & 282 446.467 & 339.04 & 9.51 & \\
		& 6 3 4 +1 & 5 3 3 +1 & 282 448.135 & 443.86 & 7.34 & \\
		& 6 3 3 -1 & 5 3 2 -1 & 282 448.135 & 443.86 & 7.34 & \\
		& 6 4 2 +1 & 5 4 1 +1 & 282 449.004 & 529.02 & 5.44 & \\
		& 6 4 3 -1 & 5 4 2 -1 & 282 449.004 & 529.02 & 5.44 & \\
		& 6 0 6 1 & 5 0 5 1 & 282 449.004 & 348.11 & 9.79 & \\
		& 6 2 5 1 & 5 2 4 1 & 282 463.121 & 447.49 & 8.72 & \\
		& 6 -1 5 1 & 5 -1 4 1 & 282 495.697 & 460.88 & 9.51 & \\
		& 6 0 6 +1 & 5 0 5 +1 & 282 531.066 & 471.12 & 9.79 & \\
		& 6 1 5 -1 & 5 1 4 -1 & 282 717.665 & 372.86 & 9.56 & \\
		& 3 2 2 -0 & 4 1 3 -0 & 291 536.562 & 51.39 & 2.86 & \\
%		& 4 -4 1 0 & 4 -3 3 0 & 293 372.489 & 111.00 & 0.60 & \\
%		& 14 2 13 1 & 13 3 11 1 & 293 626.988 & 637.23 & 0.72 & \\
		& 13 0 13 +1 & 12 1 12 +1 & 338 759.948 & 205.95 & 21.8 & \\
		& 21 3 18 -0 & 20 4 17 -0 &349 996.298 & 573.04 & 7.80 & \\
		& 1 1 1 +0 & 0 0 0 +0 & 350 103.118 & 16.80 & 32.9 & \\
		& 8 1 7 0 & 7 2 5 0 & 350 421.585 & 102.62 & 7.03 & \\
		\midrule
		CH$_3^{18}$OH 
		& 2 0 2 4 & 1 1 1 5 & 279 462.469 & 306.75 & 2.22 & CDMS\\
		& 6 1 5 0 & 5 1 4 0 & 280 450.272 & 61.46 & 9.35 & \\
		& 12 0 12 0 & 11 1 11 0 & 281 082.602 & 173.41 & 11.9 & \\
		& 11 5 7 2 & 12 4 8 2 & 291 923.895 & 283.67 & 2.64 & \\
		& 12 2 11 0 & 11 3 8 0 & 292 264.636 & 210.95 & 4.06 & \\
		& 3 2 2 0 & 4 1 3 0 & 292 611.830 & 50.83 & 2.83 & \\
		& 12 2 10 0 & 11 3 9 0 & 294 273.734 & 211.05 & 4.17 & \\
		& 18 6 13 0 & 19 5 14 0 & 294 800.838 & 574.95 & 3.47 & \\
		& 18 6 12 0 & 19 5 15 0 & 294 801.017 & 574.95 & 3.47 & \\
		& 8 1 7 2 & 7 2 5 2 & 336 743.182 & 100.85 & 6.08 & \\
		& 13 2 12 0 & 12 3 9 0 & 338 164.707 & 239.84 & 6.56 & \\
		& 10 5 6 2 & 11 4 7 2 & 338 313.086 & 259.21 & 3.67 & \\
		& 4 1 3 2 & 3 0 3 1 & 350 245.511 & 43.00 & 12.0 & \\
		& 20 1 20 0 & 19 2 17 0 & 350 673.946 & 477.07 & 7.52 & \\
		\midrule
		CH$_2$DOH 
		& 11 2 10 0 & 11 1 11 0 & 294 323.578 & 157.47 & 11.7 & JPL\\
		& 14 4 11 1 & 14 3 11 2 & 348 938.891 & 296.93 & 12.4 & \\
		& 14 4 10 1 & 14 3 12 2 & 348 990.168 & 296.93 & 12.5 & \\
		& 13 4 10 1 & 13 3 10 2 & 349 149.466 & 266.96 & 12.3 & \\
		& 13 4 9 1 & 13 3 11 2 & 349 183.827 & 266.96 & 12.3 & \\
		& 12 4 9 1 & 12 3 9 2 & 349 333.979 & 239.13 & 12.1 & \\
		& 11 4 8 1 & 11 3 8 2 & 349 495.208 & 213.44 & 11.8 & \\
		& 11 4 7 1 & 11 3 9 2 & 349 508.871 & 213.44 & 11.8 & \\
		& 10 4 7 1 & 10 3 7 2 & 349 635.597 & 189.89 & 11.5 & \\
		& 10 4 6 1 & 10 3 8 2 & 349 643.597 & 189.89 & 11.5 & \\
		& 7 4 4 1 & 7 3 4 2 & 349 951.685 & 132.08 & 10.0 & \\
		& 7 4 3 1 & 7 3 5 2 & 349 952.720 & 132.08 & 10.0 & \\
		& 5 1 4 1 & 5 0 5 0 & 350 632.072 & 48.99 & 20.7 & \\
		& 8 1 8 0 & 7 1 7 0 & 351 796.429 & 80.09 & 14.1 & \\
		\midrule
		CH$_3$OD 
		& 4 -3 -- 0 & 5 -2 -- 0 & 280 460.396 & 63.67 & 1.72 & \tablefootmark{b}\\
		& 9 -2 -- 0 & 9 -1 -- 0 & 280 630.770 & 115.62 & 13.2 & \\
		& 4 1 -- 0 & 3 0 -- 0 & 292 141.956 & 30.85 & 7.06 & \\
		& 5 1 -- 0 & 4 0 -- 0 & 338 196424 & 41.70 & 10.2 & \\
		& 6 4 -- 0 & 7 3 -- 0 & 349 427.582 & 115.61 & 3.42 & \\
		& 5 1 + 0 & 4 0 + 0 & 349 883.62 & 38.55 & 29.4 & \\
		\bottomrule
	\end{tabular}
	\tablefoot{For CH$_2$DOH only the least blended lines are listed. \tablefoottext{a}{QNs for $^{13}$CH$_3$OH, CH$_3^{18}$OH and CH$_2$DOH are (J K$_\textrm{a}$ K$_\textrm{c}$ v) and QNs for CH$_3$OD are (J K P v)} where v=0, 1, 2 refers to the three sub-states $e_0$, $e_1$ and $o_1$ of the ground state respectively. \tablefoottext{b}{\cite{Walsh2000}, and references therein}.}
\end{tiny}
\end{table*}
%-------------------END TABLE---------------------------

%--------------BEGIN FIGURE: 13C - Brogan -------------------------------
\begin{figure*}[h]
	\centering
	\includegraphics[width=0.85\textwidth, trim={0 0.9cm 1cm 2cm}, clip]{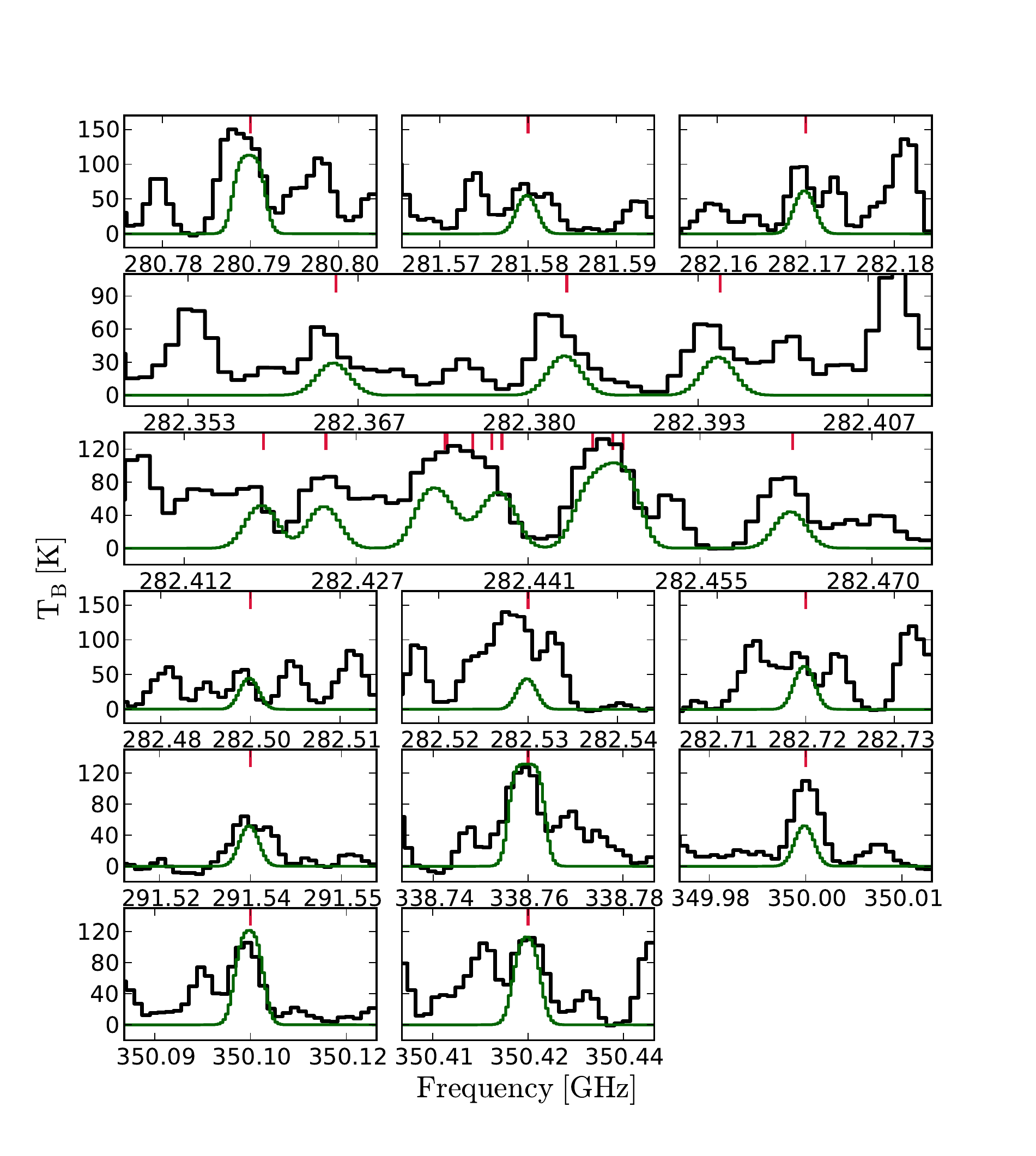} % trim={<left> <lower> <right> <upper>}
	\caption{$^{13}$CH$_3$OH transitions with $T_{\textrm{ex}}<$ 600 K and $A_{\textrm{ij}}>$ 10$^{-5}$ s$^{-1}$ detected towards NGC 6334\RN{1}. Frequencies are shifted to the rest frame of the region. The data and model are shown in black and green respectively. Rest frequencies of individual lines are indicated in red.}
	\label{fig:13C_brogan}
\end{figure*}
%--------------END FIGURE--------------------------------------

%--------------BEGIN FIGURE: 18O - Brogan -------------------------------
\begin{figure*}[h]
	\centering
	\includegraphics[width=0.85\textwidth, trim={0 3.9cm 1cm 2cm}, clip]{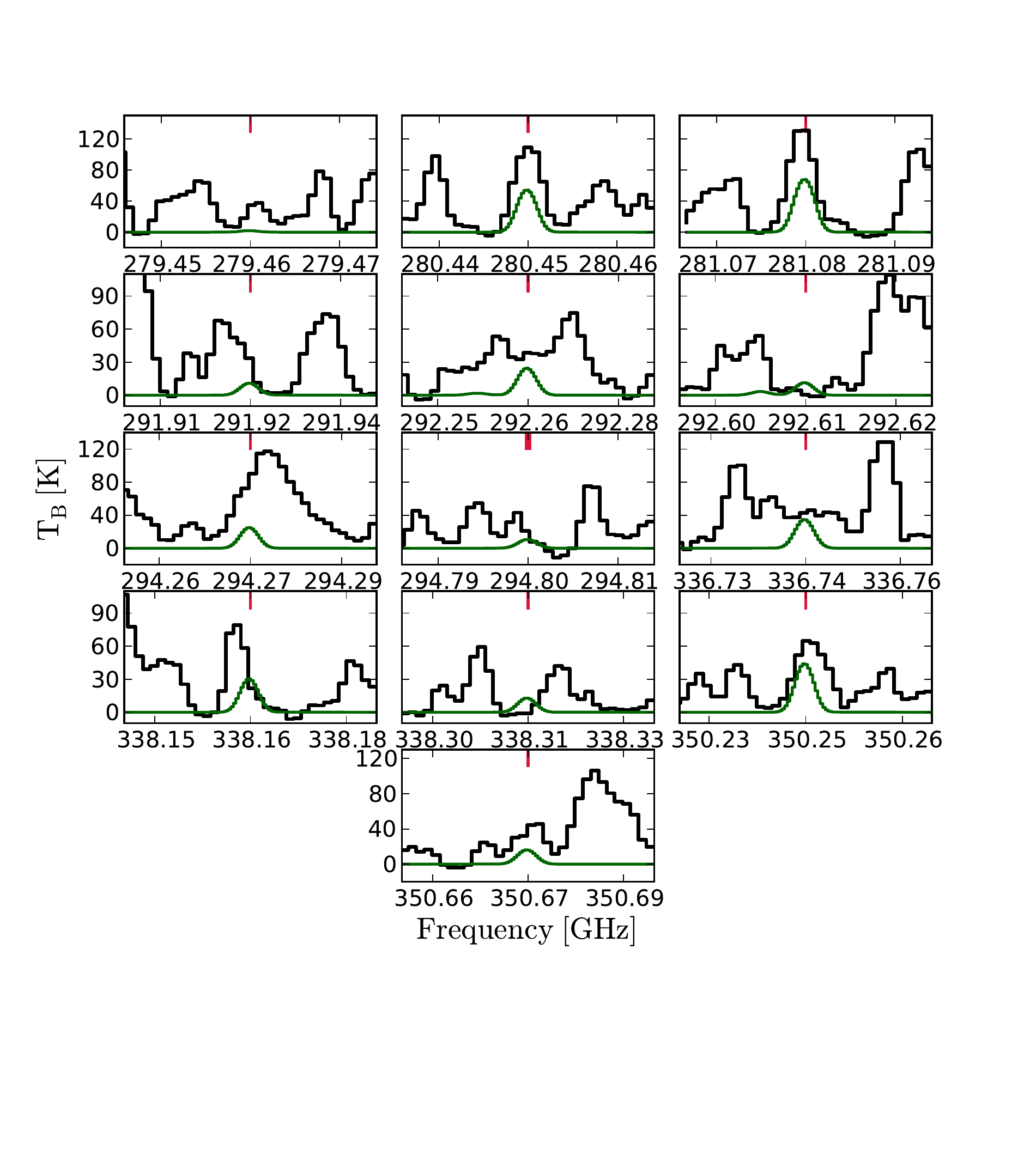} % trim={<left> <lower> <right> <upper>}
	\caption{CH$_3^{18}$OH transitions with $T_{\textrm{ex}}<$ 600 K and $A_{\textrm{ij}}>$ 10$^{-5}$ s$^{-1}$ detected towards NGC 6334\RN{1}. Frequencies are shifted to the rest frame of the region. The data and model are shown in black and green respectively. Rest frequencies of individual lines are indicated in red.}
	\label{fig:18O_brogan}
\end{figure*}
%--------------END FIGURE--------------------------------------

%--------------BEGIN FIGURE: DOH - Brogan -------------------------------
\begin{figure*}[h]
	\centering
	\includegraphics[width=0.85\textwidth, trim={0 8cm 1cm 2cm}, clip]{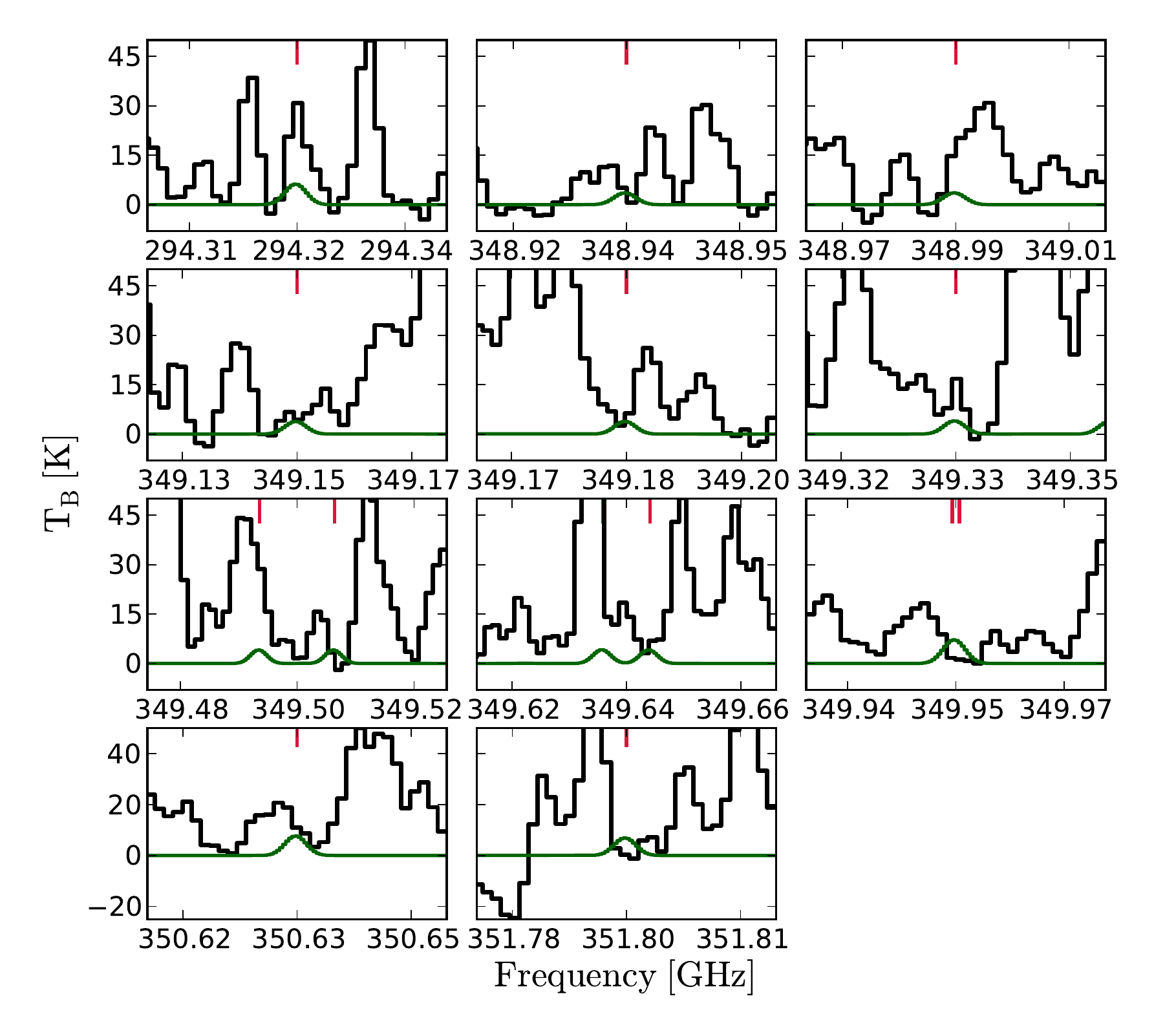} % trim={<left> <lower> <right> <upper>}
	\caption{Selection of CH$_2$DOH transitions with $T_{\textrm{ex}}<$ 400 K and $A_{\textrm{ij}}>$ 10$^{-4}$ s$^{-1}$ detected towards NGC 6334\RN{1}. Frequencies are shifted to the rest frame of the region. The data and model are shown in black and green respectively. Rest frequencies of individual lines are indicated in red.}
	\label{fig:D_brogan}
\end{figure*}
%--------------END FIGURE--------------------------------------

%--------------BEGIN FIGURE: OD - Brogan -------------------------------
\begin{figure*}[h]
	\centering
	\includegraphics[width=0.85\textwidth, trim={0 15.2cm 1cm 2cm}, clip]{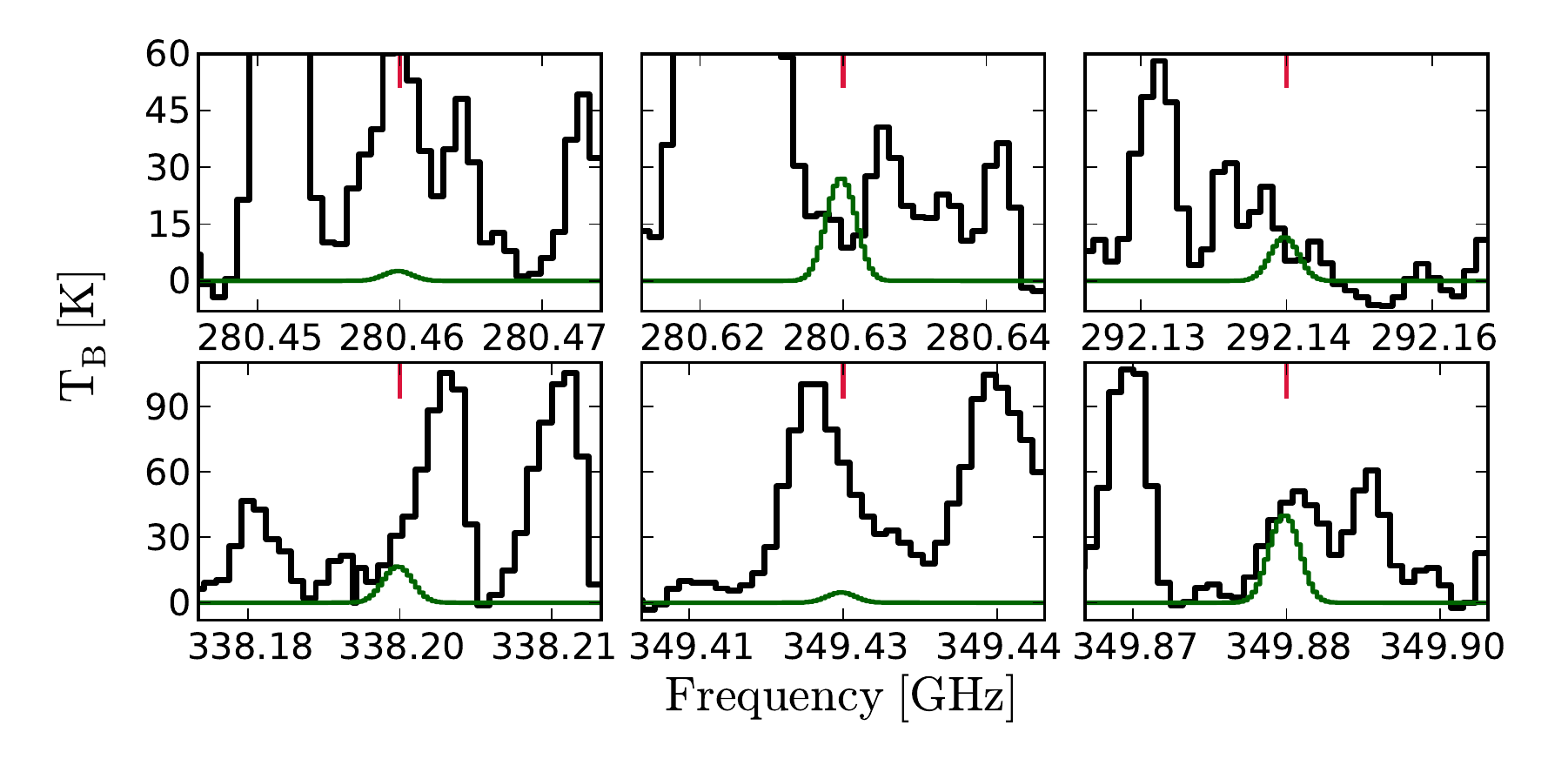} % trim={<left> <lower> <right> <upper>}
	\caption{CH$_3$OD transitions detected towards NGC 6334\RN{1}. Frequencies are shifted to the rest frame of the region. The data and model are shown in black and green respectively. Rest frequencies of individual lines are indicated in red.}
	\label{fig:OD_brogan}
\end{figure*}
%--------------END FIGURE--------------------------------------

\clearpage
\section{Full spectra} \label{app:fullspec}
%--------------BEGIN FIGURE: Full Spec MM1 -------------------------------
\begin{sidewaysfigure*}[b]
	\centering
	\begin{subfigure}[]{0.95\textwidth}
		\includegraphics[width=1.\textwidth]{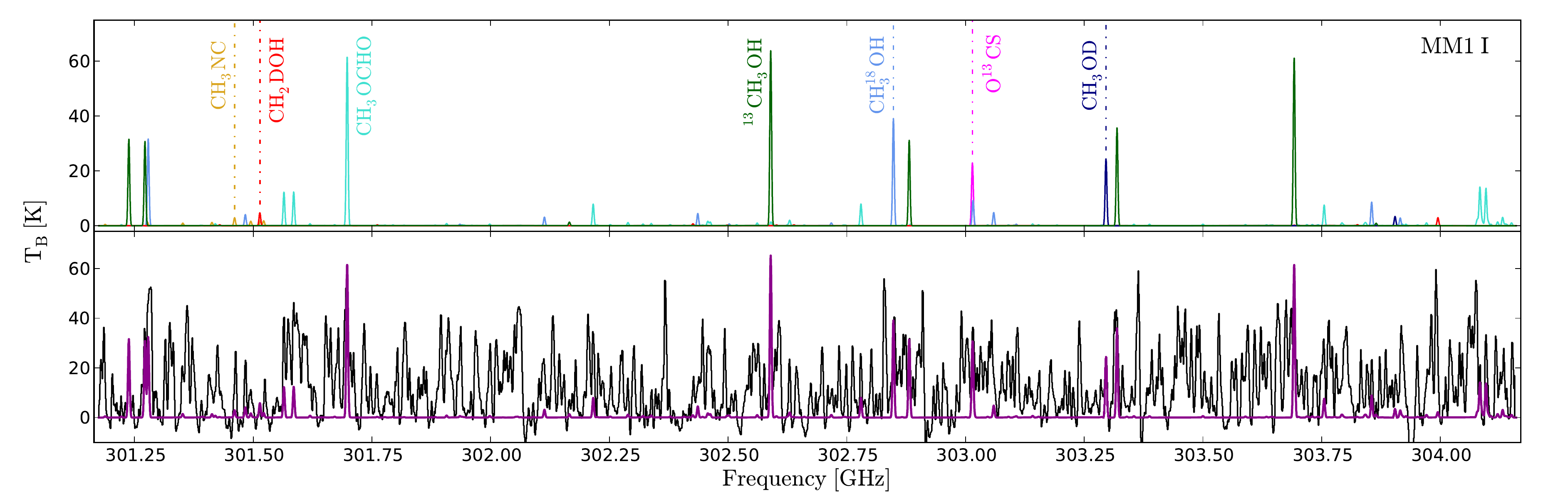}
		\caption*{}
		\label{fig:fullG}
	\end{subfigure}
	\begin{subfigure}{0.95\textwidth}
		\includegraphics[width=1.\textwidth]{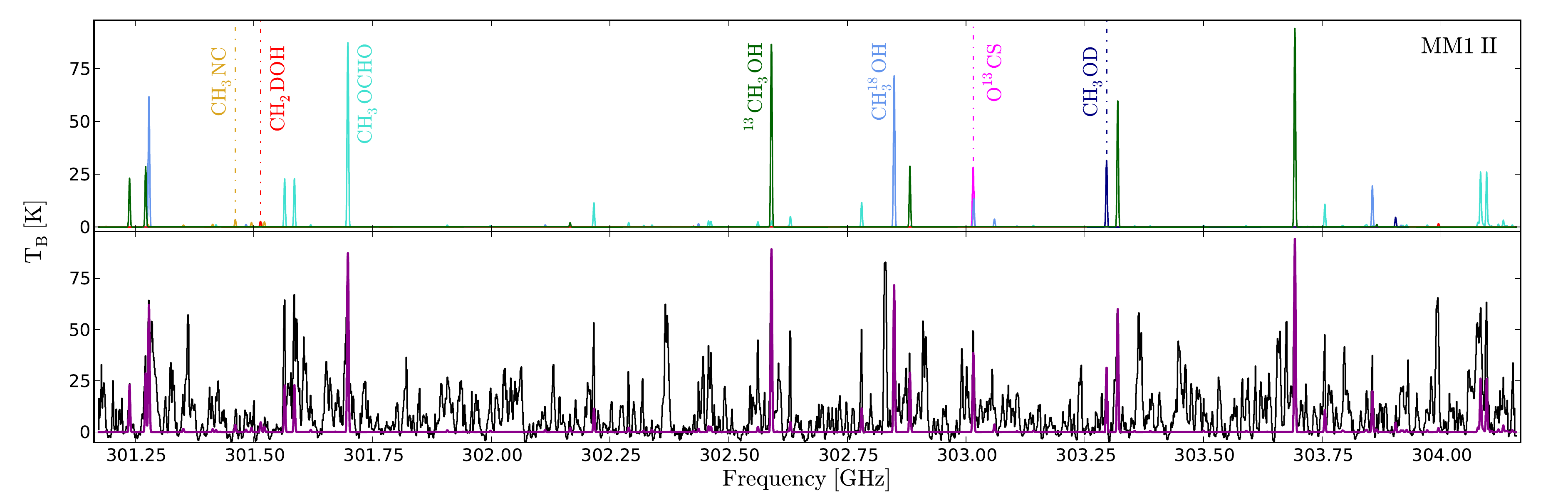}
		\caption*{}
		\label{fig:fullA}
	\end{subfigure}
\caption{MM1 \RN{1}-\RN{5} - continued on next page}
\end{sidewaysfigure*}
\clearpage
\begin{sidewaysfigure*}[]\ContinuedFloat
	\begin{subfigure}{0.95\textwidth}
		\includegraphics[width=1.\textwidth]{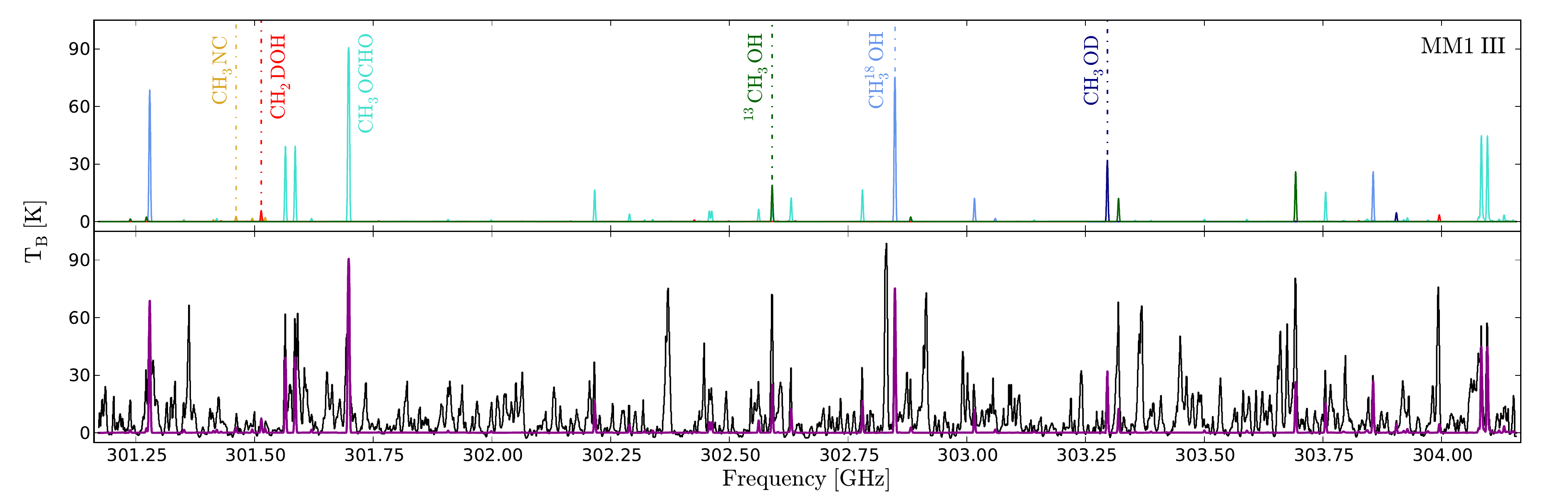}
		\caption*{}
		\label{fig:fullF}
	\end{subfigure}
	\begin{subfigure}{0.95\textwidth}
		\includegraphics[width=1.\textwidth]{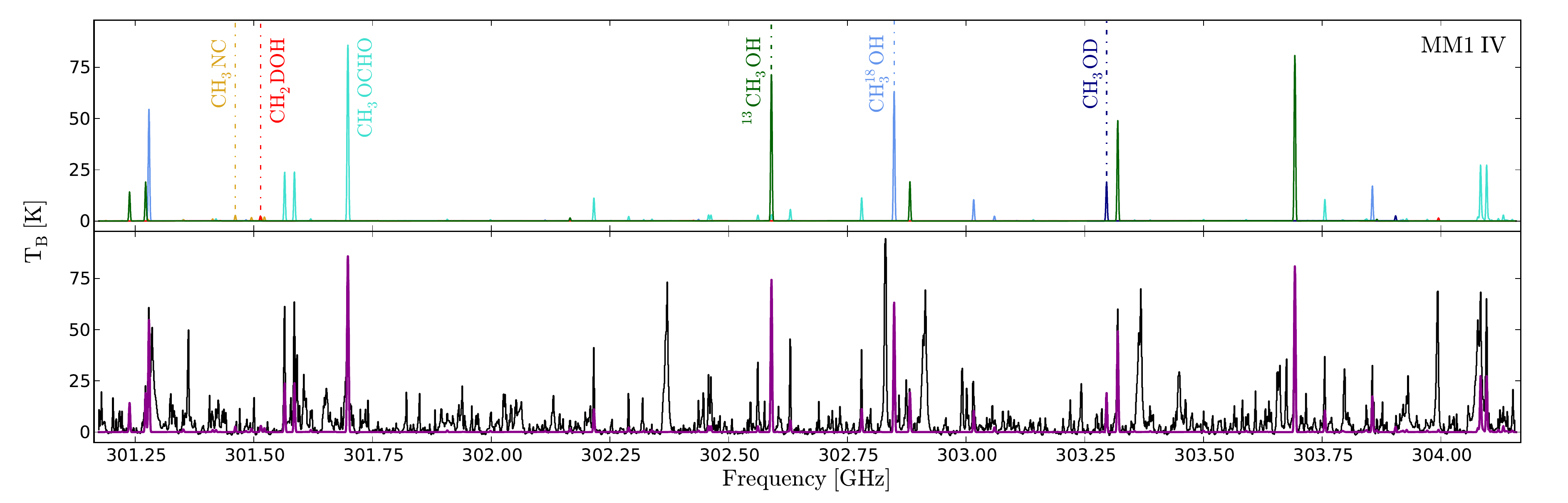}
		\caption*{}
		\label{fig:fullH}
	\end{subfigure}
\caption{MM1 \RN{1}-\RN{5} - continued on next page}
\end{sidewaysfigure*}
\clearpage
\begin{sidewaysfigure*}[]\ContinuedFloat
	\begin{subfigure}{0.95\textwidth}
		\includegraphics[width=1.\textwidth]{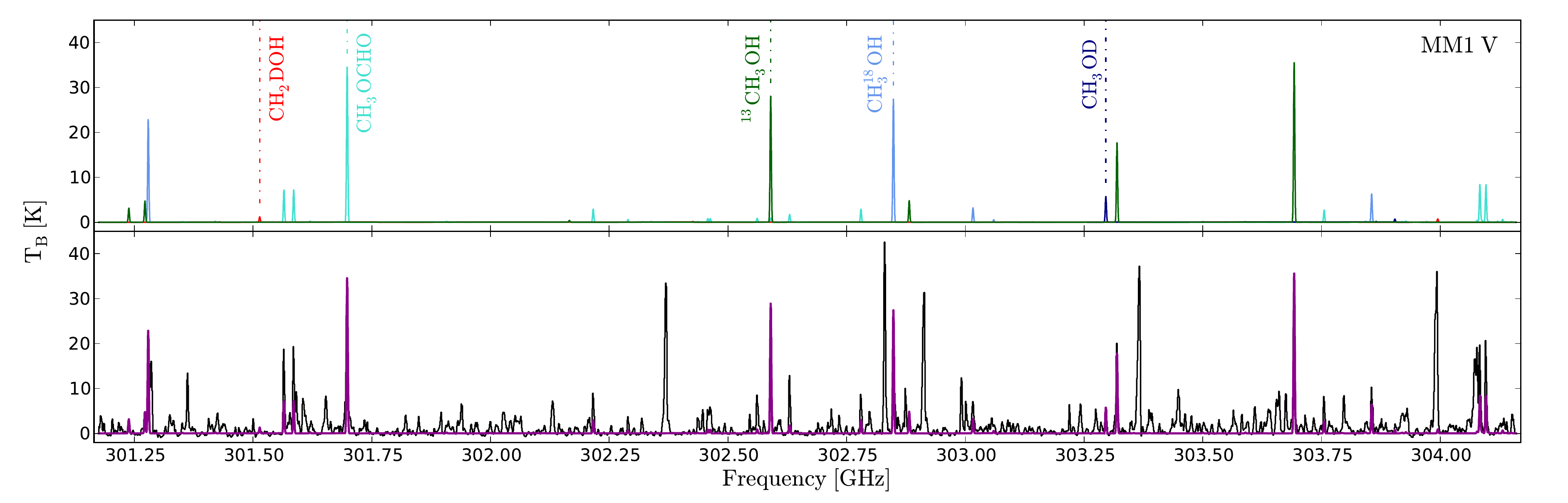}
		\caption*{}
		\label{fig:fullI}
	\end{subfigure}
\caption{MM1 \RN{1}-\RN{5}. Top panels: model components. Bottom panels: data (black) and full model (magenta), i.e., the linear combination of the synthetic spectrum for each species. Frequencies are shifted to the rest frame of the individual regions.}
\label{fig:fullMM1}
\end{sidewaysfigure*}
%--------------END FIGURE-------------------------------------

%--------------BEGIN FIGURE: Full Spec: MM2 I+II-------------------------------
\begin{sidewaysfigure*}[]
	\centering
	\begin{subfigure}[]{0.95\textwidth}
		\includegraphics[width=1\textwidth]{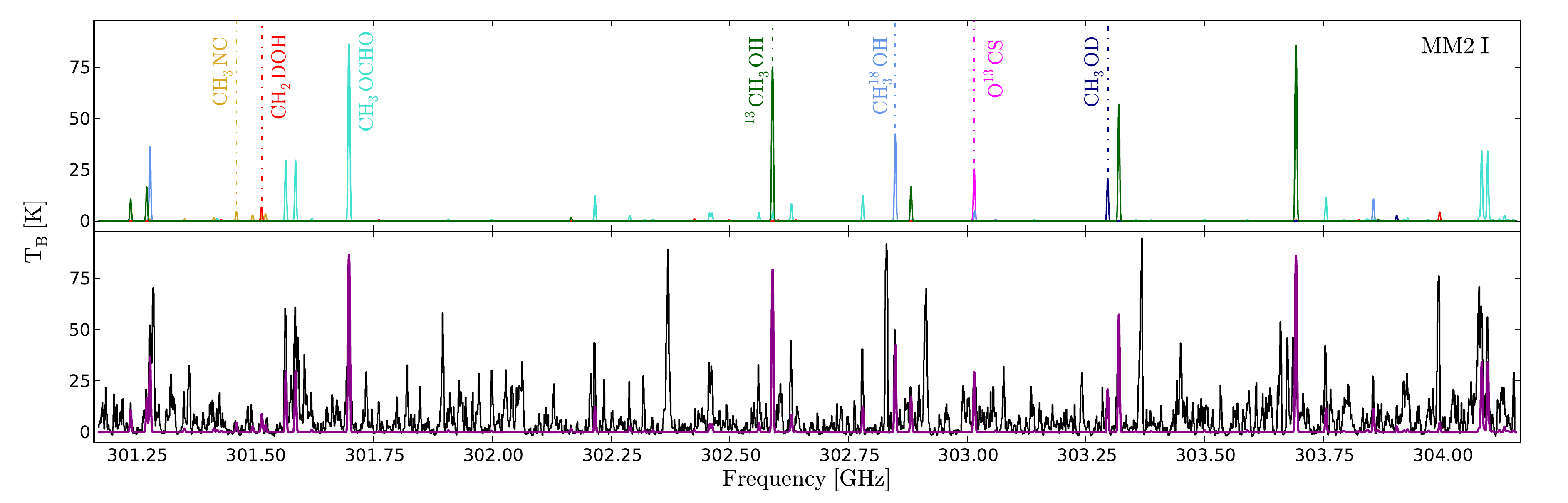}
		\caption*{}
		\label{fig:fullC}
	\end{subfigure}
	\begin{subfigure}{0.95\textwidth}
		\includegraphics[width=1\textwidth]{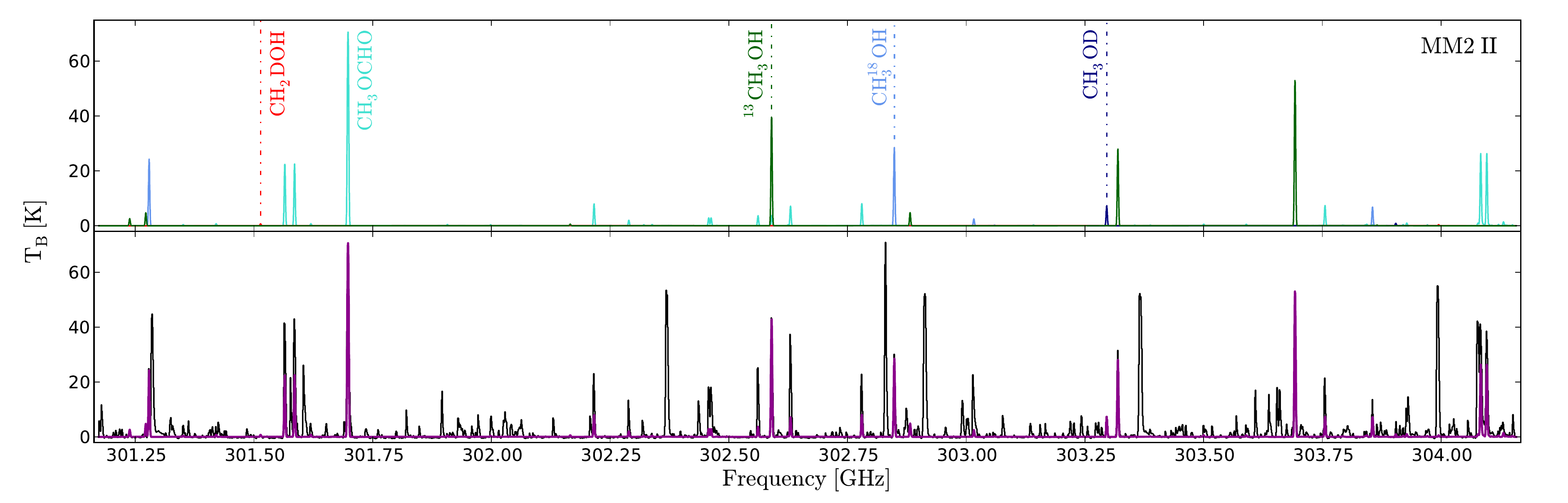}
		\caption*{}
		\label{fig:fullB}
	\end{subfigure}
\caption{MM2 \RN{1}-\RN{2}. Top panels: model components. Bottom panels: data (black) and full model (magenta), i.e., the linear combination of the synthetic spectrum for each species. Frequencies are shifted to the rest frame of the individual regions.}
\label{fig:fullMM2}
\end{sidewaysfigure*}
%--------------END FIGURE-------------------------------------

%--------------BEGIN FIGURE: Full Spec: MM3 I+II-------------------------------
\begin{sidewaysfigure*}[]
	\centering
	\begin{subfigure}[]{0.95\textwidth}
		\includegraphics[width=1\textwidth]{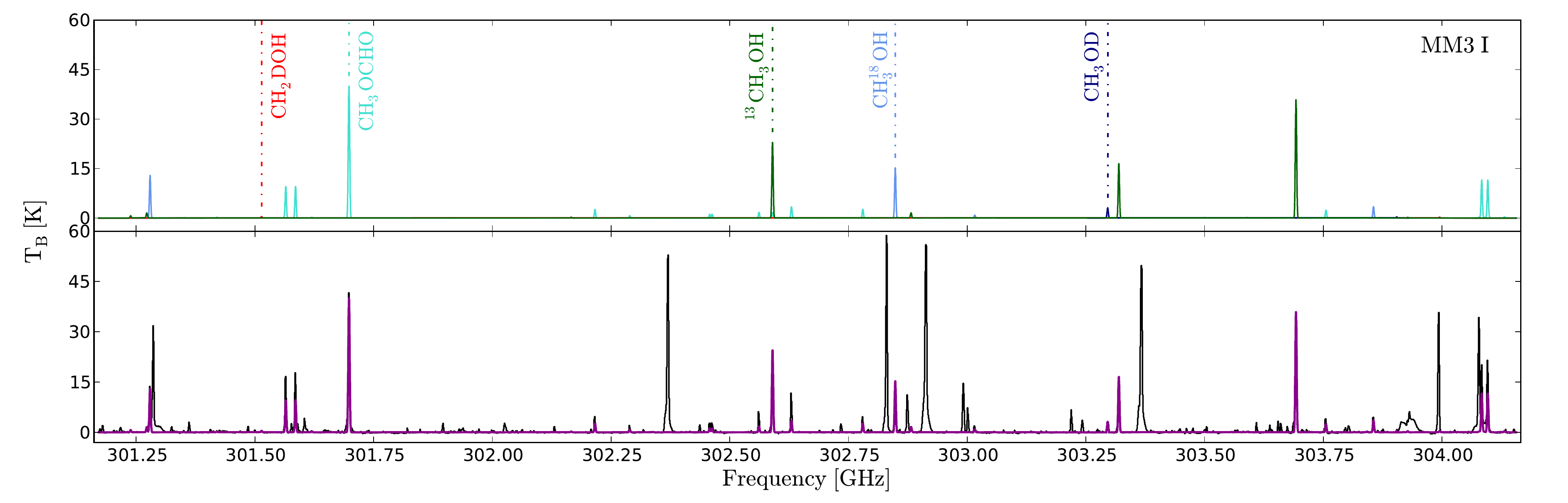}
		\caption*{}
		\label{fig:fullE}
	\end{subfigure}
	\begin{subfigure}{0.95\textwidth}
		\includegraphics[width=1\textwidth]{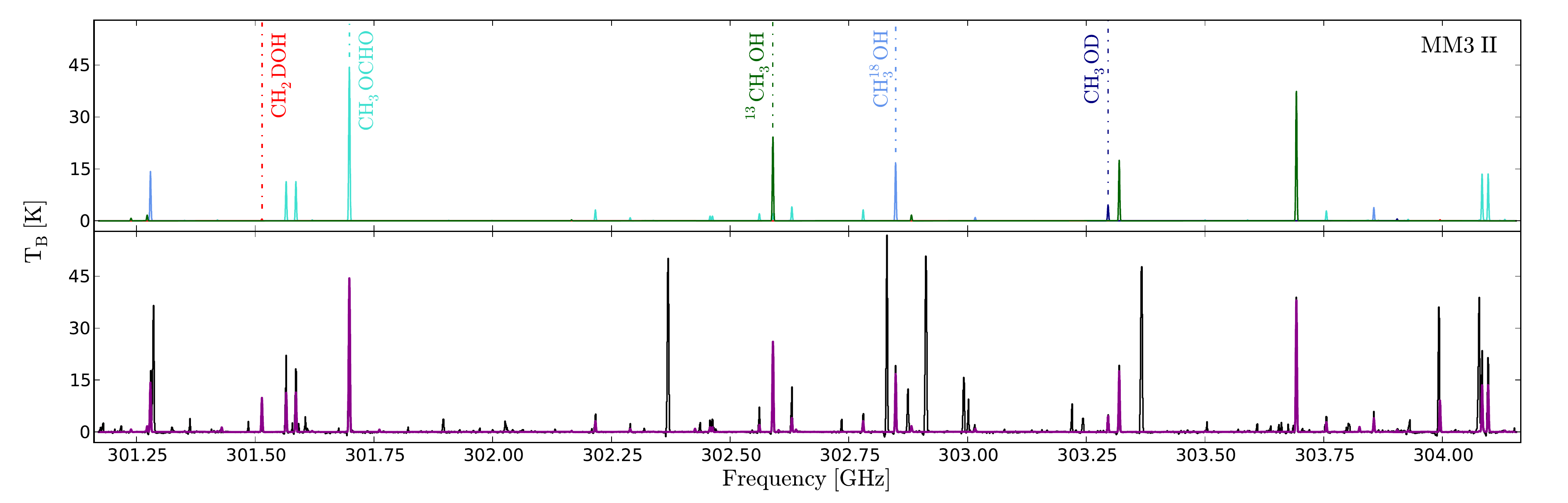}
		\caption*{}
		\label{fig:fullD}
	\end{subfigure}
\caption{MM3 \RN{1}-\RN{2}. Top panels: model components. Bottom panels: data (black) and full model (magenta), i.e., the linear combination of the synthetic spectrum for each species. Frequencies are shifted to the rest frame of the individual regions.}
\label{fig:fullMM3}
\end{sidewaysfigure*}
%--------------END FIGURE-------------------------------------

\clearpage
\section{$^{13}$CH$_3$OH transitions} \label{app:13C}
%--------------BEGIN FIGURE: All 13C lines towards MM1 -------------------------------
\begin{figure*}[b]
\centering
	\begin{subfigure}[]{0.85\textwidth}
		\centering
		\includegraphics[width=1\textwidth, trim={0 12.2cm 0 2cm}, clip]{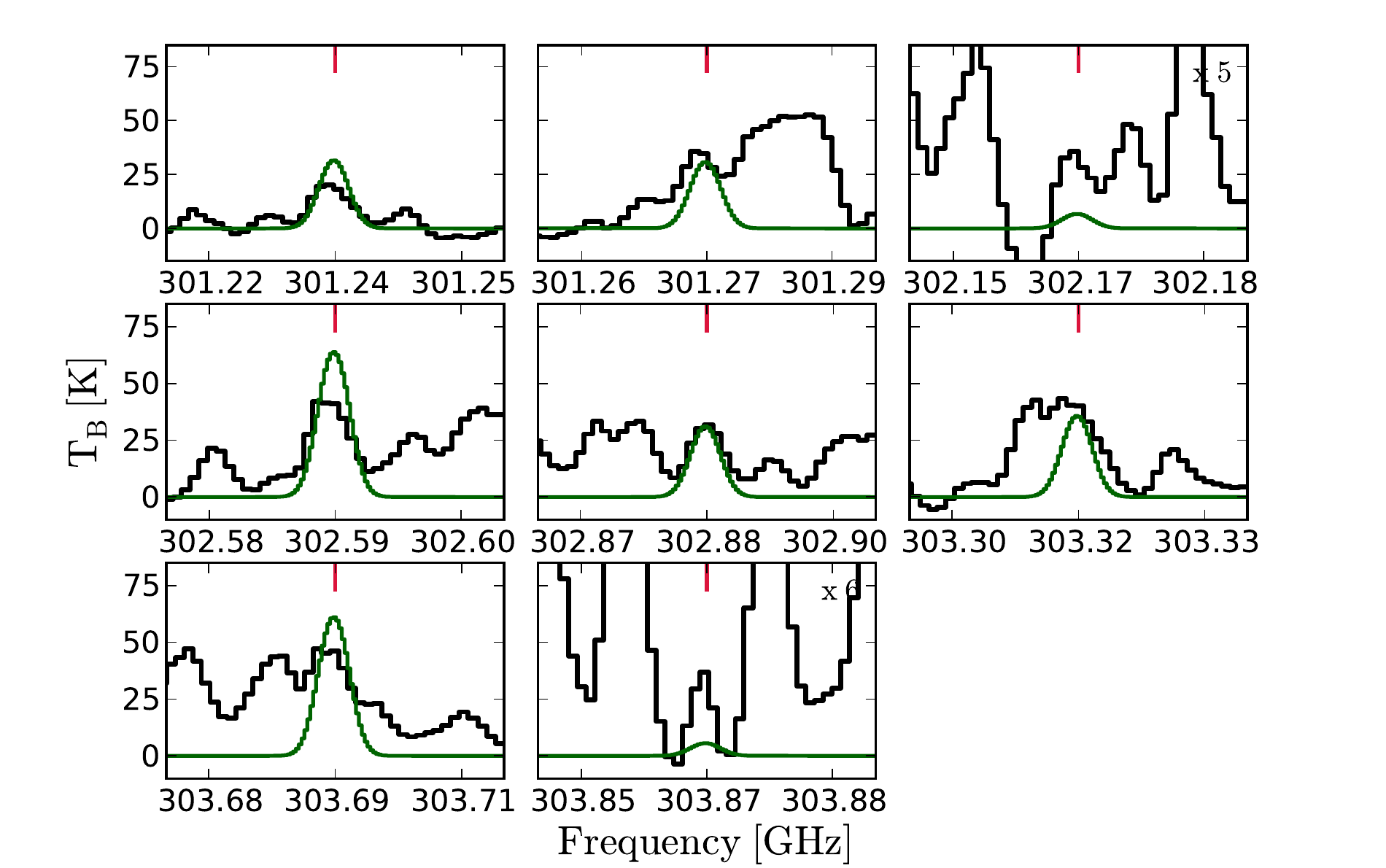}  %trim={<left> <lower> <right> <upper>}
		\caption{MM1 \RN{1}}
		\label{fig:MM1I_All13C}
	\end{subfigure}
	\begin{subfigure}{0.85\textwidth}
		\centering
		\includegraphics[width=1\textwidth, trim={0 12.2cm 0 2cm}, clip]{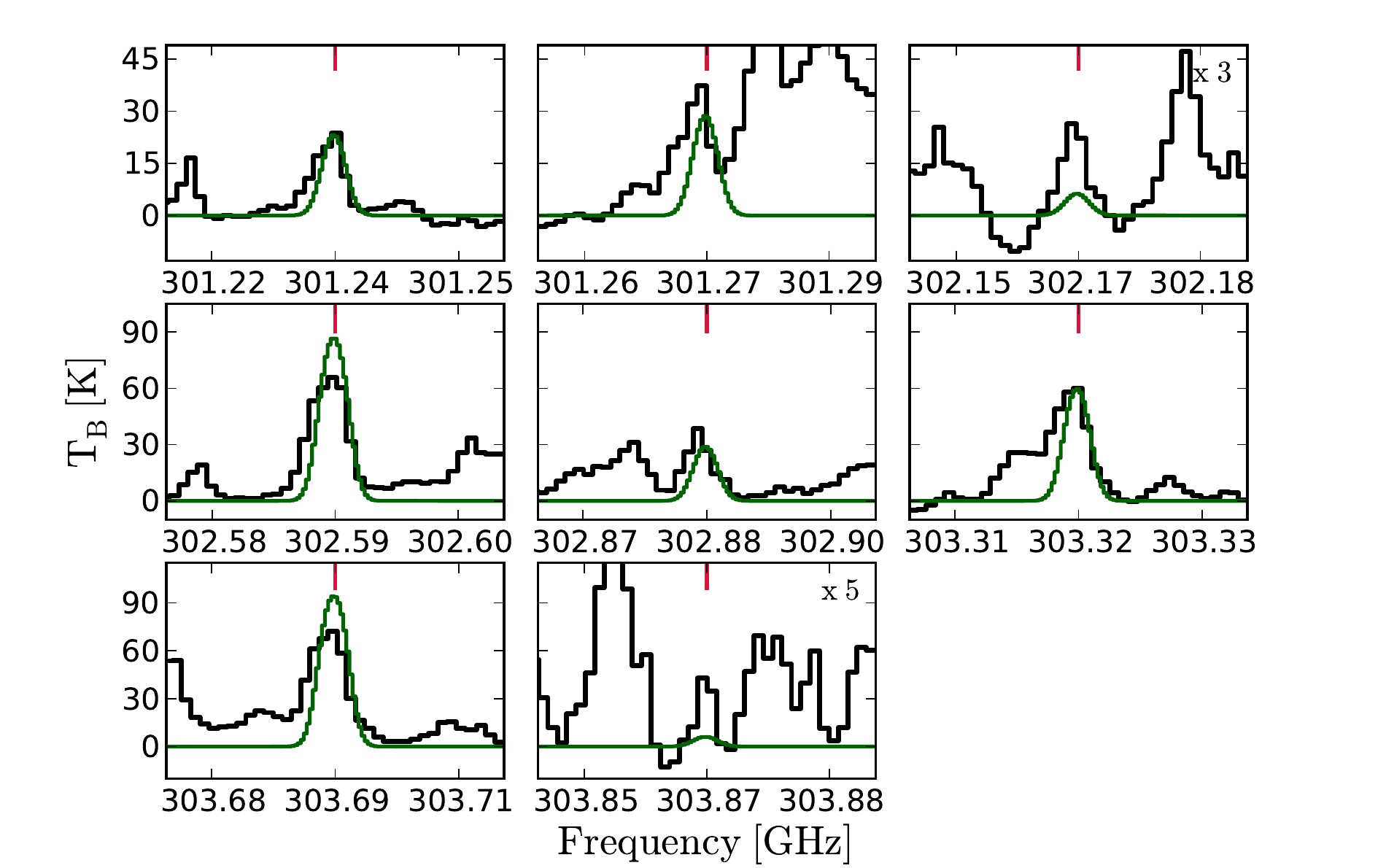}
		\caption{MM1 \RN{2}}
		\label{fig:MM1II_All13C}
	\end{subfigure}
	\caption{All $^{13}$CH$_3$OH lines detected towards NGC 6334\RN{1} MM1 \RN{1}-\RN{5} -- continued on next page}
\end{figure*}
\clearpage
\begin{figure*}[]\ContinuedFloat
\centering
	\begin{subfigure}{0.85\textwidth}
		\centering
		\includegraphics[width=1\textwidth, trim={0 12.2cm 0 2cm}, clip]{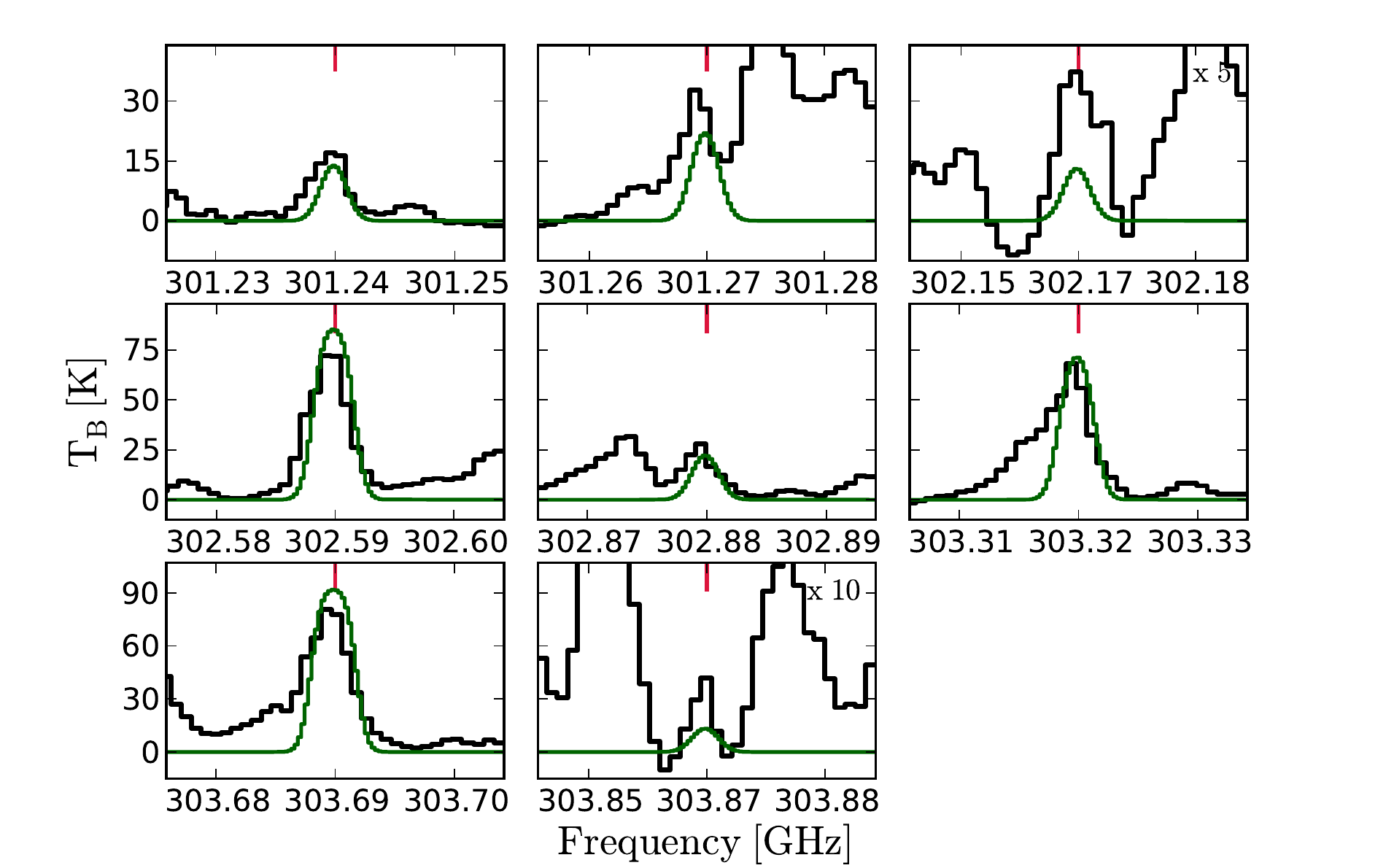}
		\caption{MM1 \RN{3}}
		\label{fig:MM1III_All13C}
	\end{subfigure}
	\begin{subfigure}{0.85\textwidth}
		\centering
		\includegraphics[width=1\textwidth, trim={0 12.2cm 0 2cm}, clip]{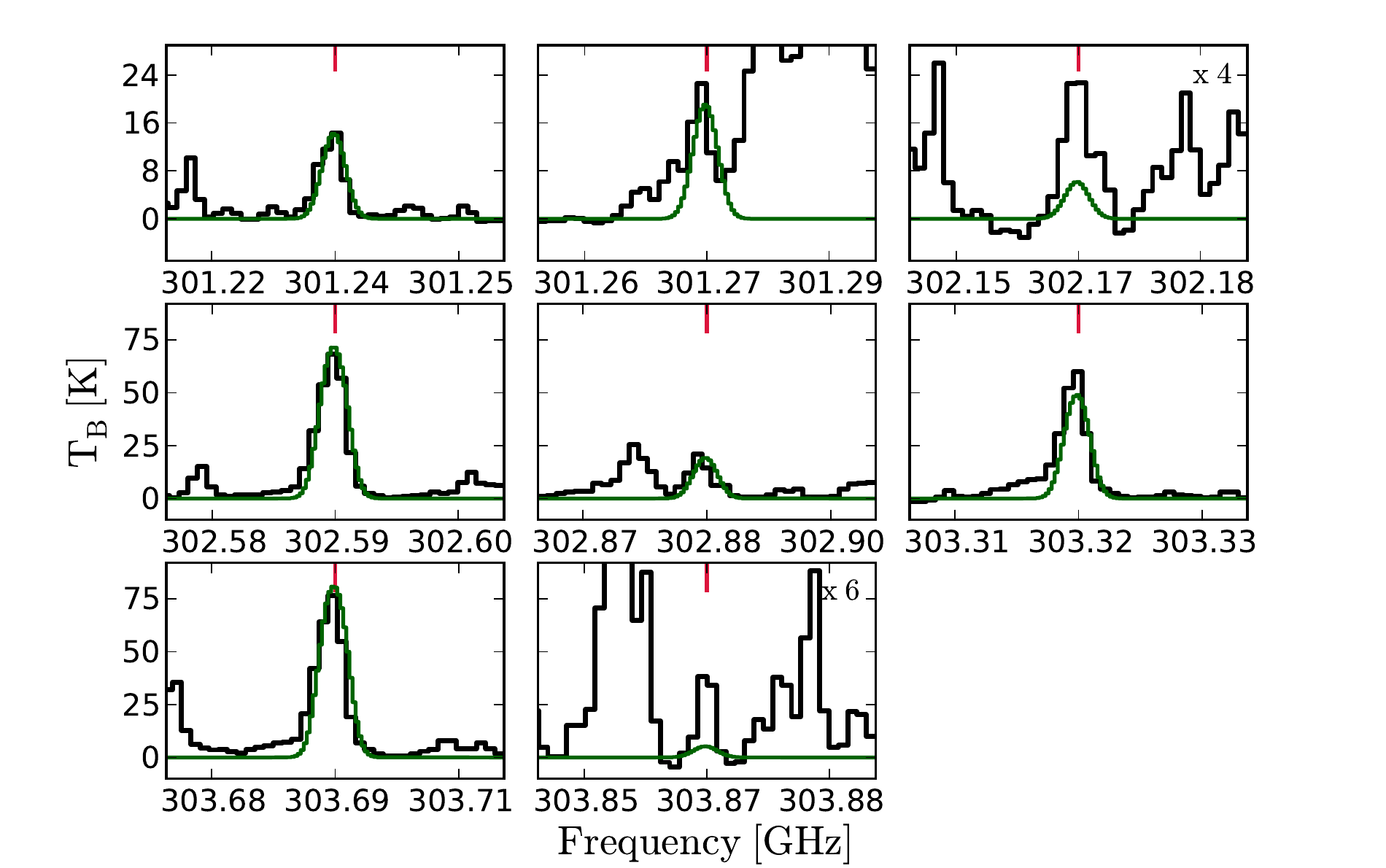}
		\caption{MM1 \RN{4}}
		\label{fig:MM1IV_All13C}
	\end{subfigure}
	\caption{All $^{13}$CH$_3$OH lines detected towards NGC 6334\RN{1} MM1 \RN{1}-\RN{5} -- continued on next page}
\end{figure*}
\clearpage
\begin{figure*}[]\ContinuedFloat
\centering	
	\begin{subfigure}{0.85\textwidth}
		\centering
		\includegraphics[width=1\textwidth, trim={0 12.2cm 0 2cm}, clip]{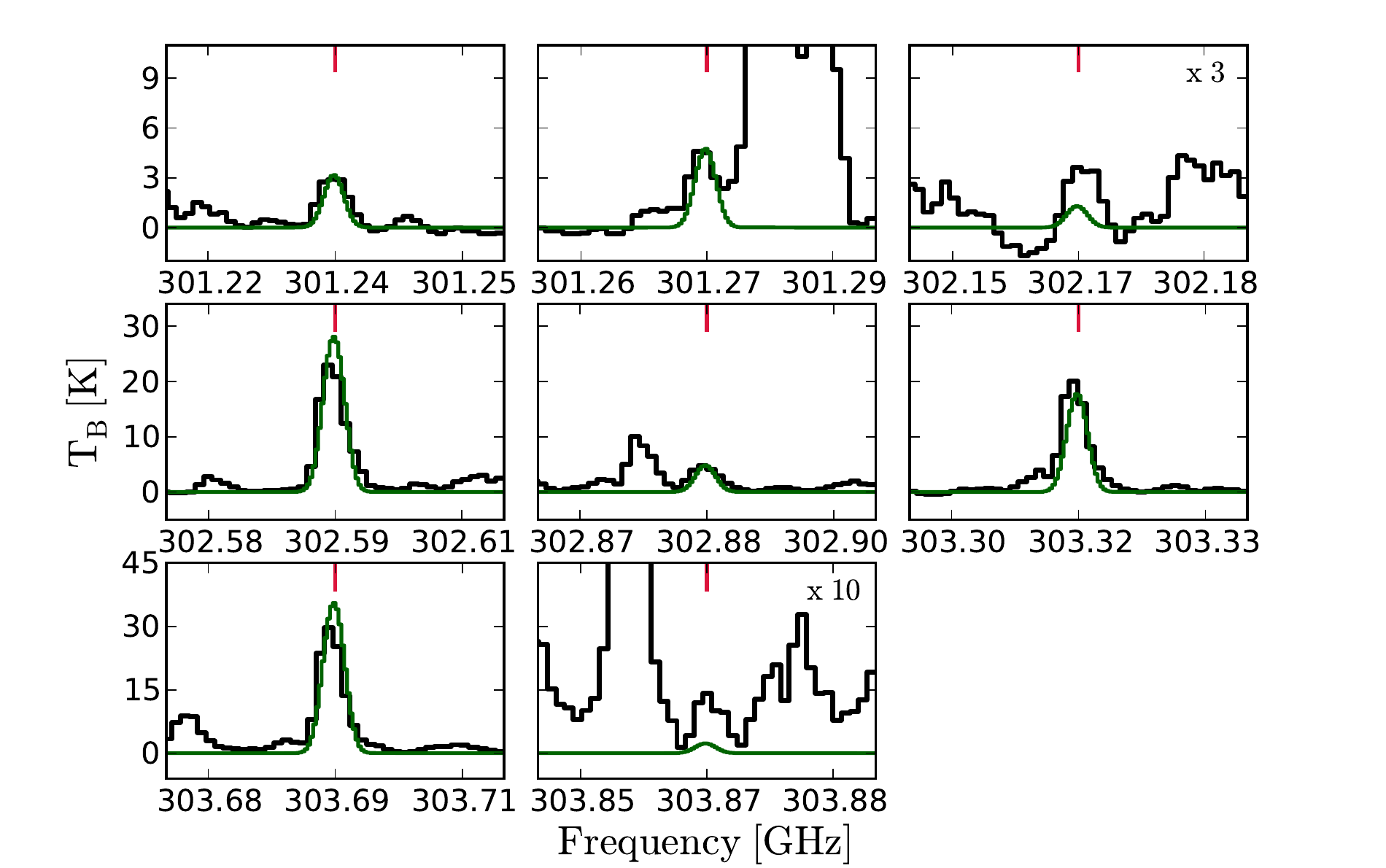}
		\caption{MM1 \RN{5}}
		\label{fig:MM1V_All13C}
	\end{subfigure}
	\caption{All $^{13}$CH$_3$OH lines detected towards NGC 6334\RN{1} MM1 \RN{1}-\RN{5}. Frequencies are shifted to the rest frame of the individual regions. Green lines represent the modelled spectra of $^{13}$CH$_3$OH without blending, i.e., excluding the contribution from CH$_3$OCHO (including the contribution from CH$_3$OCHO does not change the $^{13}$CH$_3$OH column density of the best-fit model). To enhance the readability of some panels, both data and model have been scaled up by the factor in the top right corner of the respective panel.}
	\label{fig:MM1_All13C}
\end{figure*}
%--------------END FIGURE-------------------------------------

%--------------BEGIN FIGURE: All 13C lines towards MM2 -------------------------------
\begin{figure*}[]
	\centering
	\begin{subfigure}[]{0.85\textwidth}
		\centering
		\includegraphics[width=1\textwidth, trim={0 12.2cm 0 2cm}, clip]{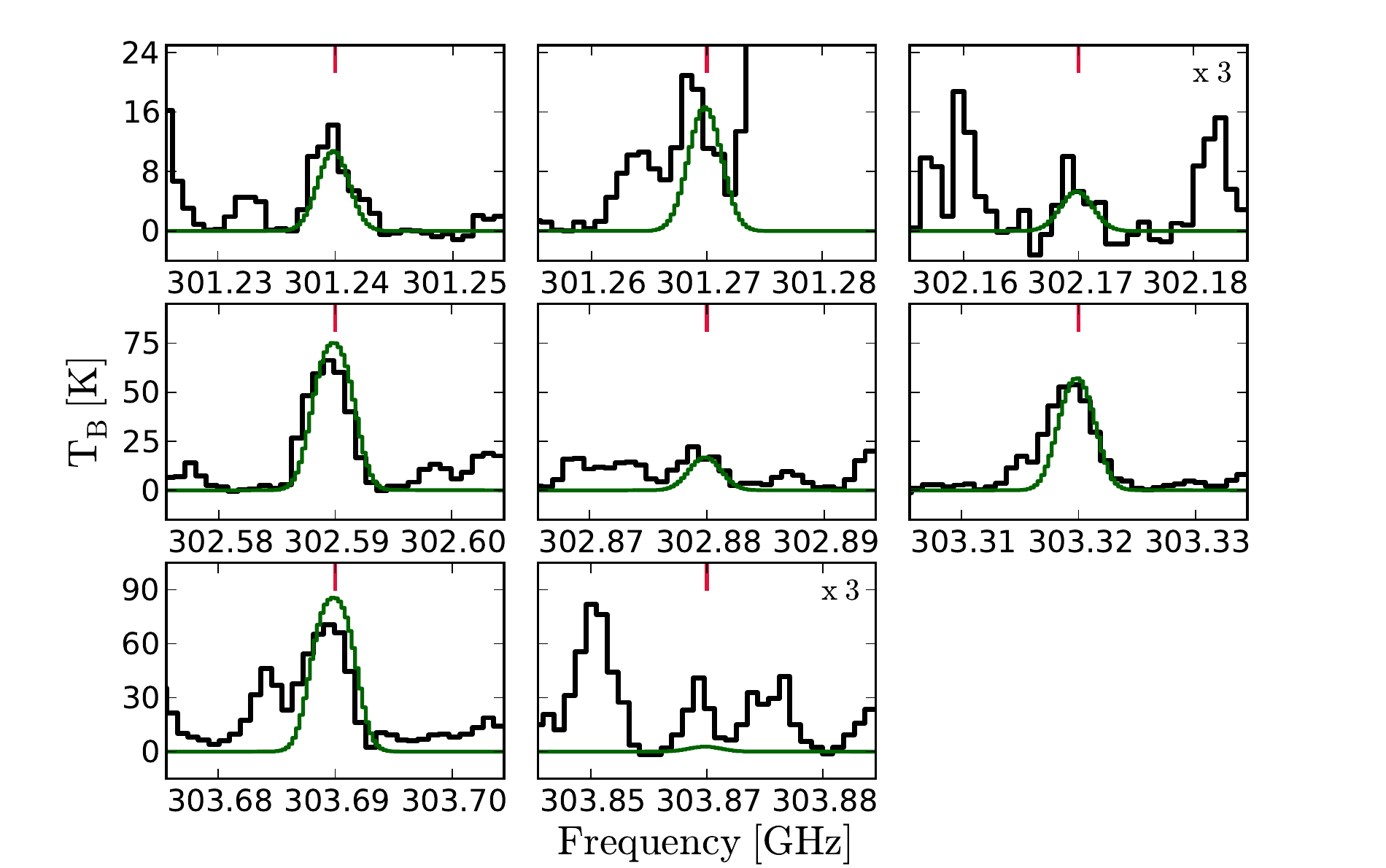}  %trim={<left> <lower> <right> <upper>}
		\caption{MM2 \RN{1}}
		\label{fig:MM2I_All13C}
	\end{subfigure}
	\begin{subfigure}{0.85\textwidth}
		\centering
		\includegraphics[width=1\textwidth, trim={0 12.2cm 0 2cm}, clip]{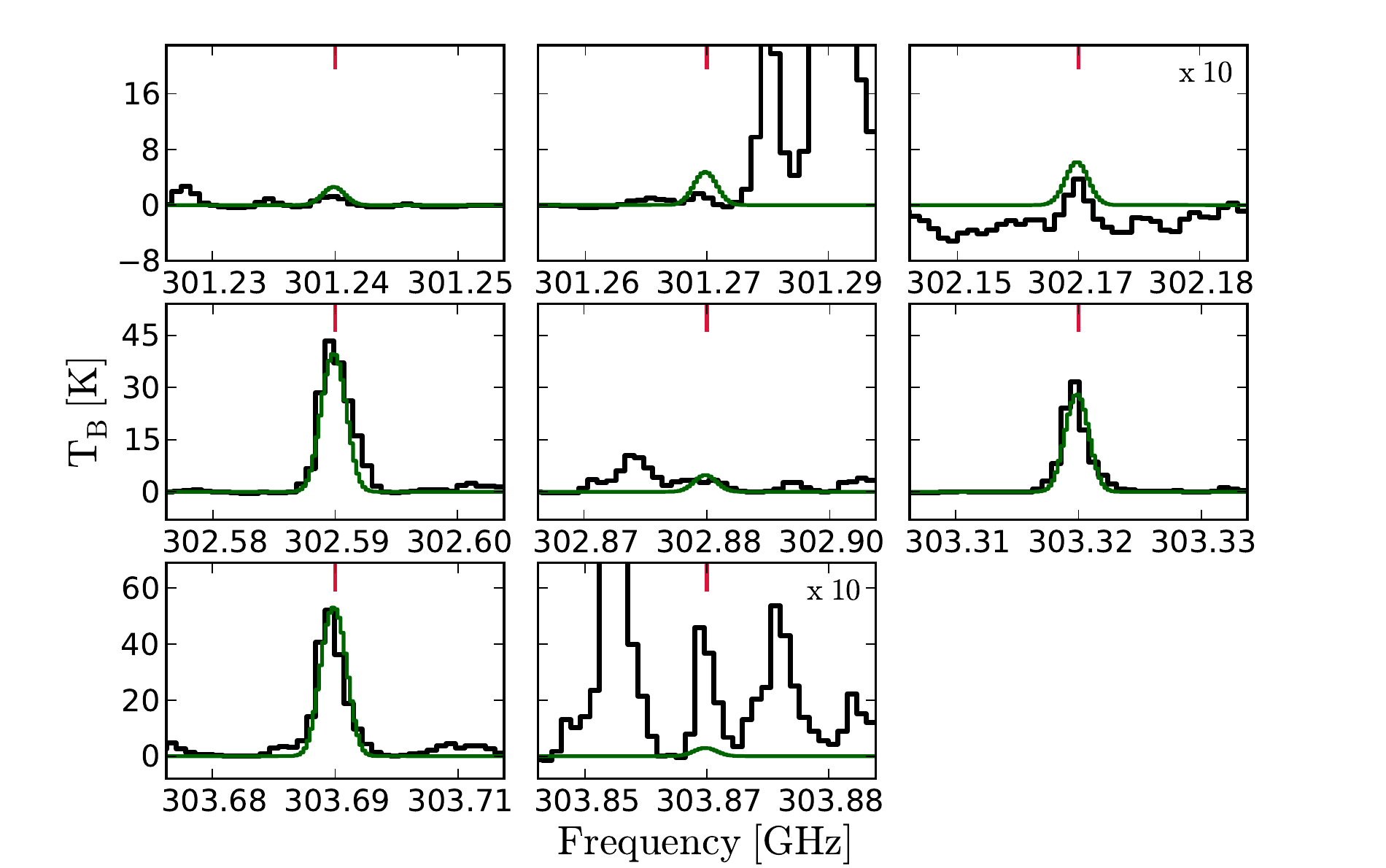}
		\caption{MM2 \RN{2}}
		\label{fig:MM2II_All13C}
	\end{subfigure}
	\caption{All $^{13}$CH$_3$OH lines detected towards NGC 6334\RN{1} MM1 \RN{1}-\RN{5}. Frequencies are shifted to the rest frame of the individual regions. Green lines represent the modelled spectra of $^{13}$CH$_3$OH without blending, i.e., excluding the contribution from CH$_3$OCHO (including the contribution from CH$_3$OCHO does not change the $^{13}$CH$_3$OH column density of the best-fit model). To enhance the readability of some panels, both data and model have been scaled up by the factor in the top right corner of the respective panel.}
	\label{fig:MM2_All13C}
\end{figure*}
%--------------END FIGURE-------------------------------------

%--------------BEGIN FIGURE: All 13C lines towards MM3 -------------------------------
\begin{figure*}[]
	\centering
	\begin{subfigure}[]{0.85\textwidth}
		\centering
		\includegraphics[width=1\textwidth, trim={0 12.2cm 0 2cm}, clip]{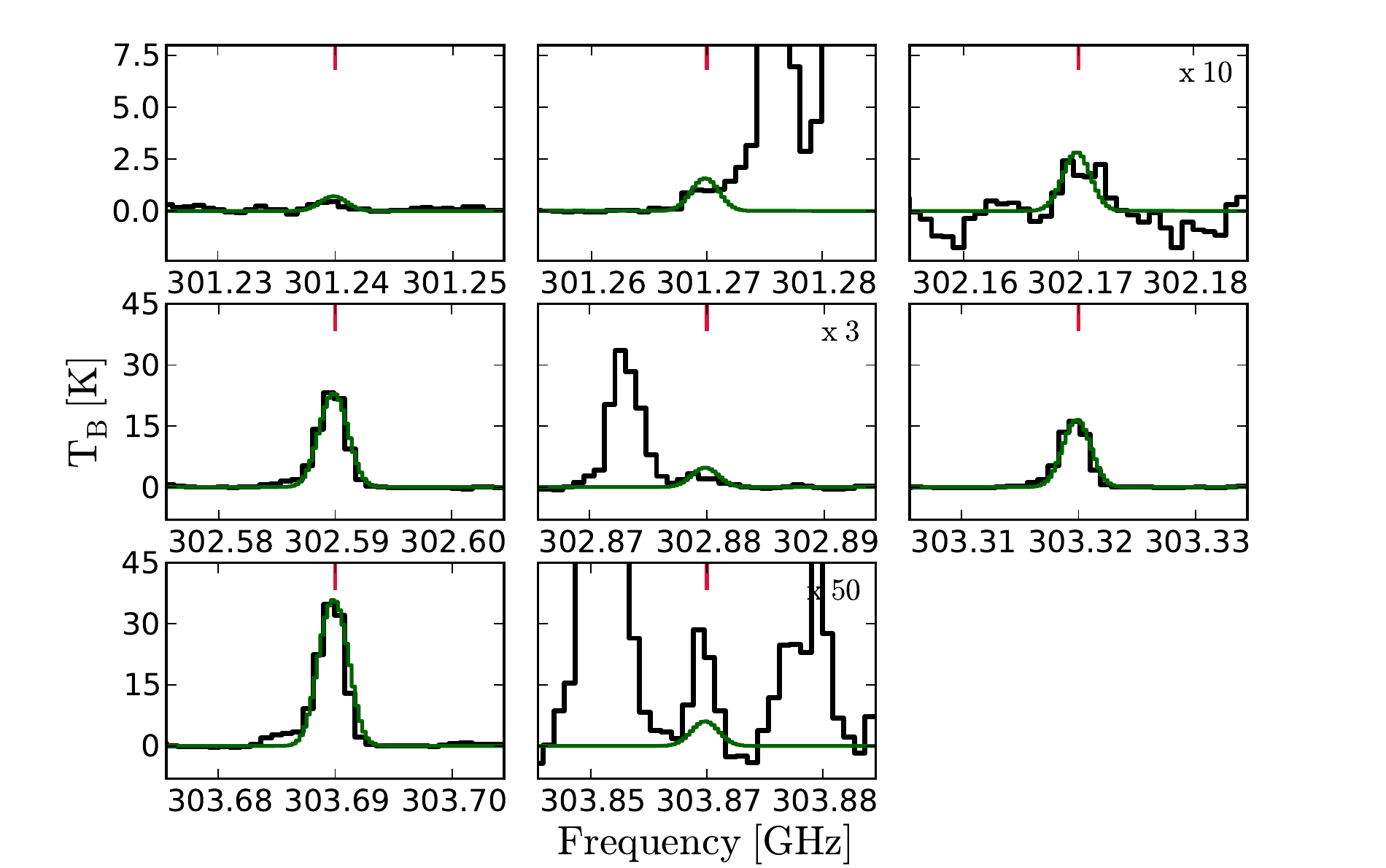}  %trim={<left> <lower> <right> <upper>}
		\caption{MM3 \RN{1}}
		\label{fig:MM3I_All13C}
	\end{subfigure}
	\begin{subfigure}{0.85\textwidth}
		\centering
		\includegraphics[width=1\textwidth, trim={0 12.2cm 0 2cm}, clip]{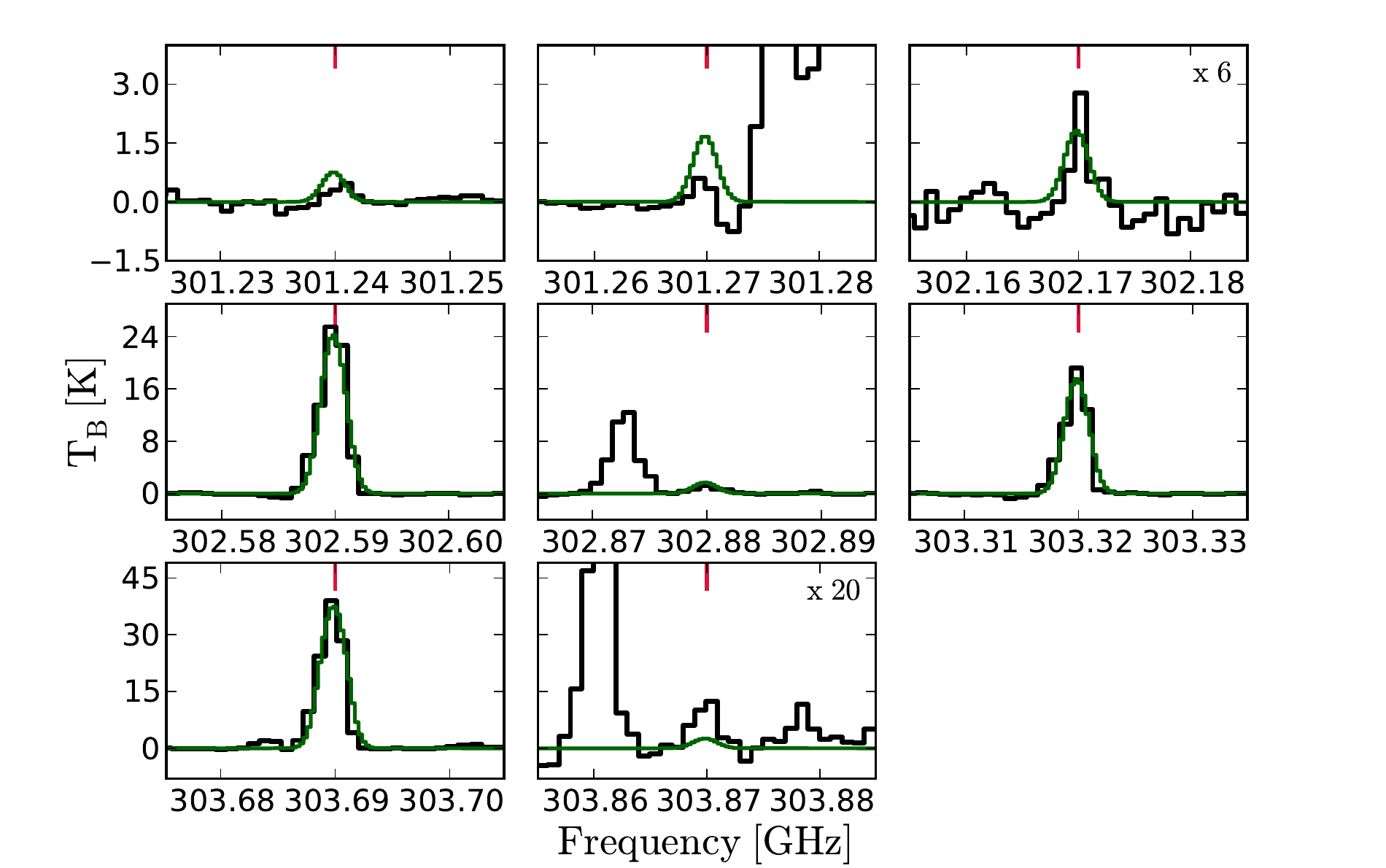}
		\caption{MM3 \RN{2}}
		\label{fig:MM3II_All13C}
	\end{subfigure}
	\caption{All $^{13}$CH$_3$OH lines detected towards NGC 6334\RN{1} MM1 \RN{1}-\RN{5}. Frequencies are shifted to the rest frame of the individual regions. Green lines represent the modelled spectra of $^{13}$CH$_3$OH without blending, i.e., excluding the contribution from CH$_3$OCHO (including the contribution from CH$_3$OCHO does not change the $^{13}$CH$_3$OH column density of the best-fit model). To enhance the readability of some panels, both data and model have been scaled up by the factor in the top right corner of the respective panel.}
	\label{fig:MM3_All13C}
\end{figure*}
%--------------END FIGURE-------------------------------------
\clearpage
\section{CH$_3^{18}$OH transitions} \label{app:18O}
%--------------BEGIN FIGURE: All 18O lines towards MM1 -------------------------------
\begin{figure*}[b]
\centering
	\begin{subfigure}[]{0.85\textwidth}
		\centering
		\includegraphics[width=0.73\textwidth, trim={0 15.6cm 5.3cm 2cm}, clip]{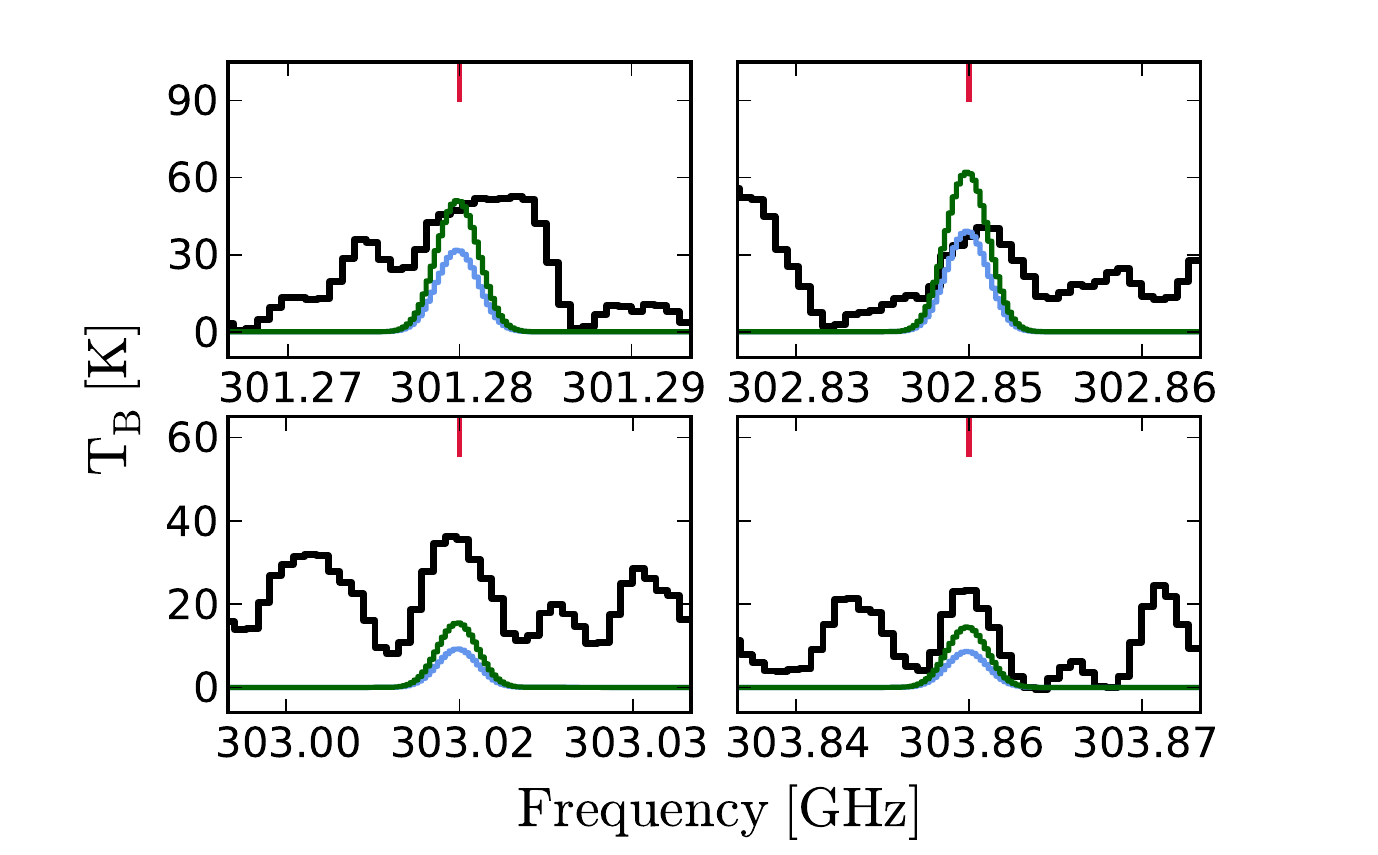}  %trim={<left> <lower> <right> <upper>}
		\caption{MM1 \RN{1}}
		\label{fig:MM1I_All18O}
	\end{subfigure}
	\begin{subfigure}{0.85\textwidth}
		\centering
		\includegraphics[width=0.73\textwidth, trim={0 15.6cm 5.3cm 2cm}, clip]{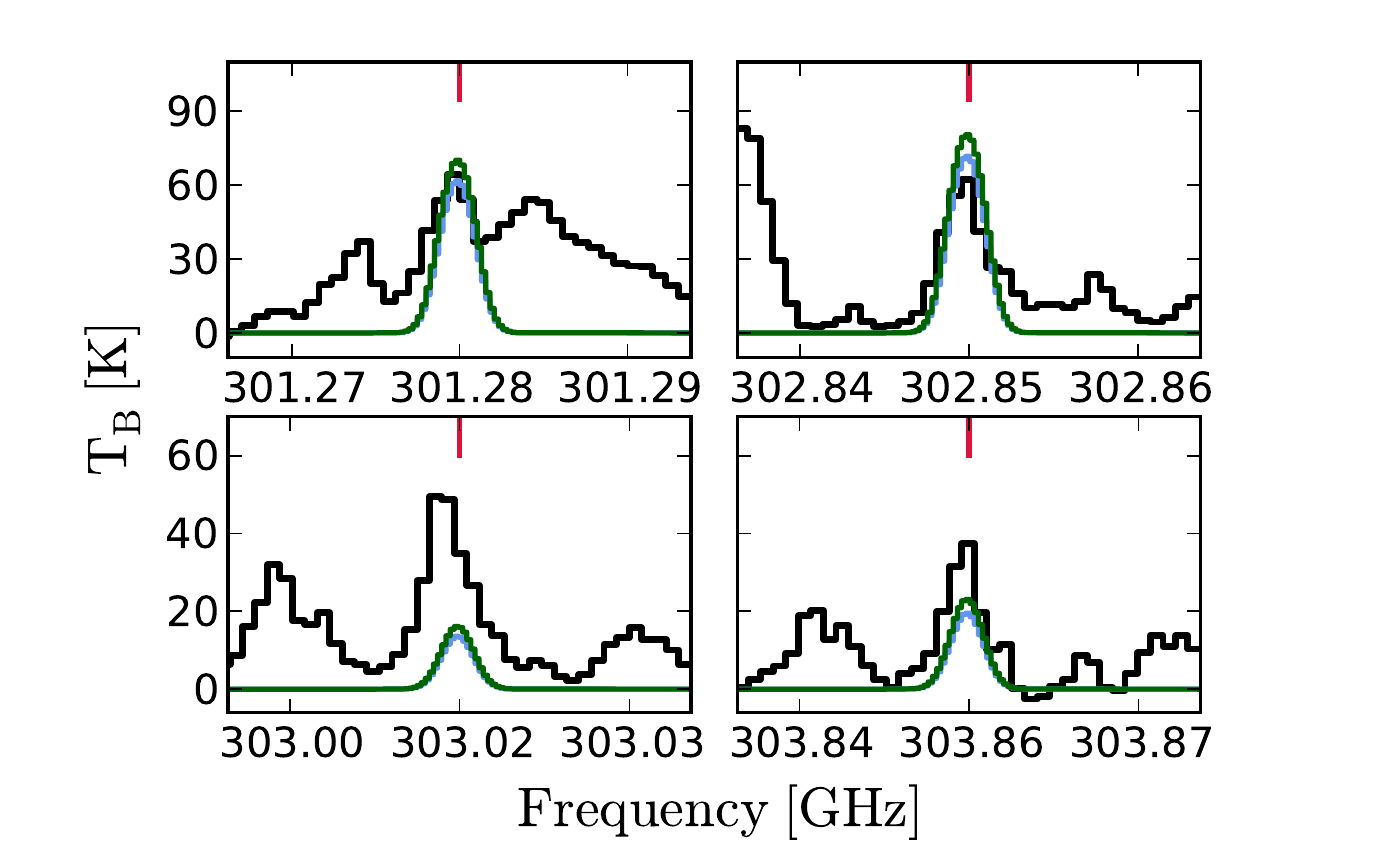}
		\caption{MM1 \RN{2}}
		\label{fig:MM1II_All18O}
	\end{subfigure}
	\caption{All CH$_3^{18}$OH lines detected towards NGC 6334\RN{1} MM1 \RN{1}-\RN{5} -- continued on next page}
\end{figure*}
\clearpage
\begin{figure*}[]\ContinuedFloat
\centering	
	\begin{subfigure}{0.85\textwidth}
		\centering
		\includegraphics[width=0.73\textwidth, trim={0 15.6cm 5.3cm 2cm}, clip]{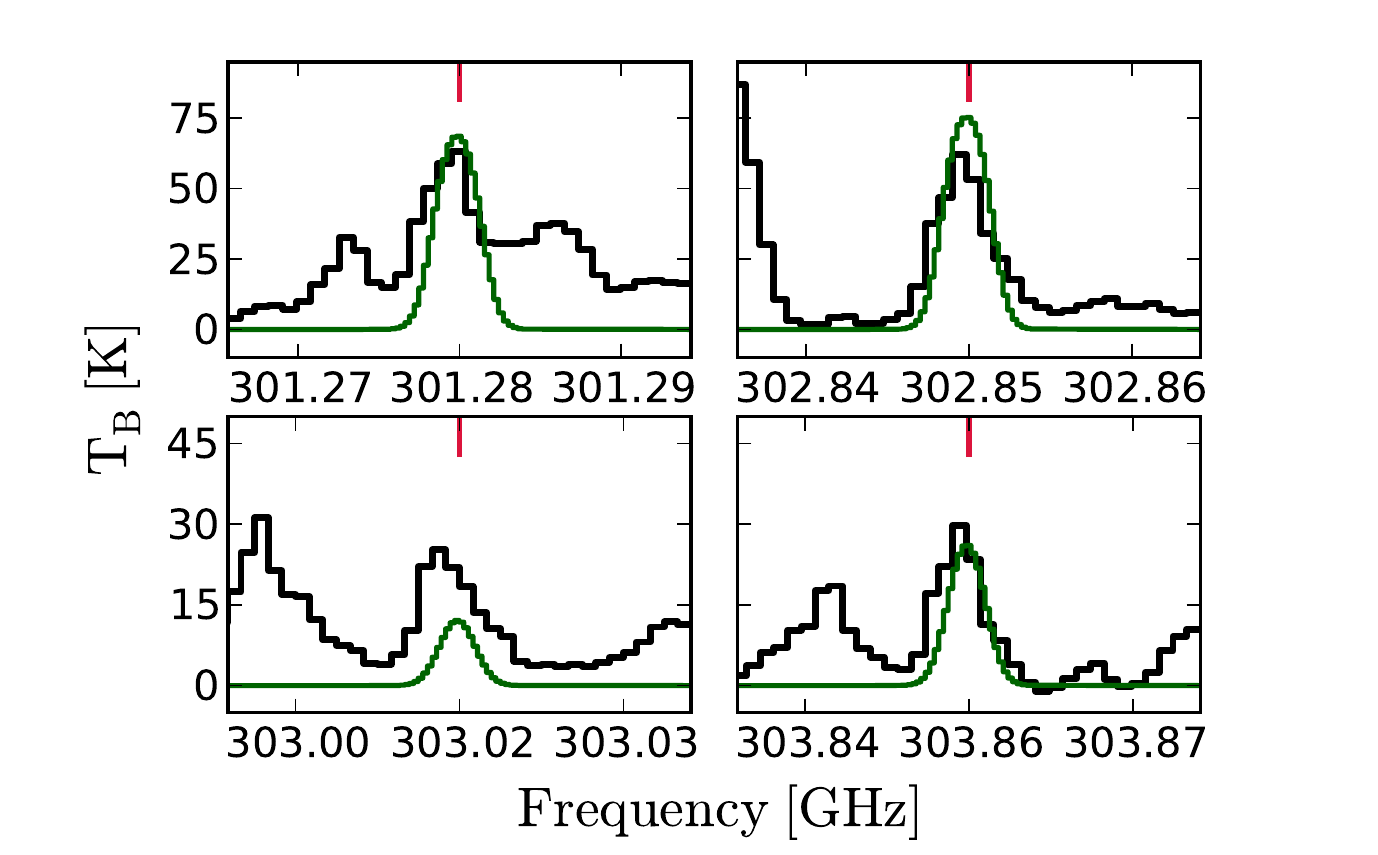}
		\caption{MM1 \RN{3}}
		\label{fig:MM1III_All18O}
	\end{subfigure}
	\begin{subfigure}{0.85\textwidth}
		\centering
		\includegraphics[width=0.73\textwidth, trim={0 15.6cm 5.3cm 2cm}, clip]{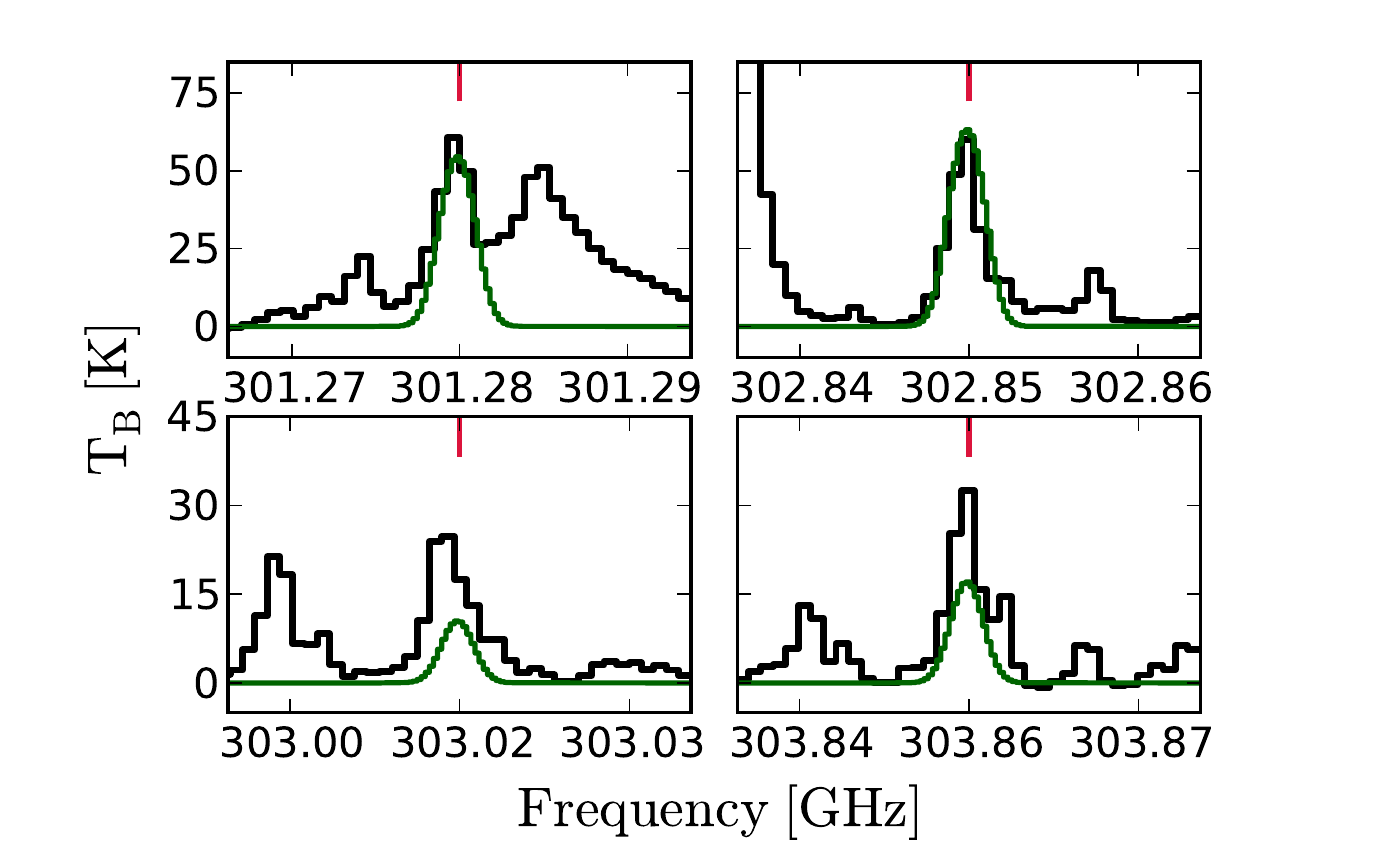}
		\caption{MM1 \RN{4}}
		\label{fig:MM1IV_All18O}
	\end{subfigure}
	\caption{All CH$_3^{18}$OH lines detected towards NGC 6334\RN{1} MM1 \RN{1}-\RN{5} -- continued on next page}
\end{figure*}
\clearpage
\begin{figure*}[]\ContinuedFloat	
\centering	
	\begin{subfigure}{0.85\textwidth}
		\centering
		\includegraphics[width=0.73\textwidth, trim={0 15.6cm 5.3cm 2cm}, clip]{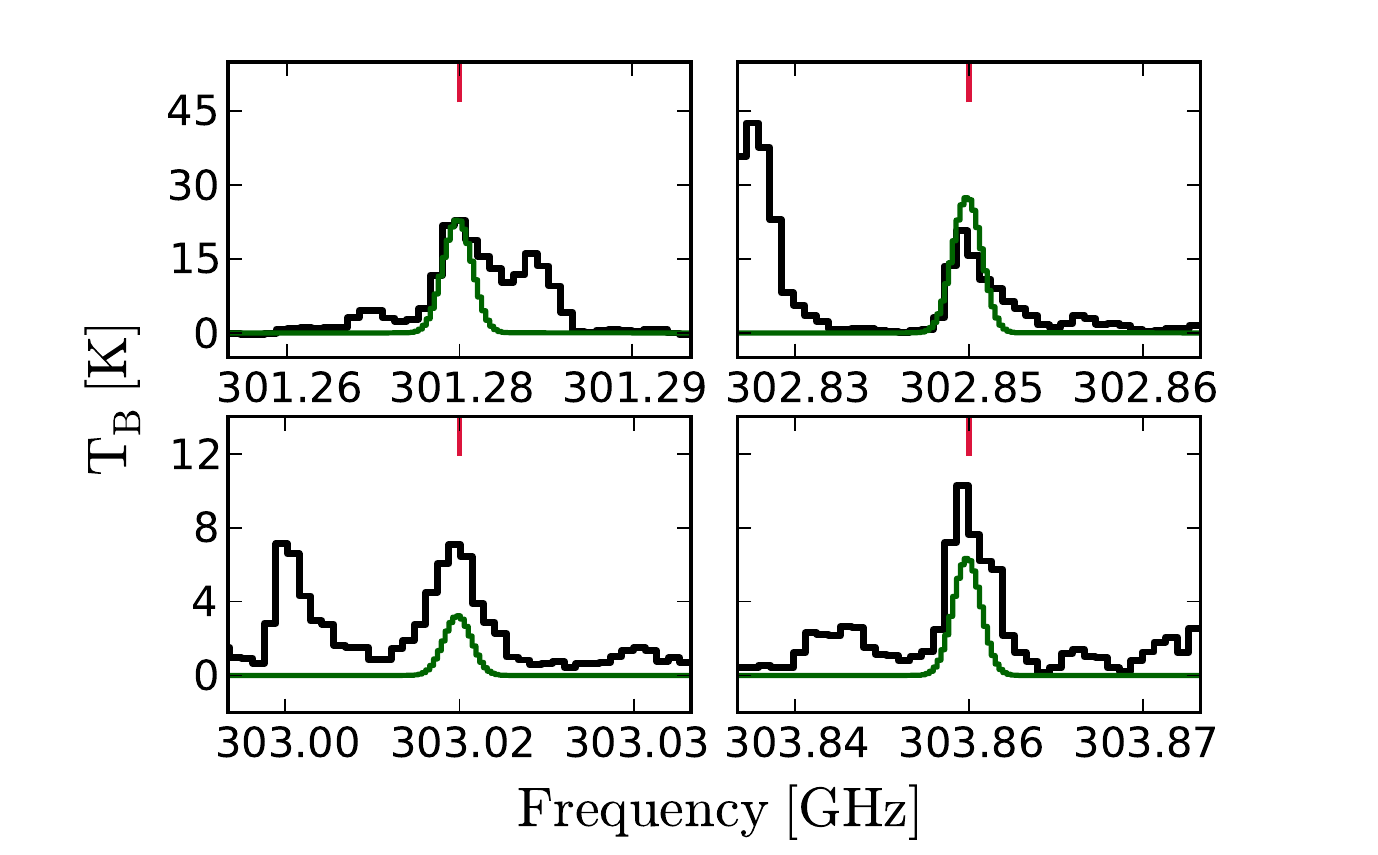}
		\caption{MM1 \RN{5}}
		\label{fig:MM1V_All18O}
	\end{subfigure}
	\caption{All CH$_3^{18}$OH lines detected towards NGC 6334\RN{1} MM1 \RN{1}-\RN{5}. Frequencies are shifted to the rest frame of the individual regions. Blue and green lines represent the modelled spectra of CH$_3^{18}$OH in each region with and without blending, i.e., including and excluding the contribution from O$^{13}$CS, respectively. For regions MM1 \RN{3}-\RN{5} only the pure-CH$_3^{18}$OH fits are shown.}
	\label{fig:MM1_All18O}
\end{figure*}
%--------------END FIGURE-------------------------------------

%--------------BEGIN FIGURE: All 18O lines towards MM2 -------------------------------
\begin{figure*}[]
	\centering
	\begin{subfigure}[]{0.85\textwidth}
		\centering
		\includegraphics[width=0.73\textwidth, trim={0 15.6cm 5.3cm 2cm}, clip]{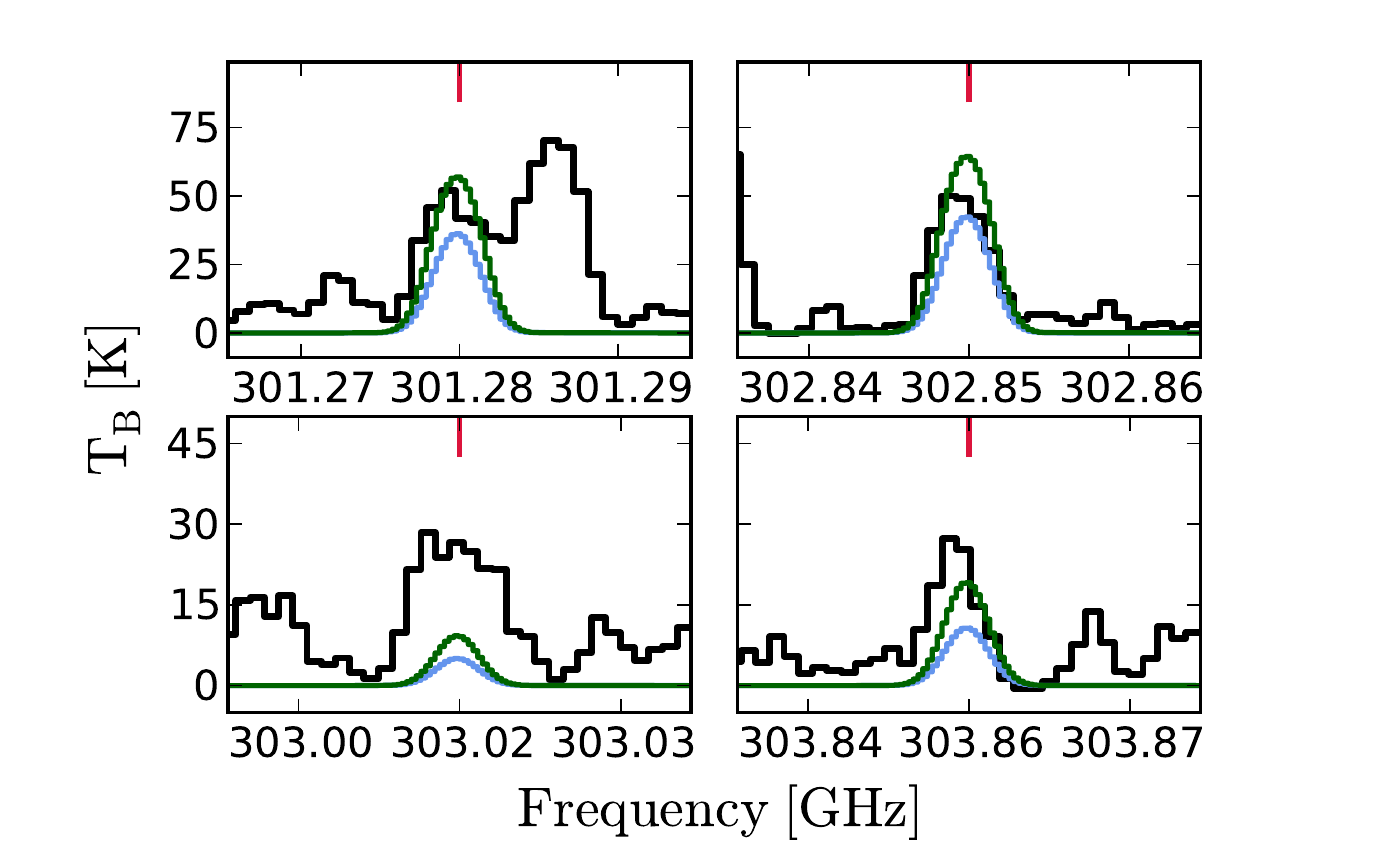}  %trim={<left> <lower> <right> <upper>}
		\caption{MM2 \RN{1}}
		\label{fig:MM2I_All18O}
	\end{subfigure}
	\begin{subfigure}{0.85\textwidth}
		\centering
		\includegraphics[width=0.73\textwidth, trim={0 15.6cm 5.3cm 2cm}, clip]{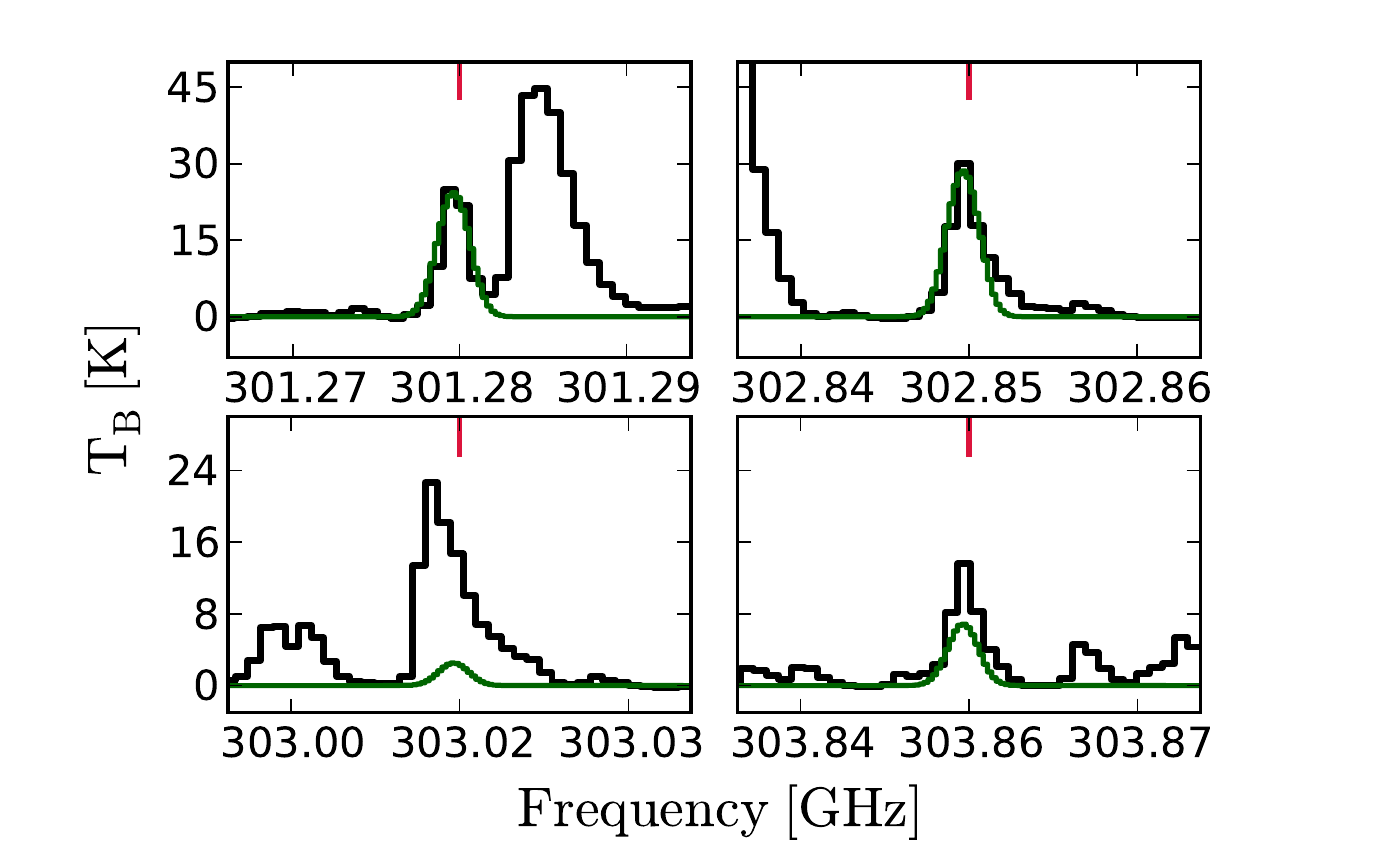}
		\caption{MM2 \RN{2}}
		\label{fig:MM2II_All18O}
	\end{subfigure}
	\caption{All CH$_3^{18}$OH lines detected towards NGC 6334\RN{1} MM2 \RN{1}-\RN{2}. Frequencies are shifted to the rest frame of the individual regions. Blue and green lines represent the modelled spectra of CH$_3^{18}$OH in each region with and without blending, i.e., including and excluding the contribution from O$^{13}$CS, respectively.  For region MM2 \RN{2} only the pure-CH$_3^{18}$OH fit is shown.}
	\label{fig:MM2_All18O}
\end{figure*}
%--------------END FIGURE-------------------------------------

%--------------BEGIN FIGURE: All 18O lines towards MM3 -------------------------------
\begin{figure*}[]
	\centering
	\begin{subfigure}[]{0.85\textwidth}
		\centering
		\includegraphics[width=0.73\textwidth, trim={0 15.6cm 5.3cm 2cm}, clip]{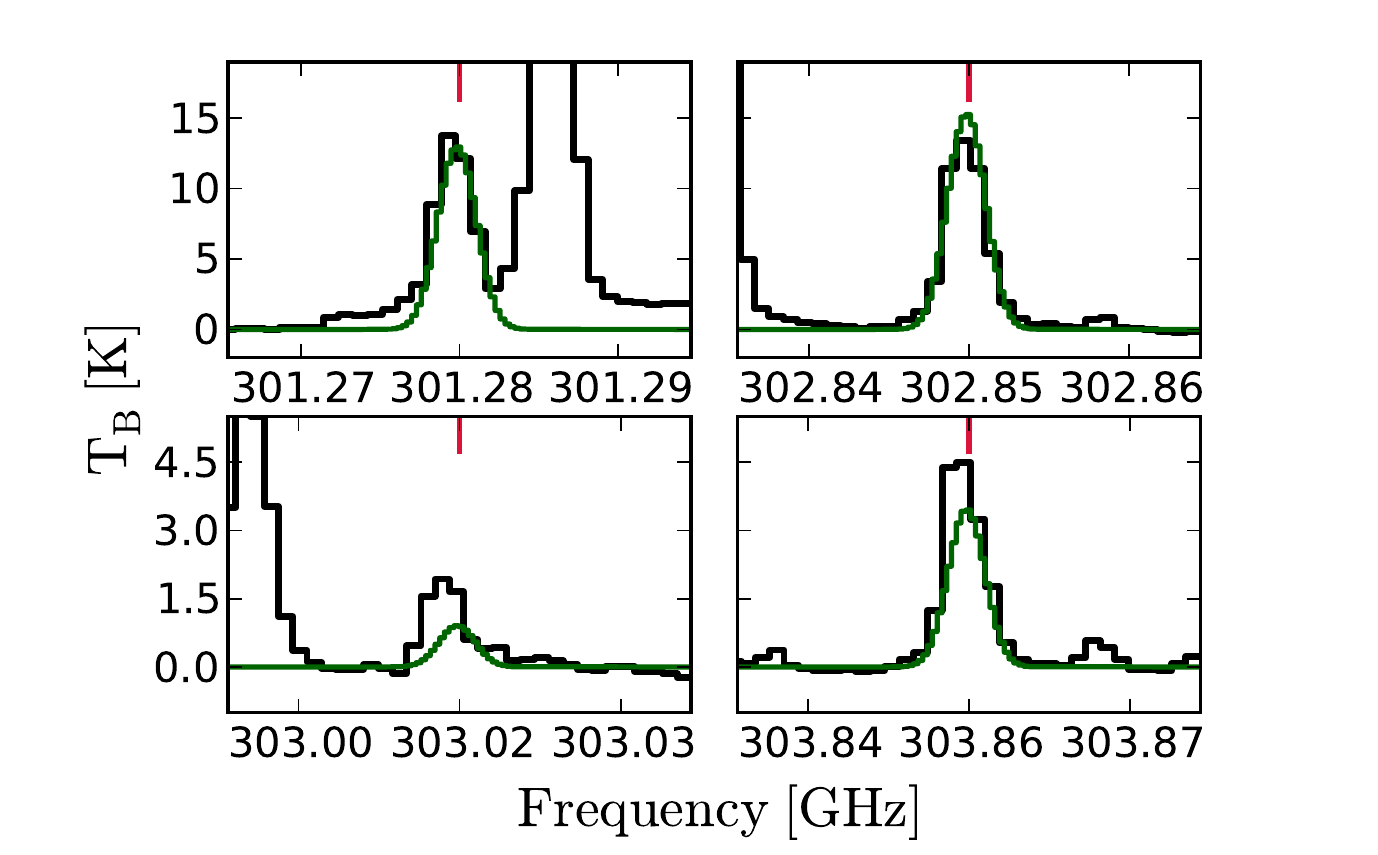}  %trim={<left> <lower> <right> <upper>}
		\caption{MM3 \RN{1}}
		\label{fig:MM3I_All18O}
	\end{subfigure}
	\begin{subfigure}{0.85\textwidth}
		\centering
		\includegraphics[width=0.73\textwidth, trim={0 15.6cm 5.3cm 2cm}, clip]{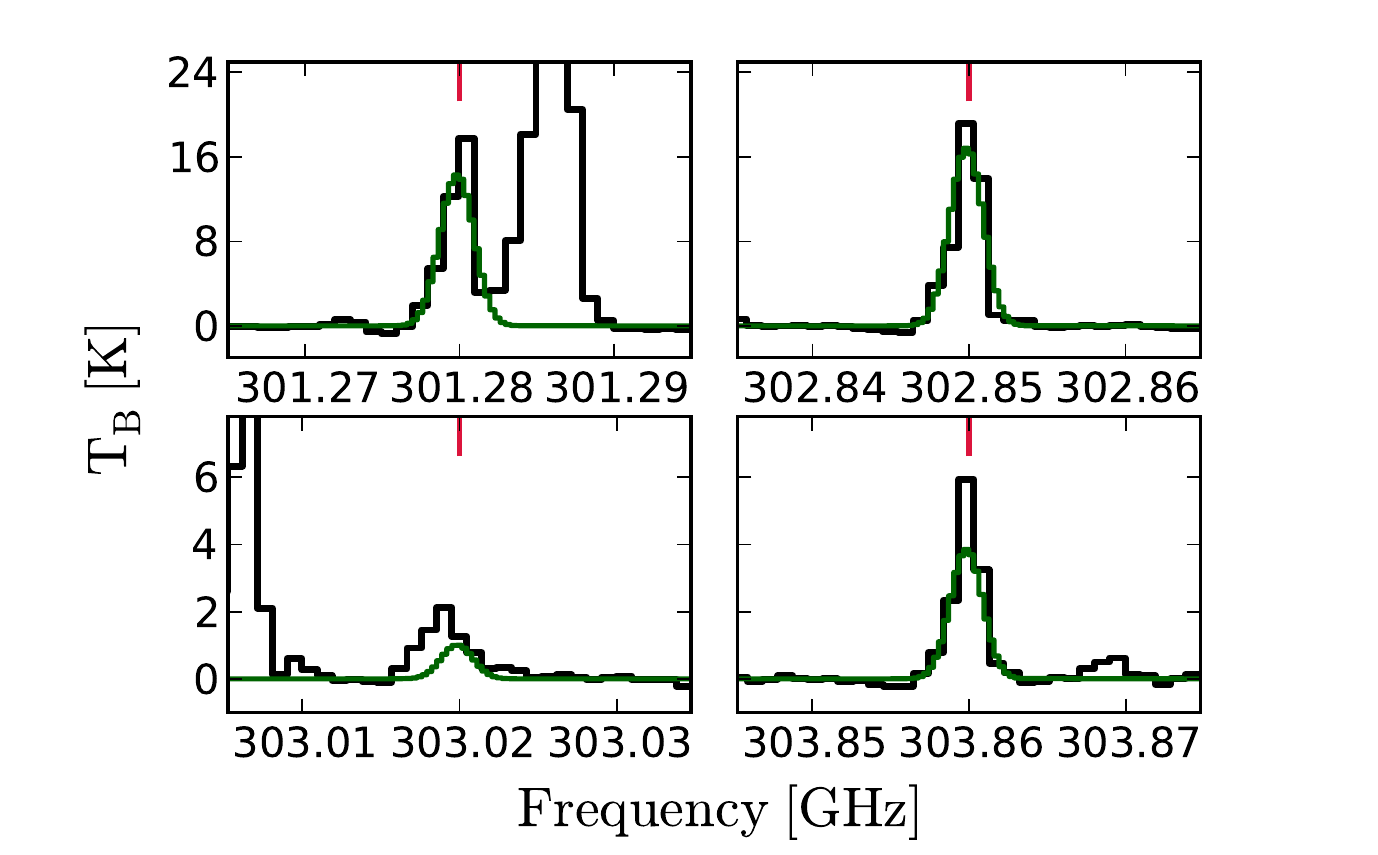}
		\caption{MM3 \RN{2}}
		\label{fig:MM3II_All18O}
	\end{subfigure}
	\caption{All CH$_3^{18}$OH lines detected towards NGC 6334\RN{1} MM3 \RN{1}-\RN{2}. Frequencies are shifted to the rest frame of the individual regions. Green lines represent the modelled spectra of CH$_3^{18}$OH in each region without blending, i.e., excluding the contribution from O$^{13}$CS.}
	\label{fig:MM13_All18O}
\end{figure*}
%--------------END FIGURE-------------------------------------
\clearpage
\section{CH$_3$OD transitions} \label{app:OD}
%--------------BEGIN FIGURE: OD lines towards MM1 -------------------------------
\begin{figure*}[b]
	\centering
	\begin{subfigure}[]{0.42\textwidth}
		\centering
		\includegraphics[width=0.95\textwidth, trim={0 15.8cm 10.2cm 2.5cm}, clip]{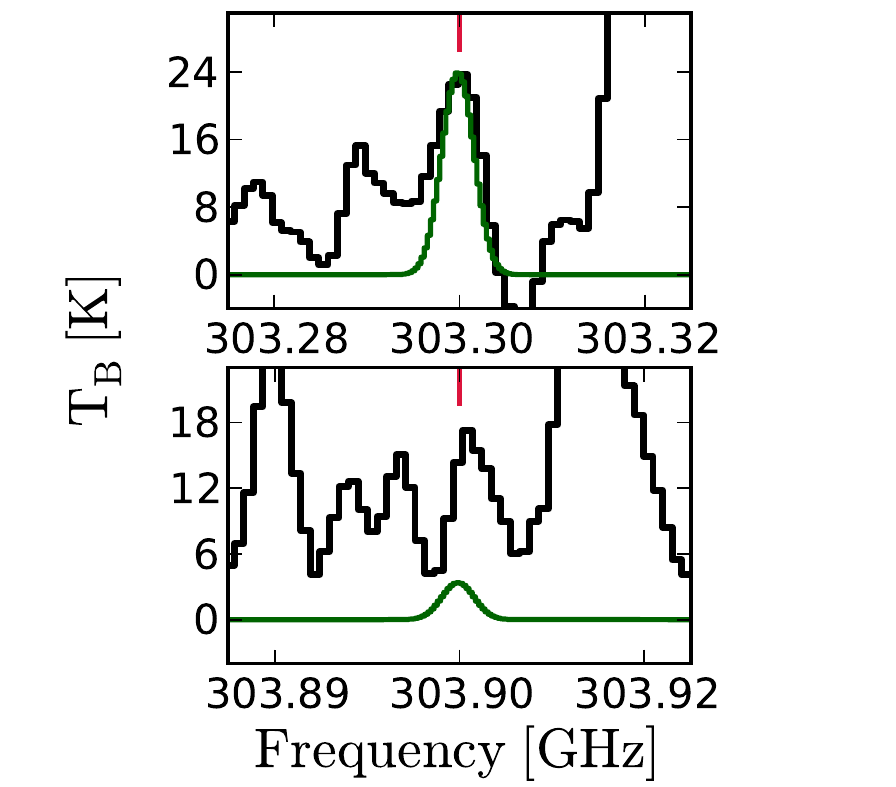}  %trim={<left> <lower> <right> <upper>}
		\caption{MM1 \RN{1}}
		\label{fig:MM1I_AllOD}
	\end{subfigure}
	\begin{subfigure}{0.42\textwidth}
		\centering
		\includegraphics[width=0.95\textwidth, trim={0 15.8cm 10.2cm 2.5cm}, clip]{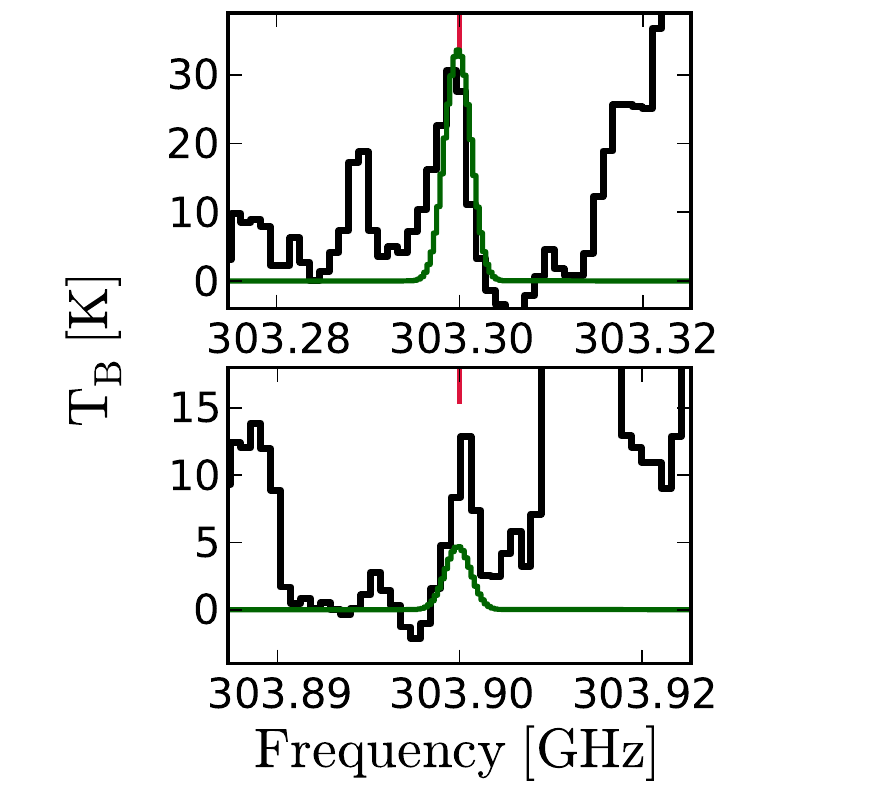}
		\caption{MM1 \RN{2}}
		\label{fig:MM1II_AllOD}
	\end{subfigure}
	\begin{subfigure}{0.42\textwidth}
		\centering
		\includegraphics[width=0.95\textwidth, trim={0 15.8cm 10.2cm 2.5cm}, clip]{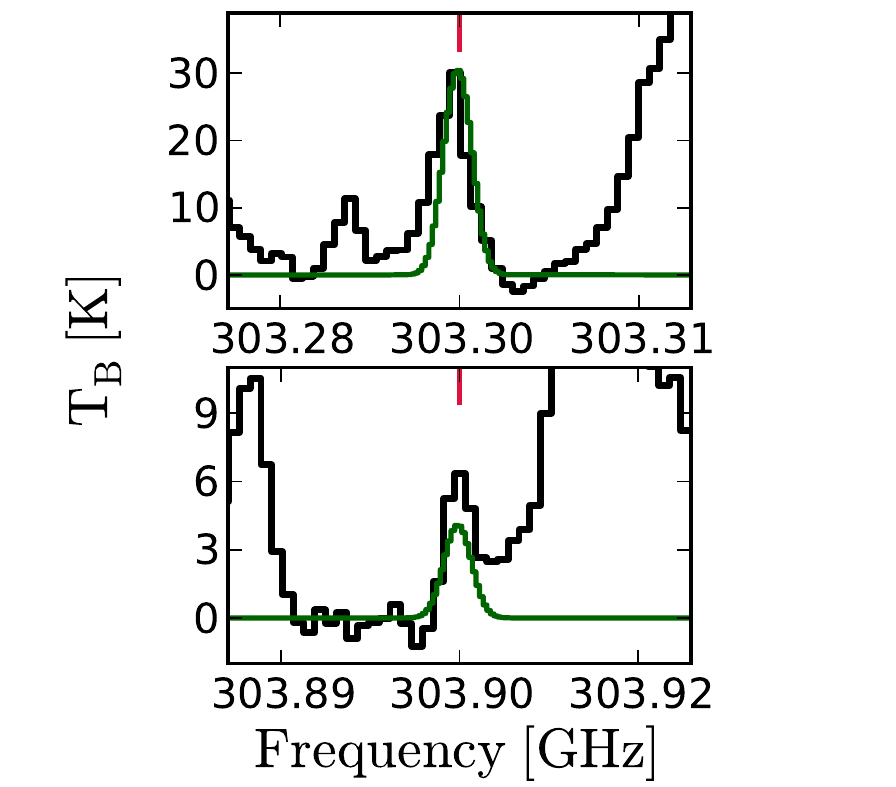}
		\caption{MM1 \RN{3}}
		\label{fig:MM1III_AllOD}
	\end{subfigure}
	\begin{subfigure}{0.42\textwidth}
		\centering
		\includegraphics[width=0.95\textwidth, trim={0 15.8cm 10.2cm 2.5cm}, clip]{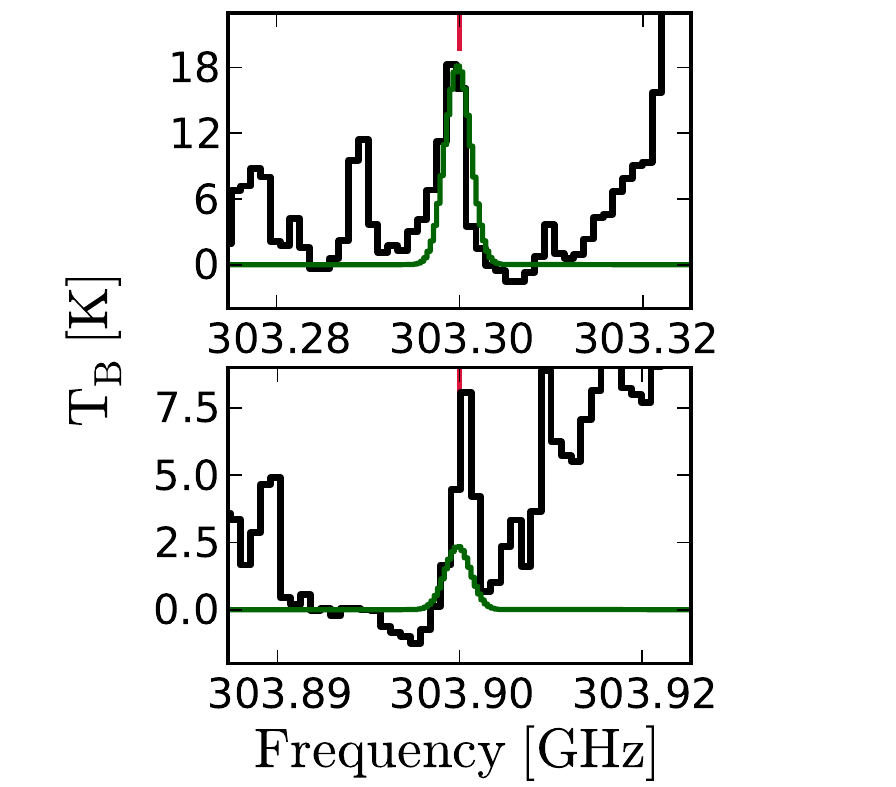}
		\caption{MM1 \RN{4}}
		\label{fig:MM1IV_AllOD}
	\end{subfigure}
	\begin{subfigure}{0.42\textwidth}
		\centering
		\includegraphics[width=0.95\textwidth, trim={0 15.8cm 10.2cm 2.5cm}, clip]{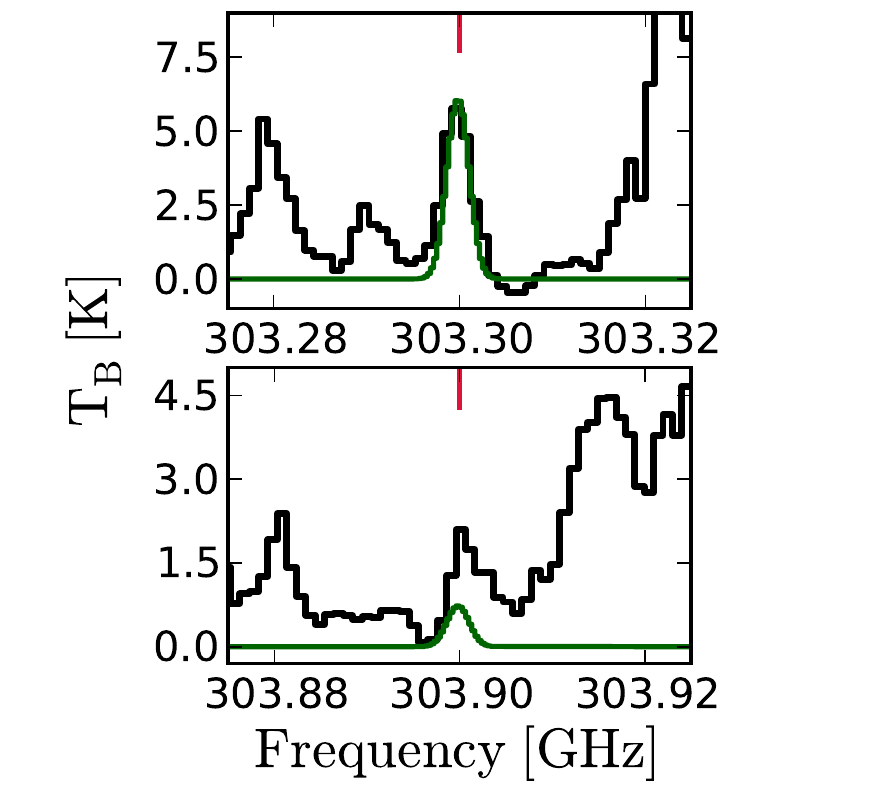}
		\caption{MM1 \RN{5}}
		\label{fig:MM1V_AllOD}
	\end{subfigure}
	\caption{All CH$_3$OD lines detected towards NGC 6334\RN{1} MM1 \RN{1}-\RN{5}. Frequencies are shifted to the rest frame of the individual regions. Green lines represent the modelled spectra of CH$_3$OD in each region.}
	\label{fig:MM1_AllOD}
\end{figure*}
%--------------END FIGURE-------------------------------------

%--------------BEGIN FIGURE: All 18O lines towards MM2 -------------------------------
\begin{figure*}[]
	\centering
	\begin{subfigure}[]{0.42\textwidth}
		\centering
		\includegraphics[width=0.95\textwidth, trim={0 15.8cm 10.2cm 2.5cm}, clip]{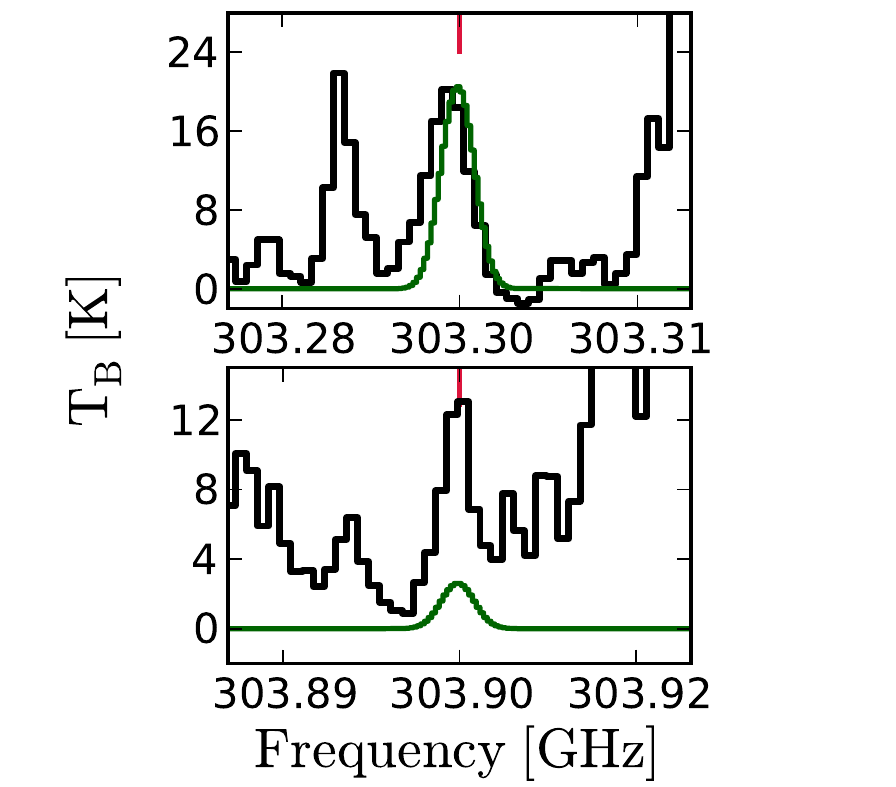}  %trim={<left> <lower> <right> <upper>}
		\caption{MM2 \RN{1}}
		\label{fig:MM2I_AllOD}
	\end{subfigure}
	\begin{subfigure}{0.42\textwidth}
		\centering
		\includegraphics[width=0.95\textwidth, trim={0 15.8cm 10.2cm 2.5cm}, clip]{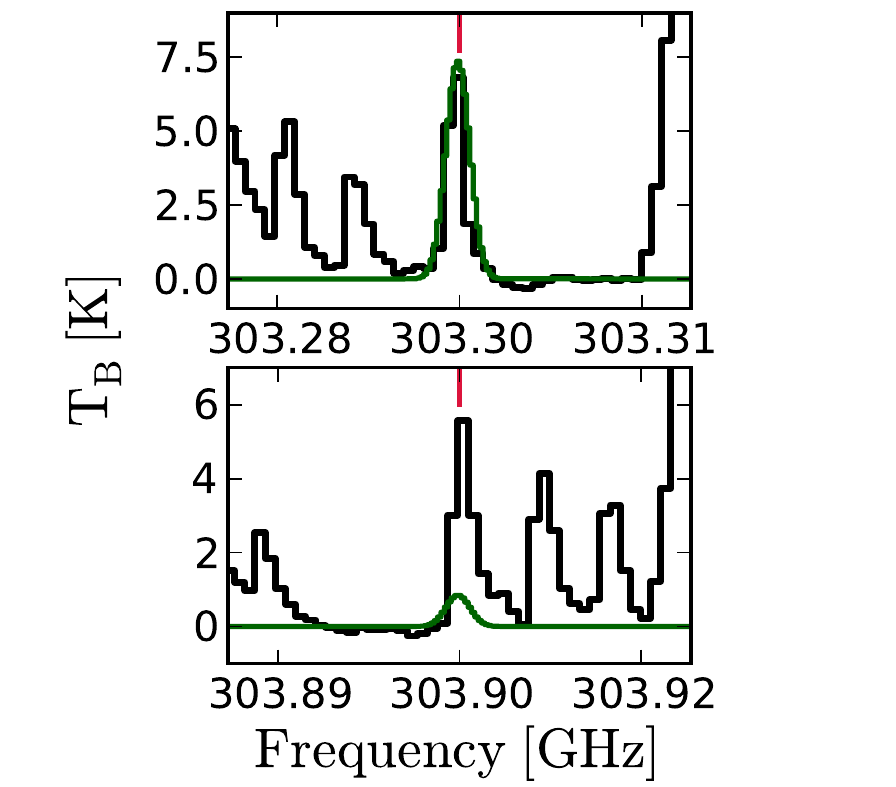}
		\caption{MM2 \RN{2}}
		\label{fig:MM2II_AllOD}
	\end{subfigure}
	\caption{All CH$_3$OD lines detected towards NGC 6334\RN{1} MM2 \RN{1}-\RN{2}. Frequencies are shifted to the rest frame of the individual regions. Green lines represent the modelled spectra of CH$_3$OD in each region.}
	\label{fig:MM2_AllOD}
\end{figure*}
%--------------END FIGURE-------------------------------------

%--------------BEGIN FIGURE: All 18O lines towards MM3 -------------------------------
\begin{figure*}[]
	\centering
	\begin{subfigure}[]{0.42\textwidth}
		\centering
		\includegraphics[width=0.95\textwidth, trim={0 15.8cm 10.2cm 2.5cm}, clip]{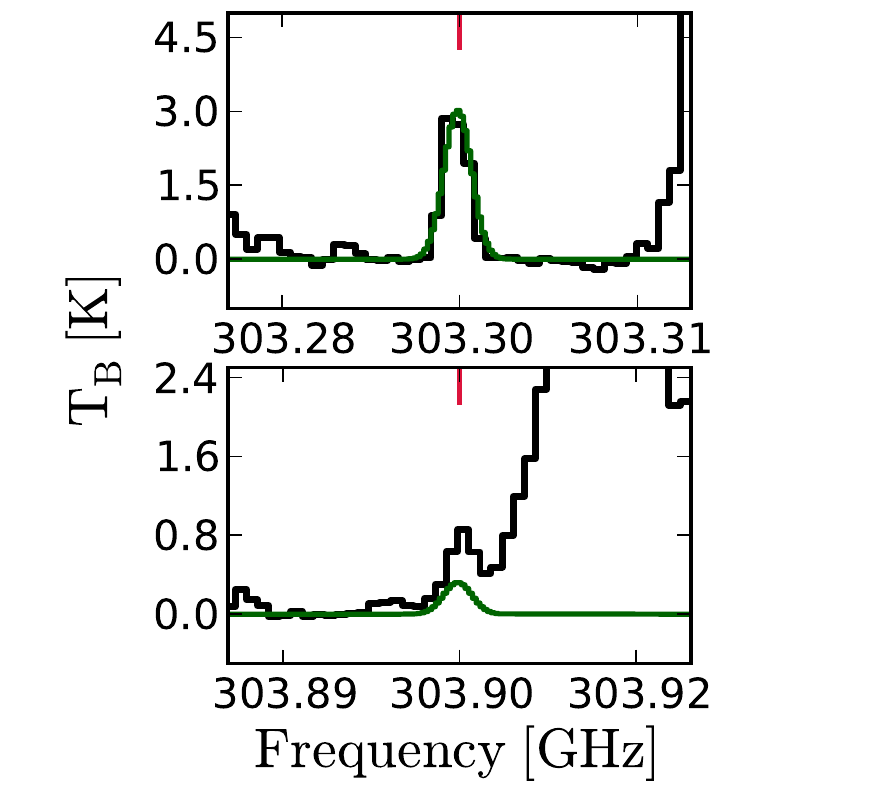}  %trim={<left> <lower> <right> <upper>}
		\caption{MM3 \RN{1}}
		\label{fig:MM3I_AllOD}
	\end{subfigure}
	\begin{subfigure}{0.42\textwidth}
		\centering
		\includegraphics[width=0.95\textwidth, trim={0 15.8cm 10.2cm 2.5cm}, clip]{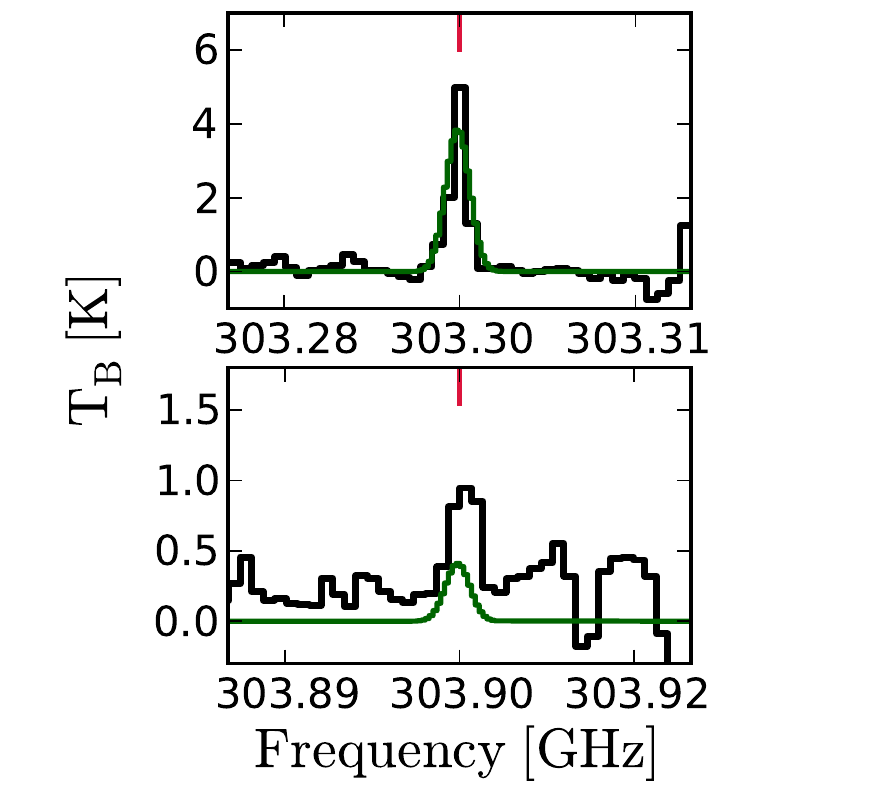}
		\caption{MM3 \RN{2}}
		\label{fig:MM3II_AllOD}
	\end{subfigure}
	\caption{All CH$_3$OD lines detected towards NGC 6334\RN{1} MM3 \RN{1}-\RN{2}. Frequencies are shifted to the rest frame of the individual regions. Green lines represent the modelled spectra of CH$_3$OD in each region.}
	\label{fig:MM3_AllOD}
\end{figure*}
%--------------END FIGURE-------------------------------------
\end{document}